\documentclass[longauth]{aa}
\usepackage[varg]{txfonts}
\usepackage[utf8]{inputenc}

\usepackage{graphicx}
\usepackage{xcolor}
\usepackage{verbatim}

\usepackage{afterpage}
\usepackage{soul}
\usepackage{natbib}
\usepackage[breaklinks=true,colorlinks=true,linkcolor=blue,citecolor=blue]{hyperref}
\usepackage{booktabs}
\usepackage{threeparttable}
\usepackage{ulem}

\newcommand{\teff}{\ensuremath{T_\textnormal{eff}}}
\newcommand{\logg}{\ensuremath{\log(g)}}
\newcommand{\logZ}{\ensuremath{\log(Z/Z_{\odot})}}
\newcommand{\mj}{\ensuremath{\textnormal{M}_{\textnormal{Jup}}}}
\newcommand{\rj}{\ensuremath{\textnormal{R}_{\textnormal{Jup}}}}
\newcommand{\mearth}{\ensuremath{\textnormal{M}_{ \oplus
}}}
\definecolor{desertsand}{rgb}{0.9, 0.6, 0.3}

\usepackage{tikz}     
\usetikzlibrary{arrows,shapes}
\usepackage[framemethod=TikZ]{mdframed}
\usepackage{marvosym}
\usepackage{tikzsymbols}

\begin{document} 

   \title{In-depth direct imaging and spectroscopic characterization of the young Solar System analog HD~95086 }
   \authorrunning{Desgrange et al.}
   \author{C.~Desgrange\inst{1}\fnmsep\inst{2}\fnmsep\inst{3}, G.~Chauvin\inst{1}\fnmsep\inst{3}, V.~Christiaens\inst{4}, 
             F.~Cantalloube\inst{5}, L.-X.~Lefranc\inst{6},  H.~Le~Coroller\inst{5}, P.~Rubini\inst{7},  G.~P.~P.~L.~Otten\inst{5}\fnmsep\inst{8},  H.~Beust\inst{1}, M.~Bonavita\inst{9}, P.~Delorme\inst{1}, M.~Devinat\inst{7}, R.~Gratton\inst{10}, A.-M.~Lagrange\inst{1}\fnmsep\inst{11}, M.~Langlois\inst{12}, D.~Mesa\inst{10}, J.~Milli\inst{1}, J.~Szul\'agyi \inst{13}, M.~Nowak\inst{14}, L.~Rodet\inst{15}, P.~Rojo\inst{3}, S.~Petrus\inst{1}, M.~Janson\inst{16}, T.~Henning\inst{2},  Q.~Kral\inst{11}, R.~G.~van~Holstein\inst{17}, F.~Ménard\inst{1},
            J.-L.~Beuzit\inst{6}, B.~Biller\inst{2}\fnmsep\inst{18}\fnmsep\inst{19}, A.~Boccaletti\inst{11}, M.~Bonnefoy\inst{1}, S.~Brown\inst{2}, A.~Costille\inst{5}, A.~Delboulbe\inst{1}, S.~Desidera\inst{10}, V.~D'Orazi\inst{10}, M.~Feldt\inst{3},  T.~Fusco\inst{20}, 
            R.~Galicher\inst{11}, J.~Hagelberg\inst{21},  C.~Lazzoni\inst{22}, R.~Ligi\inst{23}, A.-L.~Maire\inst{1}, S.~Messina\inst{24}, M.~Meyer\inst{25}, A.~Potier\inst{26},  J.~Ramos\inst{2}, D.~Rouan\inst{11}, T.~Schmidt\inst{11}, A.~Vigan\inst{5}, A.~Zurlo\inst{5}\fnmsep\inst{27}\fnmsep\inst{28}
          }
          

   \institute{
    $^{1}$ Univ. Grenoble Alpes, CNRS, IPAG, F-38000 Grenoble, France;  e-mail: celia.desgrange@univ-grenoble-alpes.fr \\
    $^{2}$ Max Planck Institute for Astronomy, K\"onigstuhl 17, D-69117 Heidelberg, Germany\\ 
    $^{3}$ Unidad Mixta Internacional Franco-Chilena de Astronom\'{i}a, CNRS/INSU UMI 3386 and Departamento de Astronom\'{i}a, Universidad de Chile, Casilla 36-D, Santiago, Chile\\
    $^{4}$ Space sciences, Technologies and Astrophysics Research (STAR) Institute, Universit\'e de Li\`ege, 19 All\'ee du Six Août, 4000 Li\`ege, Belgium\\
    $^{5}$ Aix Marseille Univ, CNRS, CNES, LAM, Marseille, France\\
    $^{6}$ Univ. Paris-Saclay, ENS de Paris-Saclay, Paris-Saclay, France \\
    $^{7}$ Pixyl, 5 Avenue du Grand Sablon, 38700 La Tronche, France\\
    $^{8}$ Academia Sinica Institute of Astronomy and Astrophysics, 11F Astronomy-Mathematics Building, NTU/AS campus, No. 1, Section 4, Roosevelt Rd., Taipei 10617, Taiwan\\
    $^{9}$ School of Physical Sciences, The Open University, Walton Hall, Milton Keynes MK7 6AA, UK\\
    $^{10}$ INAF - Osservatorio Astronomico di Padova, Vicolo dell’ Osservatorio 5, 35122, Padova, Italy\\
    $^{11}$ LESIA, Observatoire de Paris, Universit\'e PSL, CNRS, Sorbonne Université, Université de Paris, 5 place Jules Janssen, 92195 Meudon, France \\
    $^{12}$ CRAL, UMR 5574, CNRS, Universit\'e de Lyon, \'Ecole Normale Sup\'erieure de Lyon, 46 All\'ee d'Italie, F-69364 Lyon Cedex 07, France\\
    $^{13}$ Institute for Particle Physics and Astrophysics, ETH Z\"urich, Wolfgang-Pauli-Strasse 27, CH-8093, Zürich, Switzerland
    $^{14}$ Institute of Astronomy, University of Cambridge, Madingley Road, Cambridge CB3 0HA, United Kingdom \\
    $^{15}$ Cornell Center for Astrophysics and Planetary Science, Department of Astronomy, Cornell University, Ithaca, NY 14853, USA \\
    $^{16}$ Institutionen f\"or astronomi, Stockholms Universitet, Stockholm, Sweden \\
    $^{17}$ European Southern Observatory, Alonso de Cordova 3107, Casilla 19001, Vitacura, Santiago, Chile \\ 
    $^{18}$ SUPA, Institute for Astronomy, University of Edinburgh, Blackford Hill, Edinburgh EH9 3HJ, UK \\ 
    $^{19}$ Centre for Exoplanet Science, University of Edinburgh, Edinburgh, UK \\ 
    $^{20}$ DOTA, ONERA, Universit\'e Paris Saclay, F-91123, Palaiseau France \\ 
    $^{21}$ Geneva Observatory, University of Geneva, 51 ch. Pegasi, CH-1290 Versoix, Switzerland \\ 
    $^{22}$ Dipartimento di Fisica a Astronomia "G. Galilei", Universit\`a di Padova, Via Marzolo, 8, 35121 Padova, Italy \\ 
    $^{23}$ Universit\'e C\^ote d’Azur, Observatoire de la C\^ote d’Azur, CNRS, Laboratoire Lagrange, Bd de l’Observatoire, CS 34229, 06304 Nice cedex 4, France\\  
    $^{24}$ INAF–Osservatorio Astrofisico di Catania, via Santa Sofia, 78 Catania, Italy \\ 
    $^{25}$ European Southern Observatory (ESO), Karl-Schwarzschild-Str. 2, 85748 Garching, Germany \\ 
    $^{26}$ Jet Propulsion Laboratory, California Institute of Technology, 4800 Oak Grove Drive, Pasadena, CA 91109 \\ 
    $^{27}$ N\'ucleo de Astronom\'ia, Facultad de Ingenier\'ia y Ciencias, Universidad Diego Portales, Av. Ejercito 441, Santiago, Chile \\ 
    $^{28}$ Escuela de Ingeniería Industrial, Facultad de Ingenier\'ia y Ciencias, Universidad Diego Portales, Av. Ejercito 441, Santiago, Chile \\
}

   \date{}

 
  \abstract
   {HD\,95086 is a young nearby Solar System analog hosting a giant exoplanet orbiting at 57\,au from the star between an inner and outer debris belt. 
   The existence of additional planets has been suggested as the mechanism that maintains the broad cavity between the two belts.}
   {
    We present a dedicated monitoring of HD\,95086 with the VLT/SPHERE instrument to refine the orbital and atmospheric properties of HD\,95086\,b, and to search for additional planets in this system.}
   { SPHERE observations, spread over ten epochs from 2015 to 2019 and including five new datasets, were used.  
   Combined with archival observations, from VLT/NaCo (2012--2013) and Gemini/GPI (2013--2016), the extended set of astrometric measurements allowed us to refine the orbital properties of HD\,$95086$\,b. We also investigated the spectral properties and the presence of a circumplanetary disk around HD\,95086\,b by using the \texttt{special} fitting tool exploring the diversity of several atmospheric models. In addition, we improved our detection limits in order to search for a putative planet c via the \texttt{K-Stacker} algorithm. }
   {
   We extracted for the first time the JH low-resolution spectrum of HD~95086~b by stacking the six best epochs, and confirm its very red spectral energy distribution. Combined with additional datasets from GPI and NaCo, our analysis indicates that this very red color can be explained by  the presence of a circumplanetary disk around  planet b, with a range of high-temperature solutions ($1400$--$1600$ K) and significant extinction ($A_V\,\gtrsim\,10\,\rm\,mag$), or by a super-solar metallicity atmosphere with lower temperatures ($800$--$1300$ K), and small to medium amount of extinction ($A_V\,\lesssim\,10\,\rm\,mag$). We do not find any robust candidates for   planet c, but give updated constraints on its potential mass and location. 
   }
   {}

   \keywords{Instrumentation: adaptive optics, high angular resolution -- Methods: observational -- Stars: individual: HD95086 -- Planetary systems}

    \titlerunning{HD 95086}
   \maketitle

\defcitealias{Madhusudhan2011}{M11}

\section{Introduction}

Direct imaging has proven to be successful at imaging and characterizing the properties of young planetary system architectures, and young ($\leq100$ Myr) gaseous giant planets orbiting at  large distances from their host stars ($a\, \gtrsim 10$ au). 
Large-scale surveys of several hundreds of young nearby stars ($<150~\mathrm{pc}$), such as 
SHINE \citep{Chauvin2017_shine,Desidera2021_shine_paperI_sample_definition,Langlois2021_shine_paperII_observations,Vigan2021_shine_demographics} and GPIES \citep{nielsen2019gpies}, have now extended our vision of the giant planet demographics down to 10\,au. The goal is to access the bulk of the giant planet population close to the snowline and to bridge the gap with complementary indirect detection methods such as radial velocity, transit, micro-lensing, and soon astrometry (with the Gaia Data Release 4) that are sensitive to planets closer to their stars ($\lesssim 10$ au). Over the last two decades discoveries of emblematic planetary systems such as HR~8799 \citep{Marois2008_HR8799,Marois2010_8799}, $\beta$~Pictoris \citep{lagrange2010_betapictoris}, HD~95086 \citep{Rameau2013b}, 51~Eri \citep{Macintosh2015}, and PDS~70 \citep{keppler2018_PDS70,Muller2018} have offered  rich opportunities to explore the diversity of young Solar System analogs, containing giant planets and circumstellar disks shaped with cavities and belts. HR~8799 hosts four exoplanets \citep{Marois2008_HR8799,Marois2010_8799}, whereas 51 Eridani hosts one exoplanet \citep{Macintosh2015}. A second planet has been recently detected by the radial velocity technique in the $\beta$~Pictoris system \citep{2019NatAs...3.1135L} and confirmed with interferometric observations \citep{2020A&A...642L...2N},  while a second forming planet 
has been detected with MUSE in the PDS~70 system \citep{Haffert2019_pds70c}. The mass of these protoplanets is highly uncertain
, with estimates ranging from $1$ to $17$\,$\mj$ \citep{Muller2018,Christiaens2019_PDS70,Mesa2019a,Isella2019,Stolker2020,Wang2021}.
These young, planetary systems are benchmark laboratories for exploring the formation and evolution of young giant planets with the current large telescopes and instruments. They are prime targets for upcoming telescopes, such as the  James Webb Space Telescope (JWST, first light 2022) and the Extremely Large Telescope (ELT, first light 2027). From that perspective, the hunt for additional planets in the young system HD\,95086 is very interesting.

\begin{figure}[t]
     \centering \includegraphics[width=\columnwidth]{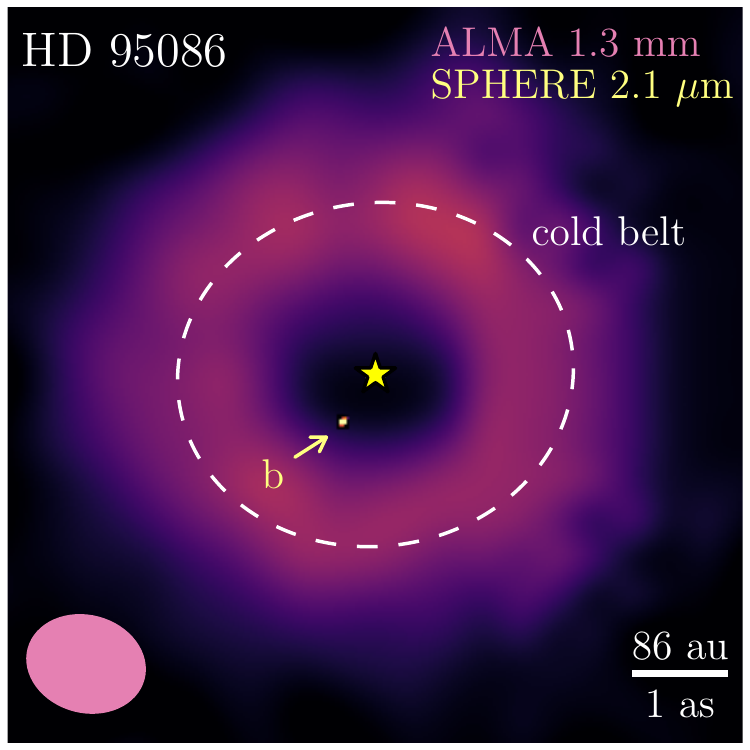}
    \caption{Composite ALMA-continuum (at $1.3$ mm) and SPHERE/IRDIS(at $2.1\,\mu$m) observations of HD\,95086 \citep[see][]{Su2017,Chauvin2018}. The exoplanet b is detected at the K1 band (red dot). The \textit{white dashed} ring at 180\,au represents the peak location of the outer cold belt located from $106\pm6$\,au to $320\pm20$\,au. The inner warm belt is not resolved with ALMA. The pink ellipse represents the ALMA synthetic beam.
    \label{fig:alma}}
\end{figure}

Since 2013, and since the discovery of a $4$--$5~\mj$ exoplanet HD\,$95086$\,b in thermal imaging using the NaCo instrument \citep{Rousset2003_naos, Lenzen_2003_naos} at the Very Large Telescope (VLT), the HD~$95086$ planetary system has become a reference to investigate the processes of planetary formation and evolution, and to characterize young planetary architectures. The star is an A-type star with an approximate effective temperature of $7 750$ K and a mass of $1.6$\,M$_\odot$. Until very recently, the star was identified as a young star located at the border of the Lower Centaurus Crux (LCC) association and thus with an age of $17 \pm 2$ Myr \citep{Pecaut2012_age}. However, based on the Gaia Data Release 2, \cite{Booth2020_Age} showed that the star could instead belong to the Carina association, which according to their age estimation would make it a few million years younger, $13.3^{+1.1}_{-0.6}~\mathrm{Myr}$. The physical parameters of the star are summarized in Table\,\ref{tab:para_sys}.

HD 95086 hosts a double-belt debris disk architecture, very similar to that of our Solar System, with an outer belt resolved by the Atacama Large Millimeter Array (ALMA) in the continuum at 1.3\,mm (\citealp{Su2017}; see Fig.\,\ref{fig:alma}). The inner warm belt is located at $8\pm2~\mathrm{au}$ ($187\pm26~\mathrm{K}$), and the large outer, colder belt between $106\pm6~\mathrm{au}$ and $320\pm20~\mathrm{au}$ ($57\pm2~\mathrm{K}$). Their existence was originally identified from the analysis of Herschel observations, in combination with the characterization of the spectral energy distribution (SED) of HD\,95086 \citep{Moor2013}.  Based on SED modeling from Herschel, Spitzer, WISE, and APEX observations, the existence of a third belt at $2~\mathrm{au}$ ($300~\mathrm{K}$), has been also proposed by \cite{Su2015}, together with a disk halo component that could extend up to $800~\mathrm{au}$ \citep{Su2017}, but this innermost belt has not been confirmed to date. 
Recently, \citet{Zapata2018} added new constraints on the structure of the outer belt at submillimeter and millimeter wavelengths with ALMA observations at $0.9$~mm and $1.3$~mm, and derived a dust-to-gas ratio $\geq 50$. 
Moreover, \citet{Zapata2018} and \citet{2019Booth} did not detect CO (J=$2$--$1$) and (J=$3$--$2$) emissions, excluding the possibility of HD 95086 being an evolved gaseous primordial disk. By using spectro-spatial filter on ALMA observations, \citet{2019Booth} found tentative evidence of CO (J=$2$--$1$) emission with an integrated line flux of $9.5\pm3.6~\mathrm{mJy\cdot km\cdot s^{-1}}$. It corresponds to a CO mass of ($1.4$--$13$)$\,\cdot\,10^{-6}~\mearth$, which they determined to be consistent with second-generation production of gas through collisional cascade \citep{Kral2017}. 
According to \citet{Su2015}, the collisions in the  HD\,95086 disk might also explain their detection of a $69~\mathrm{\mu m}$ crystalline olivine feature from the outer disk with the Spitzer telescope as the crystallization of olivine requires a high temperature,   as is the case for instance in the core of planetary bodies after their disruption. Finally, the outer belt has also been marginally detected in polarized scattered light in the near-infrared (J band) by SPHERE differential polarimetric imaging (DPI) observations \citep{Chauvin2018} colocated with the thermal emission seen by ALMA.  
The physical parameters of the debris disk architecture are given in Table\,\ref{tab:para_sys}.

At infrared wavelengths, following the discovery of HD\,$95086$\,b with VLT/NaCo \citep{Rameau2013b} in the L' band  ($3.8~\mathrm{\mu m}$), the planet was re-imaged using Gemini/GPI 
 in the H band ($1.5$--$1.8~\mathrm{\mu m}$) and in the K1 band ($1.9$--$2.2~\mathrm{\mu m}$) \citep{Rameau2016,Derosa2016}, and using VLT/SPHERE 
with IRDIS 
in H2H3 filters ($\lambda_{H2}=1.593~\mathrm{\mu m}$, $\lambda_{H3}=1.667~\mathrm{\mu m}$) and in K1K2 filters ($\lambda_{K1}=2.103~\mathrm{\mu m}$, $\lambda_{K2}=2.255~\mathrm{\mu m}$), and with the integral field spectrograph (IFS). 
in the YJ ($0.95$--$1.35\,\mu$m), and YJH ($0.97$--$1.66\,\mu$m) settings \citep{Chauvin2018}. The combination of different photometric measurements in the infrared enabled \citet{Derosa2016} and \citet{Chauvin2018} to confirm the late L spectral type of HD~95086~b, which is consistent with a dusty atmosphere of about $800$--$1300~\mathrm{K}$.

The first orbital fitting of HD~95086~b was performed by \citet{Rameau2016} from previous NaCo astrometric data \citep[epochs 2012 to 2013 from][]{Rameau2013b} and GPI astrometric monitoring between $2013$ and $2016$ \citep[published partially in][]{Galicher2014}. They found a  semimajor axis of $62^{+21}_{-8}~\mathrm{au}$, an eccentricity less than $0.21$, and an inclination of $153^{+10}_{-14}$\,$\degr$ at the  $68\%$ confidence interval by using Monte Carlo methods. \citet{Chauvin2018} updated the orbital solution using a larger orbital coverage,   this time combining NaCo and SPHERE astrometric measurements.
This recent MCMC analysis gave consistent results and showed that the planet is orbiting with a period of about $289^{+12}_{-177}$ years, a semimajor axis of $52^{+13}_{-24}~\mathrm{au}$, a relatively low eccentricity ($0.2^{+0.3}_{-0.2}$), and with an inclination of $141^{+15}_{-13}$\,\degr\ at  the $68\%$ confidence interval, compatible with a coplanar orbit with the debris disk plane. The physical parameters of the exoplanet are summarized in Table~\ref{tab:para_sys}.
In addition, \citet{Chauvin2018} used the High Accuracy Radial velocity Planet Searcher (HARPS) high-resolution optical spectrograph to search for additional exoplanets with the radial velocity (RV) technique, 
and could exclude the presence of a very massive ($>10~\mj$), coplanar inner giant planets at less than $1~\mathrm{au}$. 

\begin{table}[t!]
   \centering
    \small
         \caption{Physical properties of the HD $95086$ system from \citet{Chauvin2018,Su2017,Su2015,Rameau2013b,Derosa2016,Bailer_Jones2018_Distance,Booth2020_Age}.
         }
    \begin{tabular}{|l|l|}
       \multicolumn{2}{c}{The HD\,95086 exoplanetary system}  \\
       \noalign{\smallskip}\hline\hline\noalign{\smallskip}
        \multicolumn{2}{c}{Star}  \\
         \hline
         Spectral type  \hspace{4cm}& A        \hspace{2cm}          \\
         Teff (K)        & $7750\pm250$  \\
         log(g)  (dex)        & $4.0\pm0.5$       \\
         Distance (pc)      & $86.2\pm0.3$    \\
         Age (Myr)            & $13.3^{+1.1}_{-0.6}$  \\
         Mass  (M$_\odot$)         & $1.6\pm0.1$     \\
         Luminosity (L$_\odot$ )    & $5.7\pm1.7$    \\
         \hline\noalign{\medskip} 
         \multicolumn{2}{c}{Debris disk}          \\ 
         \noalign{\smallskip} 
         \multicolumn{2}{c}{\it -- Innermost belt (?)} \\ 
         \hline
         Location (au)   & $2$                           \\
         Temperature (K)                & $300$                          \\
         \hline
         \noalign{\smallskip} 
         \multicolumn{2}{c}{\it -- Warm belt} \\
         \hline
         Location (au)      & $7-10$                          \\
         Temperature (K)               & $187\pm26$                     \\
         \hline
         \noalign{\smallskip} 
         \multicolumn{2}{c}{\it -- Cold belt} \\
         \hline
         Location (au)      & $106-320$                     \\
         Temperature (K)                & $57\pm2$                        \\
         Inclination (\degr)    & $30\pm3$ \\  
         Position Angle  (\degr)    & $97\pm3$\\
         \hline\noalign{\smallskip} 
         \multicolumn{2}{c}{\it -- Stellar halo (?) } \\
         \hline
         
         Location (au)  & $300-800$                      \\
         \hline\noalign{\medskip} 
         \multicolumn{2}{c}{HD\,95086\,b }  \\
         \hline
         Spectral type   & L$6\pm1$                   \\
         Teff (K)          & $800-1300$ K        \\
         log(g) (dex)          & $\lesssim 4.5$             \\
         Semimajor axis (au) & $52^{\,+13}_{\,-24}$       \\
         Eccentricity    & $0.2^{\,+0.3}_{\,-0.2}$      \\
         Inclination (\degr)    & $141^{\,+15}_{\,-13}$     \\
         Period (years)         & $289^{\,+12}_{\,-177}$      \\
         Mass ($\mj$)           & $4-5$                \\
         \hline \noalign{\smallskip}\noalign{\smallskip}        
             \end{tabular} \quad
    \label{tab:para_sys}
    
\tablefoot{$\teff$ corresponds to the effective temperature when assuming the blackbody hypothesis; $\log(g)$ corresponds to the logarithm of the surface or photosphere gravity. To avoid confusion, it should be noted  that the inclination $i$ could as well be $180\degr\,-\,i$ owing to projection on the sky.}
\end{table}

Given the large cavity seen in ALMA images inside the cold outer belt (see Fig.~\ref{fig:alma}), the system HD~95086   very likely   hosts at least one or perhaps two additional planets closer to the star than HD~95086~b, which would    explain the architecture of the two debris belts.  \citet{Su2015}, \citet{Rameau2016}, and  \citet{Chauvin2018} investigated this possibility by considering various locations, eccentricities, and masses for the inner planets, the physical properties of b, and the characteristics of the inner and outer belts together with the detection performance of current planet imagers. A configuration with one or two additional inner planets between 10 and 30\,au, dynamically stable with b carving the outer belt, and participating in the replenishment of the inner belt is possible and worth investigating. 

In this paper we extend the study of \citet{Chauvin2018} to revisit the orbital and atmospheric properties of HD~95086\,b, and the presence of additional inner giant planets, considering a total of ten epochs acquired with the VLT/SPHERE instrument \citep{Beuzit2019_sphere}  between February 2015 and May 2019. These datasets include five new unpublished epochs covering January 2018 to May 2019. 
In Sect.~\ref{sec:obs} we present the data acquired, together with the archival data used for this analysis. In Sect.~\ref{sec:reduction} we describe the image processing methods used, along with our data selection and the decision to combine the different datasets considering the individual epoch contrast performance and adaptive optics (AO) correction quality. In Sect.~\ref{sec:orbit} we present the updated astrometry for the exoplanet HD~95086~b based on VLT/NaCo and VLT/SPHERE data, covering a total of seven years of monitoring between 2012 and 2019, and determine the best orbital solution. In Sect.~\ref{sec:spectro} we present for the first time the SPHERE-JH  ($1.2$--$1.6$ $\mu$m) spectroscopic observations of HD~95086~b. Using the MCMC \texttt{special} code \citep{Christiaens2021} applied to the combined spectrum, we re-analyze the physical parameters of the planet and investigate the presence of a circumplanetary disk around it. In Sect.~\ref{sec:detection_limit} we finally look for the hypothetical exoplanets c and d in the system by updating the HARPS and SPHERE combined observations, and also by applying the K-Stacker algorithm \citep{Lecoroller2015Kstacker} to the SPHERE multi-epoch datasets.

\section{Observations \label{sec:obs}}

\begin{table*}[h!]\centering \small
\caption{ Summary of all SPHERE observations of HD $95086$ in the K1K2 (IRDIS) and YJH bands (IFS), as well as the mean observational conditions when available.
 \label{tab:obslog}}  

\setlength{\tabcolsep}{6pt}
\begin{tabular}{llllllllllll} 
\hline\hline\noalign{\smallskip}\footnotesize
Epoch & Instr. & Filter & Coronagraph & Satellite  & NDIT x DIT & N$_\text{exp}$ & $\Delta \pi$ & $\epsilon$ & $\tau_0$ & Sr & airmass \\   
                 &               &           & & spots                &number x s&       & (\degr)       & (")   & (ms)          & (\%)      &           \\  
\noalign{\smallskip}\hline\hline\noalign{\smallskip} 
2015-02-03 & IFS    & YJH       & N\_ALC\_YJH\_S& no  & $1$ x $64$ & $26$   & $22.4$        & -     & -             & -         & $1.41$    \\
2015-02-03 & IRDIS  & DB K$12$  & N\_ALC\_YJH\_S& no  & $1$ x $16$ & $26$   & $22.4$        & -     & -             & -         & $1.41$   \\
\noalign{\smallskip}\hline\noalign{\smallskip}     
2015-05-05 & IFS    & YJH       & N\_ALC\_YJH\_S& no  & $4$ x $64$  & $13$  & $18.2$        & $0.79$& $2.1$         & $75$      & $1.40$   \\
2015-05-05 & IRDIS  & DB K$12$  & N\_ALC\_YJH\_S& no  & $4$ x $64$  & $13$  & $18.2$        & $0.45$& $2.1$         & $75$      & $1.40$   \\
\noalign{\smallskip}\hline\noalign{\smallskip}     
2015-05-12 & IFS    & YJH       & N\_ALC\_YJH\_S& no  & $4$ x $64$ & no     & $22.5$        & $1.07$& $3.0$         & -         & $1.41$   \\
2015-05-12 & IRDIS  & DB H$23$  & N\_ALC\_YJH\_S& no  & $4$ x $64$ & no     & $22.5$        & $1.07$& $3.0$         & -         & $1.41$   \\
\noalign{\smallskip}\hline\hline\noalign{\smallskip}     
2016-01-18 & IFS    & YJH       & N\_ALC\_YJH\_S& no  & $5$ x $64$  & $19$  & $28.1$        & $0.33$& $1.8$         & $81$      & $1.41$   \\
2016-01-18 & IRDIS  & DB K$12$  & N\_ALC\_YJH\_S& no  & $5$ x $64$  & $19$  & $28.1$        & $0.33$& $1.8$         & $81$      & $1.41$   \\
\noalign{\smallskip}\hline\noalign{\smallskip}     
2016-05-31 & IFS    & YJH       & N\_ALC\_Ks    & yes & $10$ x $64$ & $7$   & $25.2$        & $0.64$& $3.4$         & $61$      & $1.43$   \\
2016-05-31 & IRDIS  & DB K$12$  & N\_ALC\_Ks    & yes & $10$ x $64$ & $7$   & $25.2$        & $0.64$& $3.4$         & $61$      & $1.43$   \\
\noalign{\smallskip}\hline\hline\noalign{\smallskip}     
2017-05-03 & IFS    & YJH       & N\_ALC\_Ks    & no  & $7$ x $12$  & no & $2.1$         & $2.15$& $1.6$         & $47$      & $1.40$   \\
2017-05-03 & IRDIS  & DB K$12$  & N\_ALC\_Ks    & no  & $7$ x $12$  & no & $2.1$         & $2.15$& $1.6$         & $47$      & $1.40$   \\
\noalign{\smallskip}\hline\noalign{\smallskip}     
2017-05-10 & IFS    & YJH       & N\_ALC\_Ks    & yes & $10$ x $64$ & $7$   & $36.6$        & $0.89$& $3.3$         & $73$      & $1.40$   \\
2017-05-10 & IRDIS  & DB K$12$  & N\_ALC\_Ks    & yes & $10$ x $64$ & $7$   & $36.6$        & $0.89$& $3.3$         & $73$      & $1.40$   \\
\noalign{\smallskip}\hline\hline\noalign{\smallskip}     
2018-01-06 & IFS    & YJH       & N\_ALC\_Ks    & yes & $10$ x $96$ & $7$   & $41.0$        & $0.30$& $10.1$        & $83$      & $1.40$   \\
2018-01-06 & IRDIS  & DB K$12$  & N\_ALC\_Ks    & yes & $10$ x $96$ & $7$   & $41.0$        & $0.30$& $10.1$        & $83$      & $1.40$   \\
\noalign{\smallskip}\hline\noalign{\smallskip}     
2018-02-24 & IFS    & YJH       & N\_ALC\_Ks    & no  & $4$ x $96$ & $16$   & $33.4$        & $0.38$& $9.0$         & -         & $1.41$   \\
2018-02-24 & IRDIS  & DB K$12$  & N\_ALC\_Ks    & no  & $4$ x $96$ & $16$   & $33.4$        & $0.38$& $9.0$         & -         & $1.41$   \\
\noalign{\smallskip}\hline\noalign{\smallskip}     
2018-03-28 & IFS    & YJH       & N\_ALC\_YJH\_S& no  & $4$ x $96$ & $16$   & $33.3$        & $0.51$& $9.3$         & $82$      & $1.41$   \\
2018-03-28 & IRDIS  & DB K$12$  & N\_ALC\_YJH\_S& no  & $4$ x $96$ & $16$   & $33.3$        & $0.51$& $9.3$         & $82$      & $1.41$   \\
\noalign{\smallskip}\hline\hline\noalign{\smallskip}     
2019-04-13 & IFS    & YJH       & N\_ALC\_YJH\_S& yes & $9$ x $96$ & $16$   & $33.8$        & $0.76$& $3.0$         & $70$      & $1.42$   \\
2019-04-13 & IRDIS  & DB K$12$  & N\_ALC\_YJH\_S& yes & $9$ x $96$ & $16$   & $33.8$        & $0.76$& $3.0$         & $70$      & $1.42$   \\
\noalign{\smallskip}\hline\hline\noalign{\smallskip}     
2019-05-18 & IFS    & YJH       & N\_ALC\_YJH\_S& no & $4$ x $96$ & $16$.   & $33.8$        & $0.63$& $3.2$         & $72$      & $1.42$   \\
2019-05-18 & IRDIS  & DB K$12$  & N\_ALC\_YJH\_S& no & $4$ x $96$ & $16$    & $33.8$        & $0.63$& $3.2$         & $72$      & $1.42$   \\
\noalign{\smallskip}\hline\hline\noalign{\smallskip}     
\end{tabular}
\tablefoot{  NDIT represents the number of frames in the data cube, DIT the exposure time for one frame, Nexp the number of cubes, $\Delta \pi$ the variation of the parallactic angle, 
$\epsilon$ the seeing (at $\lambda = 550$ nm), $\tau_0$ the atmospheric coherence time, and SR the Strehl ratio after the AO correction. 
}
\end{table*}

\begin{table*}[h!]
\caption{Relative astrometry and photometry of the star and planet b for NaCo, GPI, and SPHERE 
at  $68\%$ confidence level.\label{tab:astrometry}}             

\centering \small
\setlength{\tabcolsep}{3.8pt}
\begin{tabular}{ccccccccccc}     
\hline\hline\noalign{\smallskip}
UT Date    & Ins.-Filter     & Algo.$^a$     & $\Delta\alpha$    & $\Delta\delta$    & Sep.          & PA            & Contrast      & True North        & Plate scale        & Ref.  \\
           &                &               & (mas)             & (mas)             & (mas)         & ($\degr$)     & (mag)         & (deg)             & (mas)             &       \\ 
\noalign{\smallskip}\hline\hline\noalign{\smallskip}\noalign{\smallskip}
12-01-2012 & NaCo-$L\!'$    & sADI  & $294\pm8$         & $-550\pm8$        & $624\pm8$     & $151.9\pm0.8$ & $9.8\pm0.4$   & $-0.57\pm0.10$    & $27.11\pm0.06$    & 1     \\ 
12-01-2012 & NaCo-$L\!'$    & sADI  & -                 & -                 & -             & -             & $9.5\pm0.2$   & -                 & -                 & 3     \\ 
14-03-2013 & NaCo-$L\!'$    & sADI  & $305\pm13$        & $-546\pm13$       & $626\pm13$    & $150.8\pm1.3$ & $9.7\pm0.6$   & $-0.58\pm0.10$    & $27.10\pm0.03$    & 1     \\ 
27-06-2013 & NaCo-$L\!'$    & sADI  & $291\pm8$         & $-525\pm8$        & $600\pm11$    & $151.0\pm1.2$ & $9.2\pm0.8$   & $-0.65\pm0.10$    & $27.10\pm0.04$    & 1     \\ 
\noalign{\smallskip}\hline\noalign{\smallskip} 
10-12-2013 & GPI-$K_1$      & LOCI  & $301\pm5$         & $-541\pm5$        & $619\pm5$     & $150.9\pm0.5$ & $12.1\pm0.5$  & $-0.10\pm0.13$    & $14.17\pm0.01$    & 2,3,7   \\  
11-12-2013 & GPI-$H$        & LOCI  & $306\pm11$        & $-537\pm11$       & $618\pm11$    & $150.3\pm1.1$ & $13.1\pm0.9$  & $-0.10\pm0.13$    & $14.17\pm0.01$    & 2,3,7   \\  
13-05-2014 & GPI-$K_1$      & LOCI  & $307\pm8$         & $-536\pm8$        & $618\pm8$     & $150.2\pm0.7$ &  -            & $-0.10\pm0.13$    & $14.17\pm0.01$    & 3,7     \\  
06-04-2015 & GPI-$K_1$      & LOCI  & $322\pm7$         & $-532\pm7$        & $622\pm7$     & $148.8\pm0.6$ &  -            & $-0.10\pm0.13$    & $14.17\pm0.01$    & 3,7     \\  
08-04-2015 & GPI-$K_1$      & LOCI  & $320\pm4$         & $-533\pm4$        & $622\pm4$     & $149.0\pm0.4$ &  $12.2\pm0.2$ & $-0.10\pm0.13$    & $14.17\pm0.01$    & 3,4,7   \\  
29-02-2016 & GPI-$H$        & LOCI  & $330\pm5$         & $-525\pm5$        & $621\pm5$     & $147.8\pm0.5$ &  $13.7\pm0.2$ & $-0.10\pm0.13$    & $14.17\pm0.01$    & 3,4,7   \\  
06-03-2016 & GPI-$H$        & LOCI  & $336\pm3$         & $-521\pm3$        & $620\pm5$     & $147.2\pm0.5$ &  -            & $-0.10\pm0.13$    & $12.17\pm0.01$    & 3,7     \\  

\noalign{\smallskip}\hline\noalign{\smallskip} 
03-02-2015 & IRDIS-$K_1$    & TLOCI     & $322\pm4$         & $-\textbf{532}\pm4$        & $621\pm4$     & $148.7\pm0.3$ & $12.2\pm0.1$  & $-1.72\pm0.06$    & $12.25\pm0.03$    & $5$   \\
03-02-2015 & IRDIS-$K_1$    & ANDR & -               & -               & $620\pm4$     & $148.9\pm0.4$ & $12.6\pm0.5$   & $-1.72\pm0.06$    & $12.25\pm0.03$    & $6$   \\
05-05-2015 & IRDIS-$K_1$    & TLOCI     & $324\pm4$         & $-530\pm6$        & $621\pm5$     & $148.5\pm0.4$ & $12.4\pm0.2$  & $-1.71\pm0.06$    & $12.25\pm0.06$    & $5$   \\
05-05-2015 & IRDIS-$K_1$    & ANDR & -               & -               & $619\pm7$     & $148.6\pm0.5$ & $12.0\pm0.5$  & $-1.71\pm0.06$    & $12.25\pm0.06$    & $6$   \\
18-01-2016 & IRDIS-$K_1$    & TLOCI     & $326\pm4$         & $-532\pm4$        & $625\pm4$     & $148.4\pm0.3$ & $12.3\pm0.2$  & $-1.74\pm0.07$    & $12.27\pm0.03$    & $5,7$   \\
18-01-2016 & IRDIS-$K_1$    & ANDR & -               & -               & $623\pm4$     & $148.7\pm0.4$ & $11.6\pm0.9$  & $-1.74\pm0.07$    & $12.27\pm0.03$    & $6$   \\
31-05-2016 & IRDIS-$K_1$    & TLOCI     & $333\pm2$                  & $-519\pm2$        & $618\pm2$     & $147.3\pm0.2$ & $12.2\pm0.2$  & $-1.81\pm0.05$    & $12.26\pm0.01$    & $5$   \\
31-05-2016 & IRDIS-$K_1$    & ANDR & -               & -               & $621\pm3$     & $147.4\pm0.3$ & $12.2\pm0.3$  & $-1.81\pm0.05$    & $12.26\pm0.01$    & $6$   \\
10-05-2017 & IRDIS-$K_1$    & TLOCI     & $341\pm2$         & $-517\pm3$        & $620\pm3$     & $146.6\pm0.2$ & $12.3\pm0.2$  & $-1.78\pm0.06$    & $12.25\pm0.02$    & $5$   \\
10-05-2017 & IRDIS-$K_1$    & ANDR & -               & -               & $624\pm3$     & $146.7\pm0.3$ & $12.2\pm0.2$  & $-1.78\pm0.06$    & $12.25\pm0.02$    & $6$   \\
06-01-2018 & IRDIS-$K_1$    & TLOCI     & $351\pm2$         & $-514\pm2$        & $622\pm2$     & $145.6\pm0.2$ & $12.3\pm0.1$  & $-1.83\pm0.05$    & $12.26\pm0.01$    & $5$   \\
06-01-2018 & IRDIS-$K_1$    & ANDR & -               & -               & $625\pm4$     & $145.6\pm0.3$ & $12.2\pm0.1$  & $-1.83\pm0.05$    & $12.26\pm0.01$    & $6$   \\
24-02-2018 & IRDIS-$K_1$    & TLOCI     & $353\pm3$         & $-517\pm3$        & $627\pm3$     & $145.6\pm0.3$ & $12.2\pm0.1$  & $-1.75\pm0.06$    & $12.25\pm0.01$    & $5,7$ \\
24-02-2018 & IRDIS-$K_1$    & ANDR & -               & -               & $625\pm4$     & $145.7\pm0.4$ & $12.1\pm0.1$  & $-1.75\pm0.06$    & $12.25\pm0.01$    & $6,7$ \\
28-03-2018 & IRDIS-$K_1$    & TLOCI     & $357\pm4$         & $-521\pm4$        & $632\pm4$     & $145.5\pm0.3$ & $12.3\pm0.1$  & $-1.73\pm0.07$    & $12.26\pm0.02$    & $5,7$ \\
28-03-2018 & IRDIS-$K_1$    & ANDR & -               & -               & $631\pm4$     & $145.6\pm0.4$ & $12.3\pm0.1$  & $-1.73\pm0.07$    & $12.26\pm0.02$    & $6,7$ \\
13-04-2019 & IRDIS-$K_1$    & TLOCI     & $368\pm3$         & $-508\pm3$        & $623\pm3$     & $144.1\pm0.2$ & $12.3\pm0.2$  & $-1.78\pm0.07$    & $12.25\pm0.02$    & $5$   \\
13-04-2019 & IRDIS-$K_1$    & ANDR & -               & -               & $627\pm3$     & $144.1\pm0.3$ & $12.2\pm0.2$  & $-1.78\pm0.07$    & $12.25\pm0.02$    & $6$   \\
18-05-2019 & IRDIS-$K_1$    & TLOCI     & $368\pm4$         & $-509\pm4$        & $630\pm4$     & $144.1\pm0.3$ & $12.2\pm0.1$  & $-1.78\pm0.07$    & $12.26\pm0.01$    & $5,7$   \\
18-05-2019 & IRDIS-$K_1$    & ANDR & -               & -               & $628\pm4$     & $144.3\pm0.4$ & $12.1\pm0.1$  & $-1.77\pm0.07$    & $12.26\pm0.01$    & $6$   \\
\noalign{\smallskip}\hline\noalign{\smallskip} 

\end{tabular}
\tablefoot{($^a$): Algo. refers to the algorithms used for the reduction: smart ADI \citep[sADI;][]{Rameau2013a}, TLOCI \citep{Lafreniere2007_tloci}, and ANDROMEDA \citep[ANDR;][]{mugnier2009}. 
}
\tablebib{(1) Astrometric and
  photometric results from \citet{Rameau2013b}; 
  (2, 3) Astrometric results processed by \citet{Rameau2016} and photometric results from \citet{Galicher2014};  
  (4) Photometric results reported by \citet{Derosa2016}; 
  (5,6) This work: astrometric and photometric results  done automatically with  pipeline (5)  SpeCal-TLOCI or (6) ANDROMEDA. The 2018 and 2019 data have not been reported and exploited yet. The 2015, 2016, and 2017 data are  reported by \citet{Chauvin2018}, but they  reduced the data with the IPAG-ADI pipeline; 
  (7) These epochs are not used for the orbital fit of the HD $95086$ b exoplanet (see Section \ref{sec:selection}). 
  }
\end{table*}

\subsection{VLT/SPHERE data}

\begin{figure*}[t]
    \centering
    \includegraphics[width=1\linewidth]{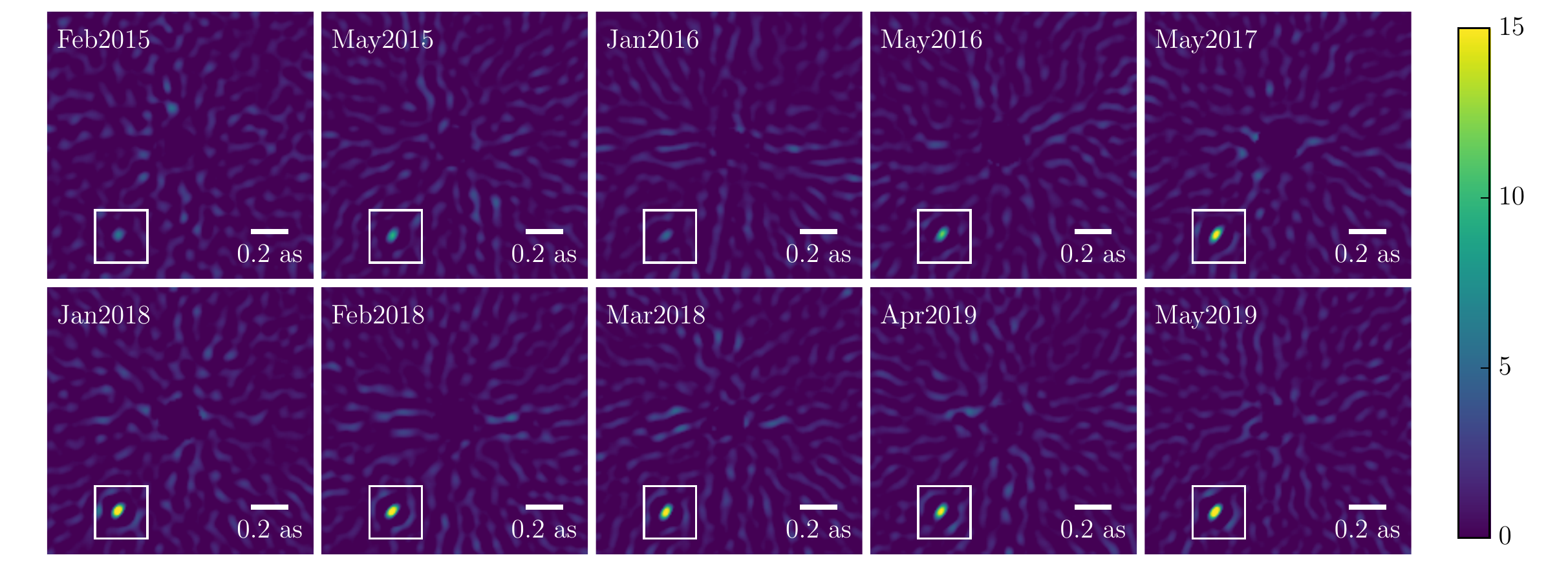} 
    \vspace{0.1cm}
    \includegraphics[width=1\linewidth]{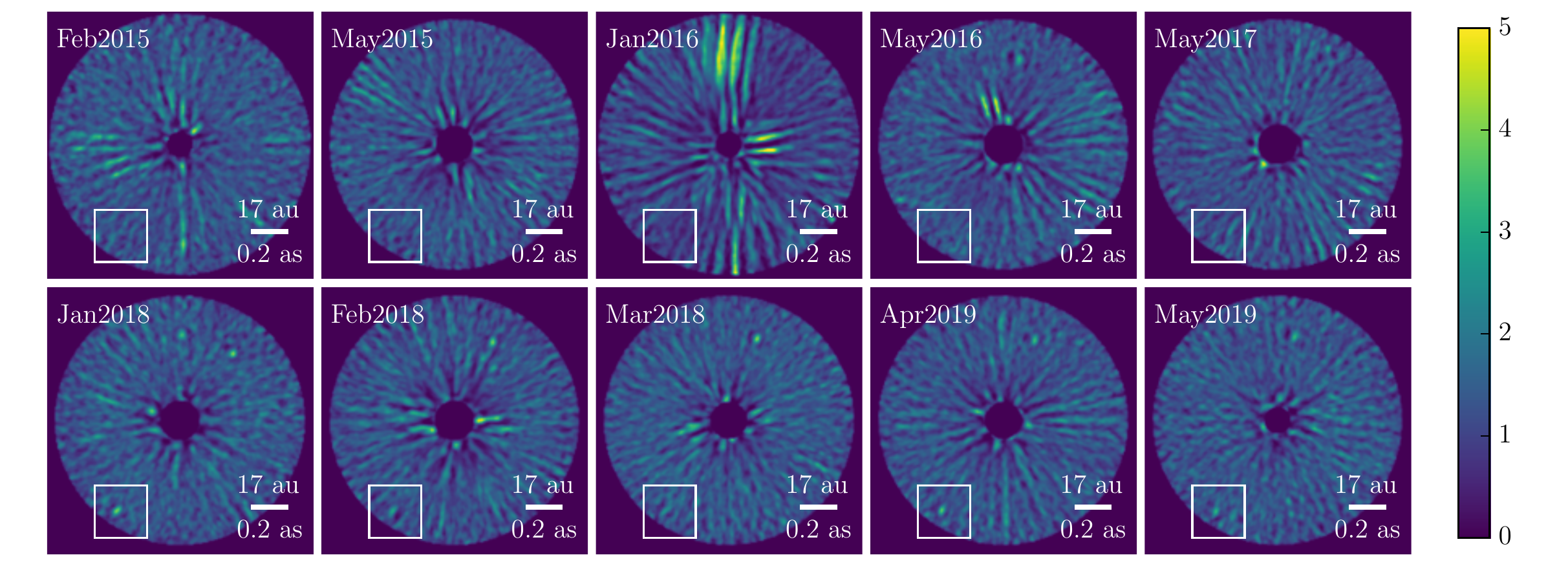}
    \caption{
    Signal-to-noise ratio maps for all the SPHERE epochs. At the \textit{top}, SPHERE-IRDIS in the K1 band and at the \textit{bottom}, SPHERE-IFS in the YJH bands, both reduced by the pipeline ANDROMEDA. The color bar corresponds to the signal-to-noise ratio, $0$ to $15$ for IRDIS and $0$ to $5$ for the IFS. Planet b is located in the white square. The region below the inner working angle of the coronagraph is masked.  In the top right corner of the  IFS data, the one (or two) point-like feature(s)   correspond to the remanent of the star on the IFS detector.
    \label{fig:mosaic_epochs_andromeda}}
\end{figure*}

The HD\,$95086$ system was  monitored during the SHINE survey 
at $13$ different epochs between February 2015 and May 2019 (see  Table~\ref{tab:obslog}) using the VLT/SPHERE high-contrast instrument \citep{Beuzit2019_sphere}. The observations were obtained with the modes IRDIFS (3 epochs) and IRDIFS-EXT (11 epochs) that combine simultaneously the IRDIS \citep{Dohlen2008} and IFS instruments \citep{Claudi2008}. The IRDIFS-EXT mode combines IRDIS in dual-band imaging \citep[DBI;][]{Vigan2010dbi} mode with the K1K2 filter doublet  
$\lambda_{K1}=2.103\pm0.102~\mathrm{\mu m}$, $\lambda_{K2}=2.255\pm0.109~\mathrm{\mu m}$, and IFS in the YJH ($0.97$--$1.66\,\mu$m) setting. The IRDIFS mode combines  IRDIS in DBI with H2H3 filters ($\lambda_{H2}=1.593\pm0.055~\mathrm{\mu m}$, $\lambda_{H3}=1.667\pm0.056~\mathrm{\mu m}$), and IFS in the YJ ($0.95$--$1.35\,\mu$m) setting. 
Each observing sequence was performed with the pupil-tracking mode. This  combination  enables  the  use  of  angular \citep{Marois2006_ADI} and/or spectral differential imaging techniques \citep{Racine1999_speckle_noise,Sparks2002} to reach higher contrast at subarcsecond separations. The details of the observations are reported in Table~\ref{tab:obslog}.

In this work we focused our analysis on data acquired with the IRDIFS-EXT mode as HD\,$95086$\,b, and any expected inner planet in the system, are L-type planets, expected to be particularly red and therefore easier to detect at longer wavelengths. A total of ten epochs are considered as the data acquired on May $3$, 2017, are not exploitable owing to very poor observational conditions. The observing conditions are summarized in Table~\ref{tab:obslog} and Fig.~\ref{fig:histo_sparta_dimm} of Appendix~A. 
The Strehl ratio (SR) 
and the wind parameters are measured by the SPHERE eXtreme AO \citep[SAXO,][]{Petit2014} real-time computer named Standard Platform for Adaptive optics Real Time Applications \citep[SPARTA,][]{Fedrigo2006sparta}, while the seeing ($\epsilon$) 
and the atmospheric coherence time parameters ($\tau_0$) were obtained by the Differential Image Motion Monitor (DIMM) and the Multi-Aperture Scintillation Sensor \citep[MASS,][]{Kornilov2007massdimm} turbulence monitor at the Paranal Observatory.

\subsection{Archival data: VLT/NaCo and Gemini-South/GPI }

To revisit the orbital and spectral properties of HD\,95086\,b, we analyzed archival data from the VLT/NaCo imager obtained in 2012 and 2013 \citep{Rameau2013b,Rameau2013a}, together with Gemini-S/GPI observations obtained between 2013 and 2016 \citep{Rameau2016,Derosa2016}. A summary of these observations and the astrometric and spectro-photometric results used for this work are reported in Table~\ref{tab:astrometry}. 

\section{Data reduction and analysis\label{sec:reduction}}

\subsection{Pre-processing}

All SPHERE observations of HD\,95086 were reduced by the SPHERE Data Center \citep{delorme2017_red1step}, using the SPHERE Data Reduction and Handling pipeline \citep{Pavlov2008_red1step}, following the same approach as described by \cite{Chauvin2018}. To summarize, the pre-processing corrects the non-coronagraphic point spread function (PSF) and the coronagraphic image cube for bad pixels, dark current, flat non-uniformity; the sky background for both  IRDIS and IFS; and  the wavelength and cross-talk between spectral channel calibration for IFS.
A normalization is applied to calibrate the coronagraphic images in intensity relative to the star (i.e., in terms of contrast).
The coronagraphic images are centered by using the four satellite spots to accurately determine the position of the star behind the coronagraphic mask, as also described in \cite{Chauvin2018}. To calibrate the astrometry of both IRDIS and IFS on the sky, a
 star-crowded field (47 Tuc) is regularly observed as part of the long-term analysis of the SPHERE guaranteed time observation (GTO) astrometric calibration described in \citet{Maire2016,Maire2021} to measure the detector plate scale, true north, and distortion. The plate scale and true north solutions at each epoch are reported in Table~\ref{tab:astrometry}. 

\subsection{Image processing}

To detect and characterize potential planetary signals in the images, we used two dedicated pipelines, namely ANDROMEDA \citep{Cantalloube2015_andromeda} and SpeCal \citep{Galicher2014Specal}. Both are based on the angular differential imaging (ADI) technique, which  removes the starlight residuals in the coronagraphic images. 
ANDROMEDA is a forward-modeling approach based on a maximum likelihood estimator \citep{mugnier2009}. It first performs a simple pair-wise subtraction. Then, it searches for the specific signature that would appear in the presence of an unresolved point-source in the residual image, and estimates its probability, jointly for all pairs of subtracted images. 
The SpeCal pipeline combines a set of different algorithms like classical ADI (cADI, \citealt{Marois2006_ADI}), locally optimized combination of images (LOCI/TLOCI, \citealt{Lafreniere2007_tloci}), and principal component analysis (PCA, \citealt{Amara2012_pca,Soummer2012_pca}). To exploit the spectral diversity given by the IFS and the IRDIS-DBI modes, in addition to the temporal dimension as done in ADI, SpeCal has been developed to apply LOCI/TLOCI and PCA in angular and spectral differential imaging (ASDI) (\citealt{Mesa2015}). For the analysis of the SHINE survey, the reference algorithms benchmarked with various blind tests are the TLOCI-ADI and the PCA-ASDI algorithms \citep{Langlois2021_shine_paperII_observations}. For IRDIS, the contrast performance showed that ANDROMEDA-ADI performs better than TLOCI-ADI (see Figs. \ref{fig:mosaic_epochs_andromeda} and  \ref{fig:contrast_ifs_irdis}), and  in addition provides a more robust estimate of the statistical threshold for the candidate detection. The signal-to-noise ratio maps (S/N maps) 
obtained with TLOCI-ADI for IRDIS and PCA-ASDI for IFS using SpeCal are shown in Fig.~\ref{fig:mosaic_epochs_specal}. The S/N maps obtained for ANDROMEDA in ADI for IRDIS and IFS (consisting of a combination of the IFS channels after processing individual each channel) are shown  in Fig~\ref{fig:mosaic_epochs_andromeda}.

\subsection{Multi-epoch selection and combination\label{sec:selection}}

Given the very red spectrum of HD~95086~b, the planet is clearly detected for each individual epoch (with $S/N>10$) in the K1 and K2 bands with SPHERE, and in the K band with GPI. This band (K1, where the background noise is lower than in K2) is therefore used to extract the planet's relative astrometry at each epoch, and derive its updated orbital properties (see below). The detection becomes more challenging at the H band ($S/N\sim3$--$7$) for both SPHERE and GPI, but remains possible for good-quality observation  (January 2018, February 2018, March 2018, April 2019 and May 2019). At the J band the planet is currently detected  only by stacking reduced SPHERE IFS images taken at various epochs to optimize the speckle cancelation, as done by \cite{Chauvin2018}. Following a similar multi-epoch strategy, we first selected the best observing epochs obtained by SPHERE between 2015 and 2019, then re-aligned each final IFS datacubes correcting for the planet's orbital motion, and stacking them to extract the JH spectrum of the planet. 

\begin{figure*}[t]
    \centering
    \includegraphics[height=7cm]{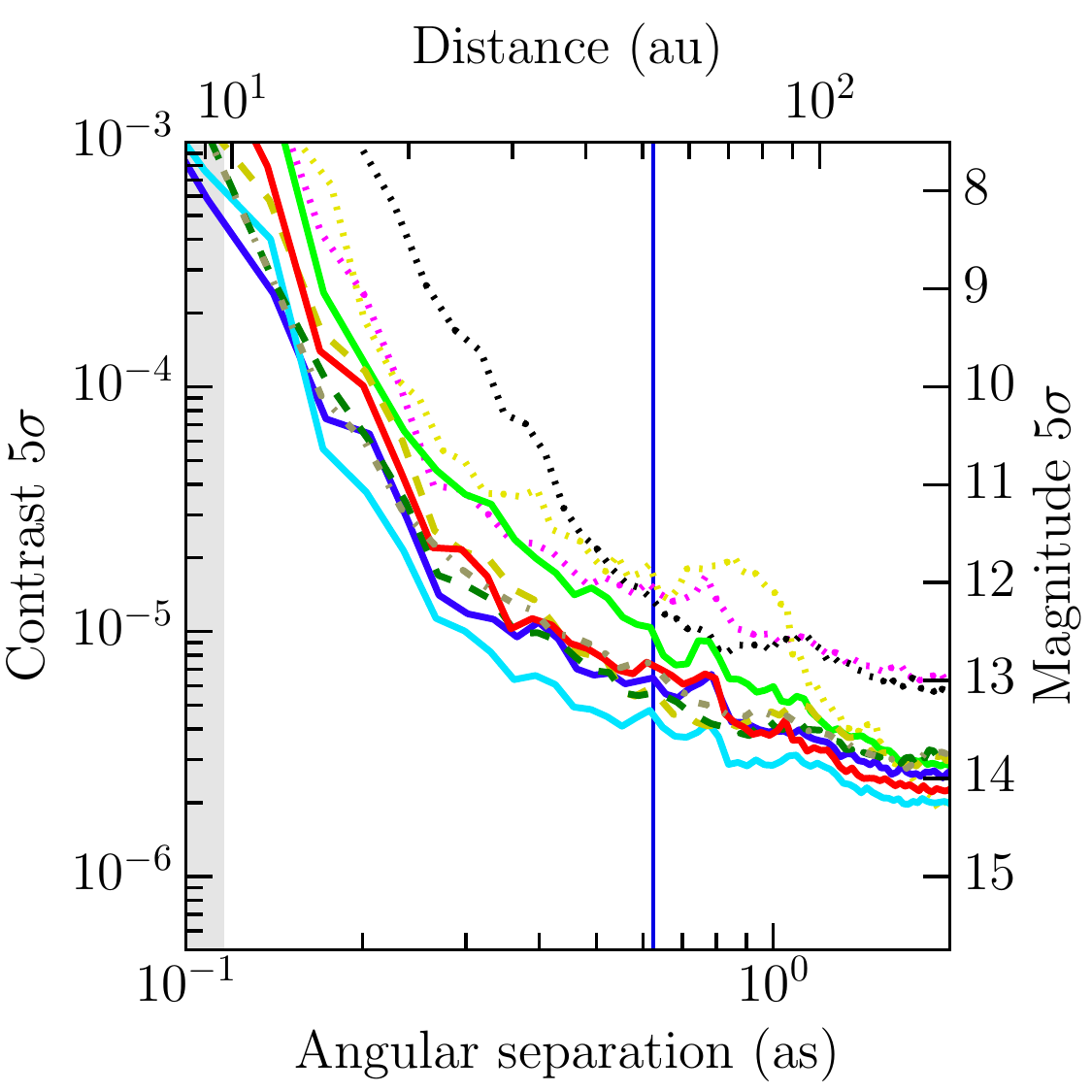} \qquad \includegraphics[height=7cm]{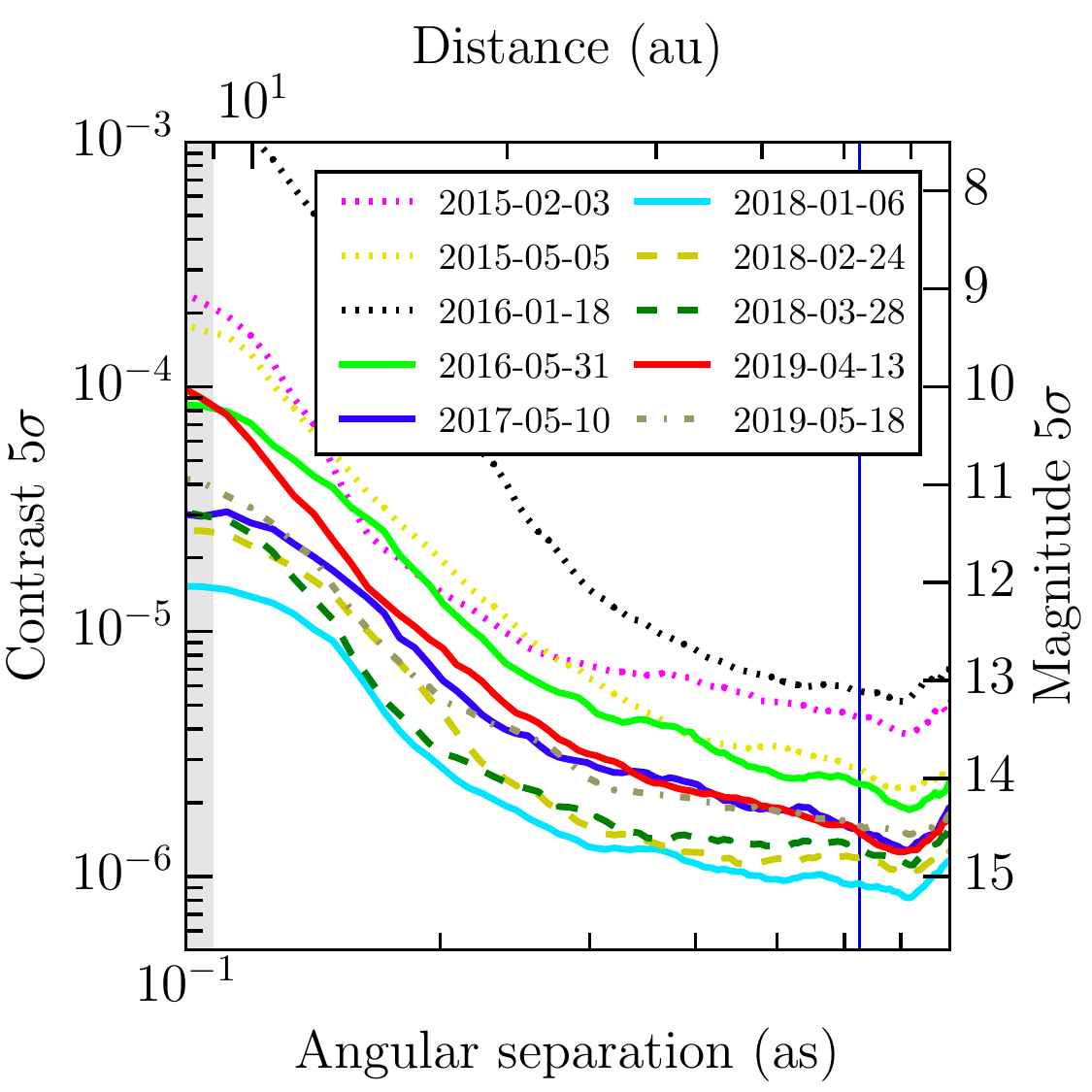}\medskip
    
    \includegraphics[height=7cm]{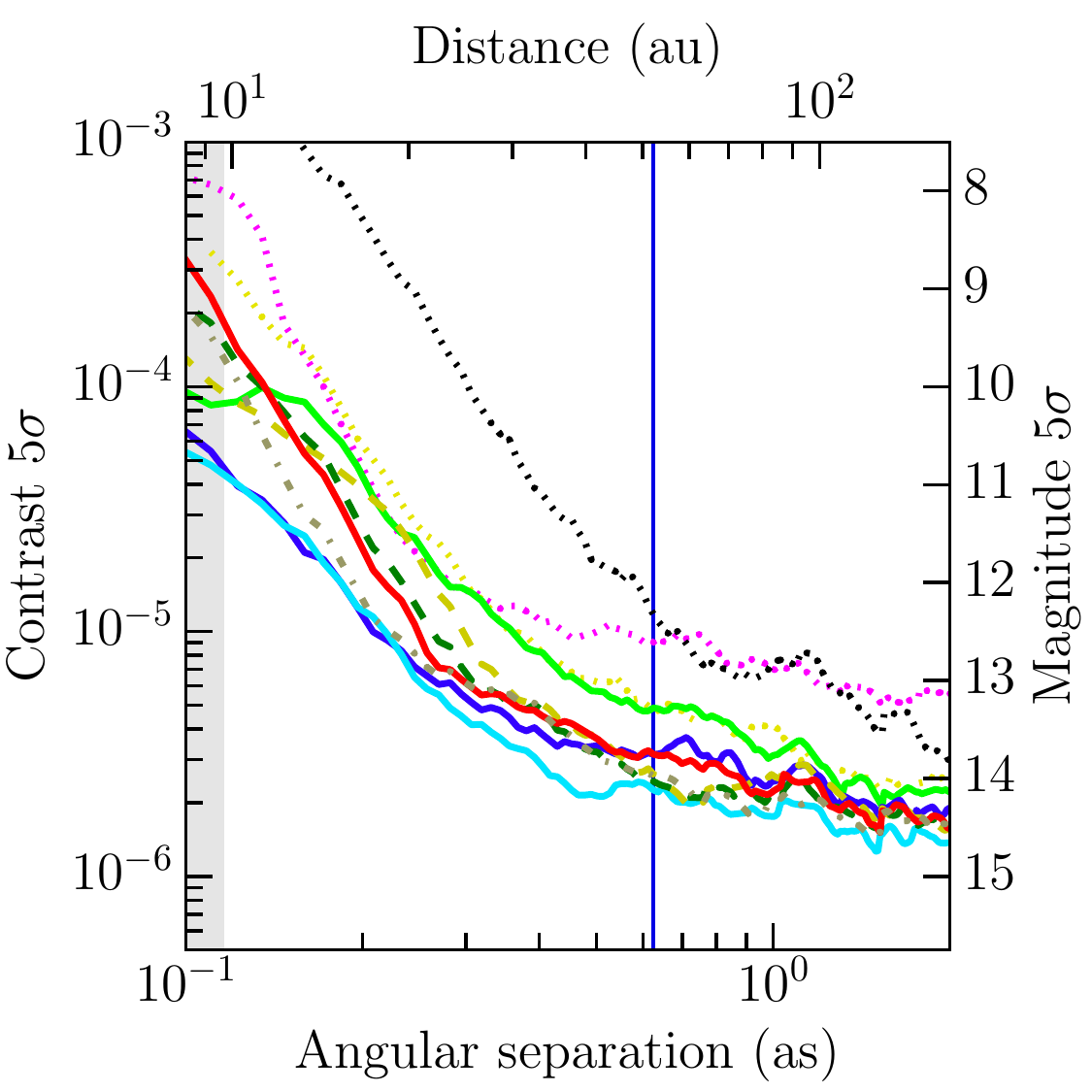} \qquad
    \includegraphics[height=7cm]{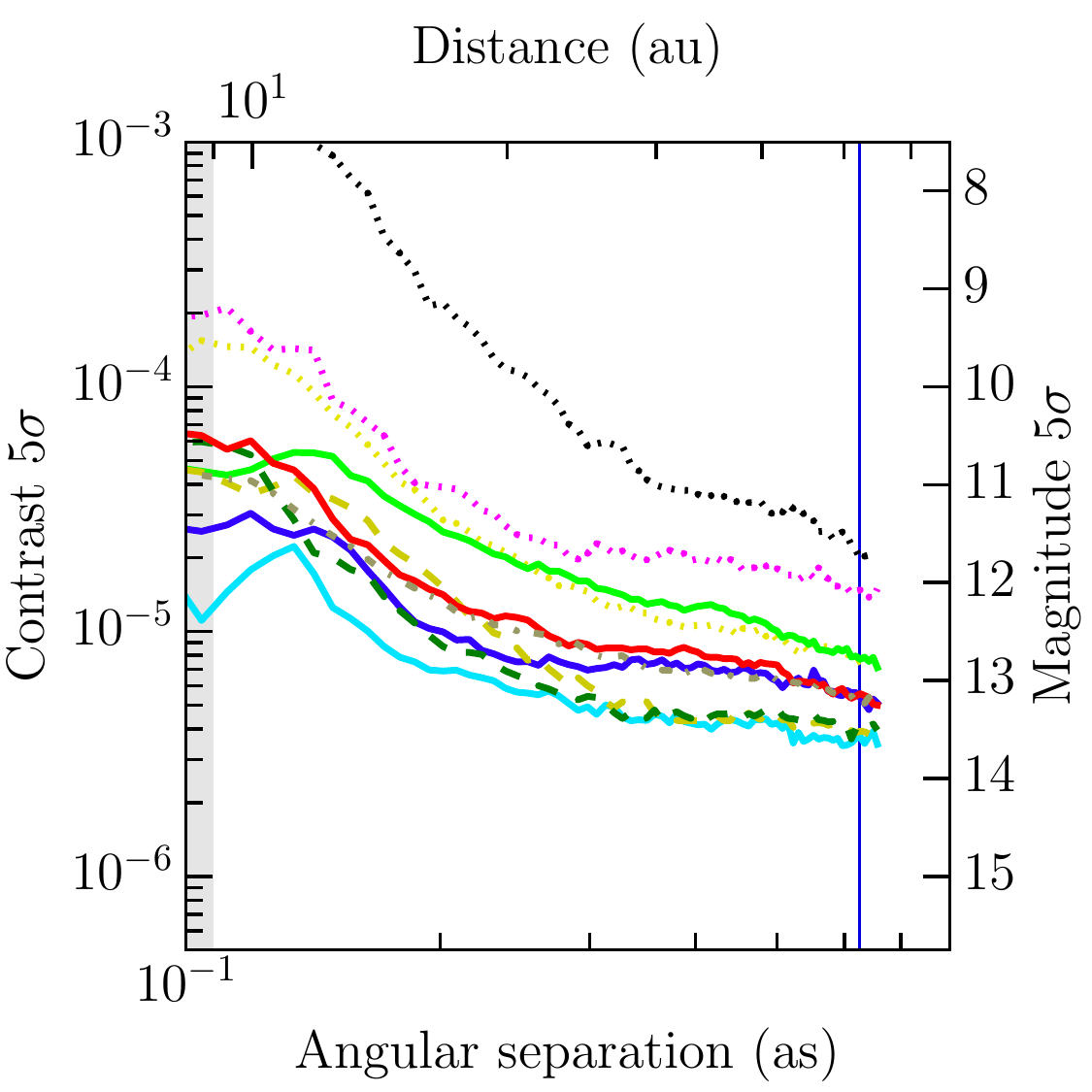}
    \caption{Detection limits at $5\sigma$ for IRDIS in the  K1 band on the \textit{left} and IFS in the  YJH bands on the \textit{right}. Data are reduced at the \textit{top} by the pipeline SpeCal-TLOCI (for IRDIS) or SpeCal-PCAPad (for IFS) and at the \textit{bottom} by the pipeline ANDROMEDA. The gray area corresponds to the area hidden by the larger coronagraph used for all the epochs (i.e., N\_ALC\_Ks) which has an inner working angle of $0.107\,"$ and $0.116\,"$ for IFS and IRDIS, respectively. The blue vertical line corresponds to the position of the planet HD~95086~b at the best epoch January 2018 (cyan, $0.625\,\mathrm{"}$).}
    \label{fig:contrast_ifs_irdis}
\end{figure*}

To investigate the contrast performance  at each epoch, we considered as the first criterion the contrast curves determined with TLOCI-ADI and ANDROMEDA-ADI for IRDIS, and PCA-ASDI and ANDROMEDA-ADI for IFS (Fig.~\ref{fig:contrast_ifs_irdis}). The observing conditions (atmospheric turbulence conditions and AO telemetry, summarized in Table~\ref{tab:obslog} and in Fig.~\ref{fig:histo_sparta_dimm}) were used as a consistency check. From  a total of ten epochs, a subsample clearly emerges of six very good epochs. The best epoch is January 2018, followed by February 2018, March 2018, April 2019, May 2019, and May 2017. The worst epochs are January 2016 and February 2015. Despite the very good Strehl ratio ($81$\%) and seeing ($0.33$ as) for the January 2016 epoch, the atmospheric coherence time is very fast ($1.8$ ms), producing a wind-driven halo, which degrades the contrast performance by about one order of magnitude at the planet location \citep{Cantalloube2020_WDH}. 

The $5\sigma$ detection limits are $1\cdot10^{-6}$ and $2\cdot10^{-6}$ at $0.6$" respectively with the IFS and IRDIS for the best epoch (January 2018). IFS performs better in contrast at close separations as PCA-ASDI exploits the spectral diversity of $39$ spectral channels to remove the speckles, while TLOCI-ADI does not for the IRDIS dual-band K12. We  note that the IFS and IRDIS contrast curves are coherent with each other: the best contrast curves with IFS are the same as with IRDIS (see Fig. \ref{fig:contrast_ifs_irdis}). In summary, the best epochs for extracting the spectrum of the exoplanet HD~95086~b are the last six: May 2017, January 2018, February 2018, March 2018, April 2019, and May 2019. 
To extract the astrometry, 
we preferred  the epochs acquired with  the  satellite spots 
for the whole sequence of observation (and not only at the beginning and/or the end of the observation)  to ensure a correct astrometric calibration for the whole sequence of observation  (May 2016, May 2017, January 2018, and April 2019), as well as the two epochs in 2015 (February 2015 and May 2015), as in 2015 no observation with satellite spots for the whole sequence was   acquired. We note  that the position of the star on the detector can marginally evolve during the sequence of observation due to optomechanical and thermal variations that  slightly move the optics. The Differential Tip Tilt Sensor (DTTS) aims to ensure its centering, but a centroid variation can be expected. We adopted the conservative error on the centroid position of $0.2$ pixel corresponding to $2.5$ mas for the epochs acquired without the continuous satellite spots mode, as done in \citet{Chauvin2018}. This value is taken into account in the astrometric calibration error budget, as well as the errors on the true north, plate scale, and pupil offset, following \citet{Maire2021_jatis_budget_error}. In addition, it  should be noted that the epoch of March 2018 suffers from a higher centroid variation, which is a rare problem, but even so occurred in a few observations over the whole seven years of exploitation of the SPHERE instrument. This higher centroid variation is about $5$ mas both in declination and right ascension based on the expected position of HD 95086\,b in March 2018 considering the epochs better calibrated in astrometry, as of January 2018 and April 2019.


The relative astrometry and photometry of the exoplanet HD~95086~b to its host star are gathered in Table~\ref{tab:astrometry} for all the epochs obtained with SPHERE-IRDIS in the K1 band, using ANDROMEDA and SpeCal-TLOCI image processing, along with the measurements from archival NaCo and GPI observations. The ANDROMEDA and SpeCal-TLOCI astrometric and photometric measurements from 2015 to 2019 are consistent with each other (see Table \ref{tab:astrometry}).

\subsection{The 1.2--3.8\,\texorpdfstring{$\mu$}{mu}m spectrum of HD 95086 b }

\begin{figure}[h!]
    \centering
    \includegraphics[width=\linewidth]{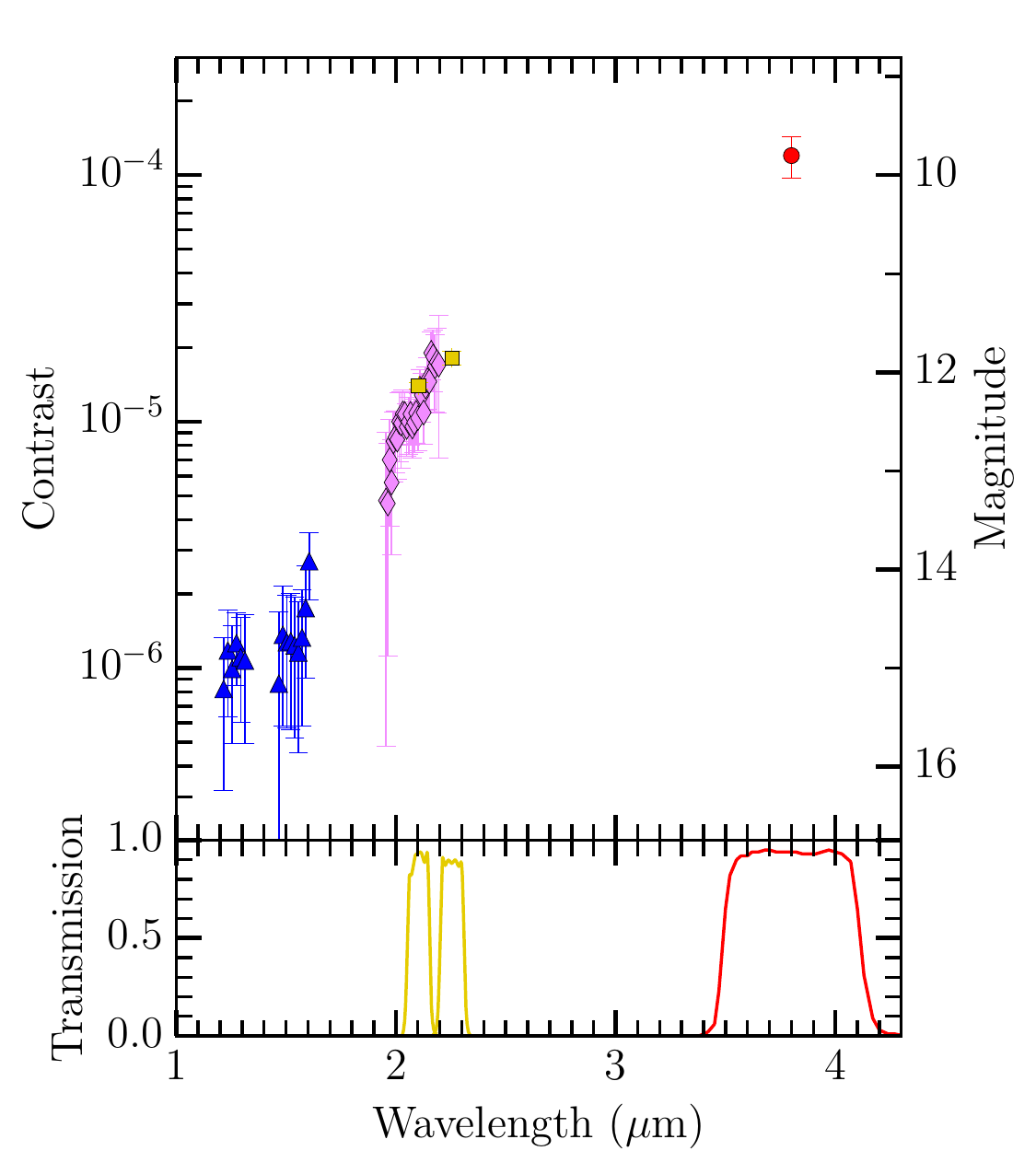}
    \caption{Spectroscopic and photometric values of the exoplanet HD~95086~b expressed in contrast relative to the host star from $1.2\,\mu$m to $3.8\,\mu$m with SPHERE-IFS (blue triangles), SPHERE-IRDIS (yellow squares), GPI (pink diamonds), and NaCo (red circle)  in the bands JH, K, K12, and L' at the \textit{top}. At the \textit{bottom}, the transmission curves of the filters K1, K2, and L' are shown\protect\footnotemark. 
    Results are extracted from an averaged stacking of the six best epochs for  the SPHERE data, and for one single epoch for the GPI and NaCo data.}
    \label{fig:spectrum_complete}
\end{figure}

\begin{figure}[t]
    \centering
    \includegraphics[width=\columnwidth]{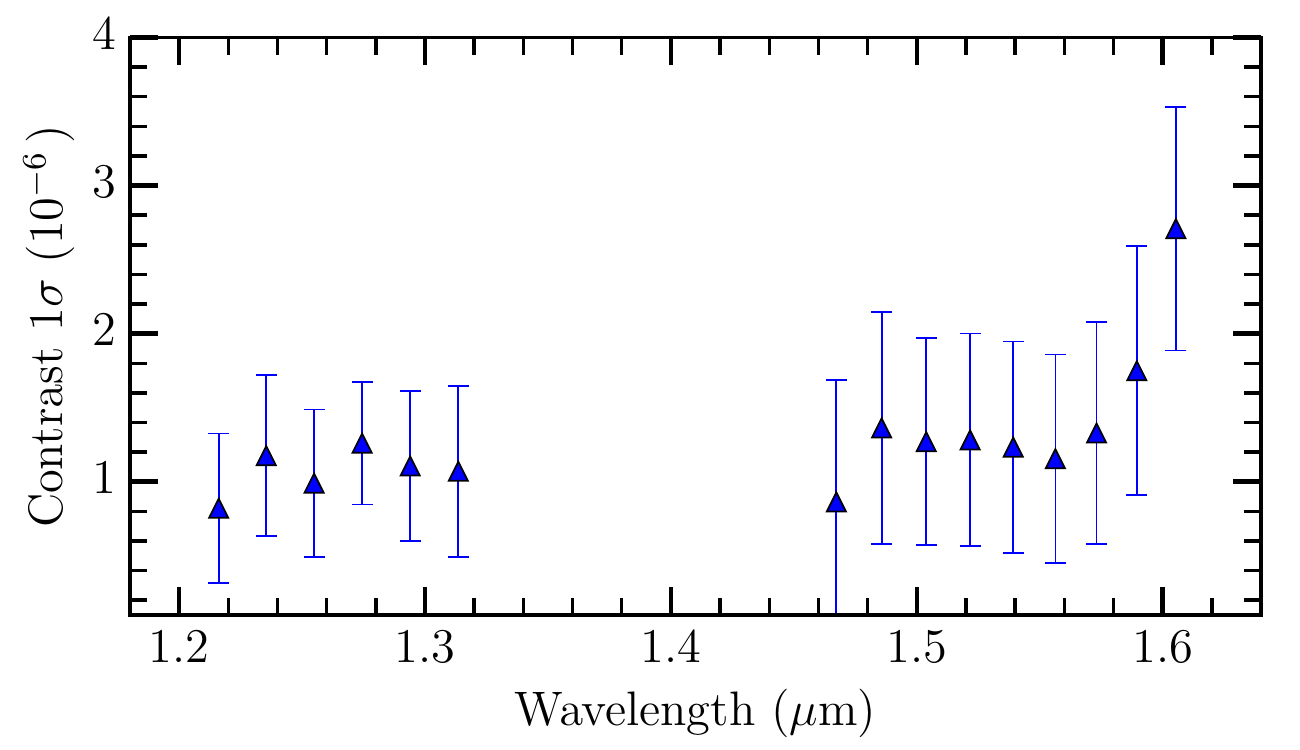}
    \caption{Spectrum of the exoplanet HD~95086~b expressed in contrast relative to the host star from SPHERE-IFS data in the J band ($1.21$--$1.32~\mathrm{\mu m}$) and in the H band ($1.46$--$1.62~\mathrm{\mu m}$) at a 68\% confidence level. The contrast is given in units of $10^{-6}$. The data are reduced by the pipeline SpeCal-PCA ADI where a six-epoch averaged stack was made before extracting the spectrum to boost the S/N.}
    \label{fig:spectrum_yjh}
\end{figure}

For the first time, we have extracted the spectrum of the planet HD~95086~b in the J and H bands. This was achieved on the SpeCal-PCA ADI reduced images by measuring the contrast of the planet in an aperture of $1$ FWHM of the six-epoch averaged image in which we corrected for the exoplanet's orbital motion by using its precise astrometry through time before stacking. 
The exoplanet is still not detectable in the Y band, hardly in the J band, and with a signal-to-noise ratio of about $5$ per spectral channel in the H band. By using the BT-NEXTGEN synthetic spectrum of the star HD~95086~A, we converted the contrast values to flux values. We completed the SPHERE-IFS YJH ($0.96$--$1.64~\mathrm{\mu m}$, resolution $30$) and SPHERE-IRDIS K12 photometric points ($2.10~\mathrm{\mu m}$ and $2.26~\mathrm{\mu m}$) with archival data from GPI, which provides the spectrum of the exoplanet HD~95086~b in the K1 band at low resolution ($1.95$--$2.20~\mathrm{\mu m}$, resolution $66$), and from NaCo, which provides a photometric measurement in the L' band ($3.80~\mathrm{\mu m}$, bandwidth $0.62~\mathrm{\mu m}$). 
The complete spectrum of HD~95086~b is shown in Fig.~\ref{fig:spectrum_complete};  a zoomed-in image of the J and H bands is shown in Fig.~\ref{fig:spectrum_yjh}. The increasing slope in the spectroscopic K1 and photometric K12 and L' points acknowledges the redness of  exoplanet b \citep{Derosa2016} and seems to be verified in our H-band spectrum as well, even though the uncertainties are significant.
We empirically estimated the spectral correlation matrix as in \citet{Greco2016} and \citet{Derosa2016} for the measurements obtained with the IFS of SPHERE and GPI (see Appendix~\ref{app:SpectralCovariance}), and used it to compute the covariance matrix used for spectral fitting of
HD~95086~b  (see Sect.~\ref{sec:spectro}).

\section{Orbital analysis \label{sec:orbit}}

\begin{figure}[t]
    \centering
    \includegraphics[width=1\linewidth]{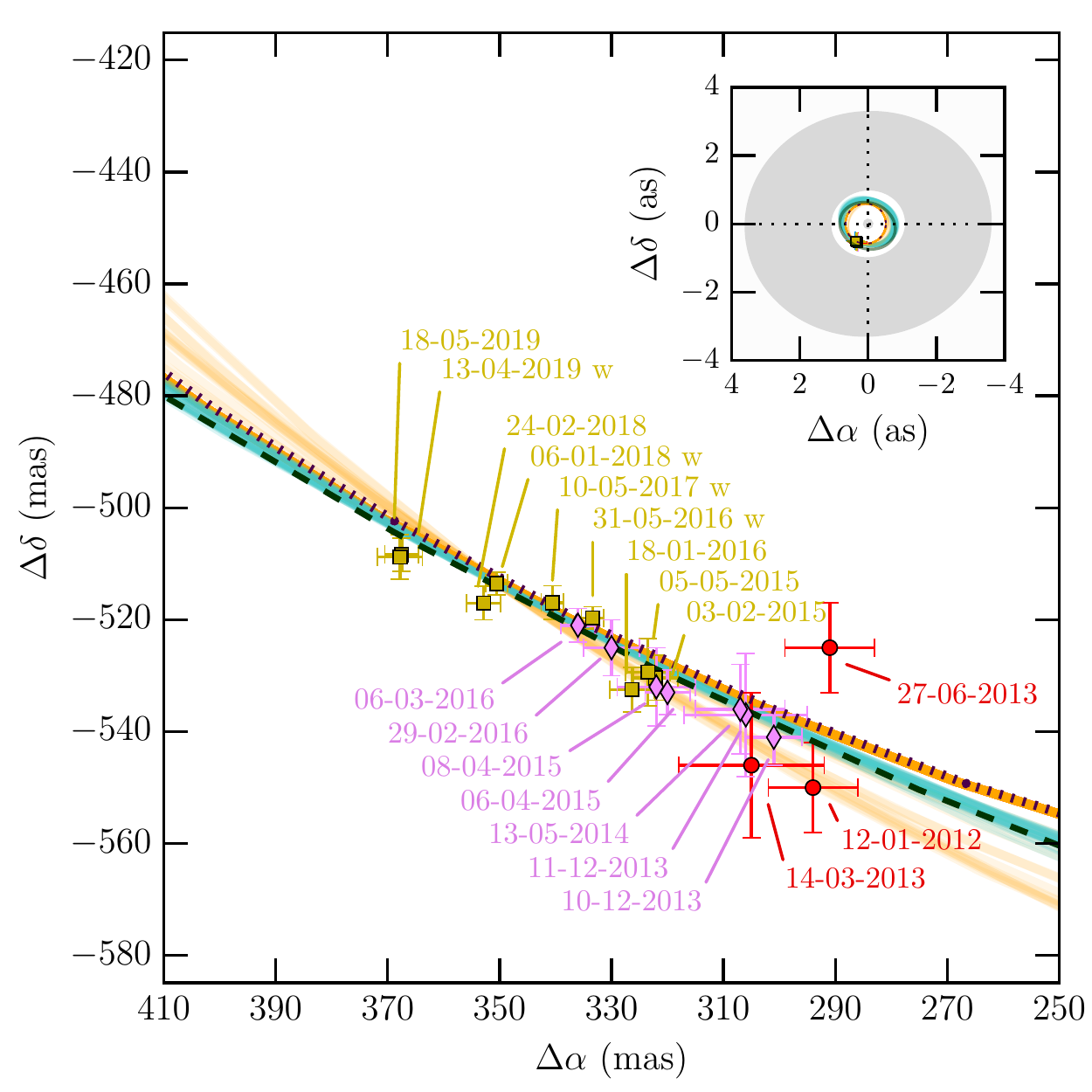}
    \caption{Astrometric positions of the planet HD $95086$ b between $2012$ and $2020$ with three different instruments: SPHERE (yellow squares), GPI (pink diamonds), and NaCo (red circles) 
    with $1\sigma$ error bars. The astrometric positions from SPHERE were   computed with the SpeCal-TLOCI pipeline.  The letter ``w'' in the legend indicates the observations imaged with satellite spots that enable   finding the exact position of the star during the whole sequence and later recenter the frames if necessary. The green and orange solid lines correspond to a sample of the orbital solutions found by the MCMC orbital fit and the K-Stacker tools, respectively. The black dashed and dotted lines  respectively represent the MCMC and K-Stacker orbits for which the corresponding orbital parameters are given in Table\,\ref{tab:mcmcfit_solutions}. 
    For the MCMC tool, the orbit corresponds to the MAP (see text), whereas for K-Stacker it is the orbit closest to the mean of the orbital solutions. Both of these orbits are by construction true orbital solutions. 
    Notes: The Feb2015 point from SPHERE is hidden by the two 2015 points from GPI; the two 2019 points are very close to each other. 
    The insert (in the top right corner) shows the location of the planet relative
    to the cold outer belt and warm inner belt from \citet{Su2017}.
    }
    \label{fig:astrometry}
\end{figure}

We ran the Markov chain Monte Carlo (MCMC) orbital fit \citep{Ford2005,Ford2006}, as done in \citet{Chauvin2012_betapic}. We used previous NaCo astrometric data \citep[epochs 2012 to 2013 from][]{Rameau2013b} and SPHERE measurements \citep[epochs February 2015, May 2015, May 2016, and May 2017 from][]{Chauvin2018} including two new astrometric points from the SPHERE-IRDIS images obtained in the K1 band (January 2018 and April 2019) with the updated distance value of $86.2\rm\,pc$ from \citet{Bailer_Jones2018_Distance}. 
In this run the stellar mass was fixed to its mean value $1.6\,\mathrm{M}_\odot$ and to the stellar distance (86.2\,pc). It would have been technically possible to leave the stellar mass free and to let the code redetermine it, but this turned out to be inaccurate. The reason is that only a tiny part of the orbital period is covered by the observations, leading to a degeneracy between the central mass and the inclination. 

The priors assumed for this run were logarithmic between $1\,\mathrm{yr}$ and $4000\,\mathrm{yr}$; linear for the eccentricity $e$ between $0$ and $1$; $\propto\sin i$ for the inclination $i$ between $0$ and $180\degr$; and linear between $-180\degr$ and $180\degr$ for the longitude of ascending node $\Omega$, the argument of periastron $\omega$, and for the mean anomaly at the time of the first observation epoch (related to the time of periastron passage $T_p)$. We note that the MCMC run is represented by taking the orbital period $P$ as variable instead of the semimajor axis $a$. Both approaches are equivalent as $P$ and $a$ are linked via Kepler's third law.

In addition, when dealing with pure relative astrometric data as we do  here, it is well known that there is a $\pm180\degr$ degeneracy between solutions in the longitude of ascending node $\Omega$ and the argument of periastron $\omega$. Each solution with $(\Omega,\omega)$ yields exactly the same projected orbit as the same solution, but with $(\Omega+180\degr,\omega+180\degr)$. To overcome this difficulty, as explained in \citet{Chauvin2012_betapic}, the code actually fits $\Omega+\omega$ and $\omega-\Omega$ rather than $\Omega$ and $\omega$ directly. The former angles are indeed unambiguously determined contrary to $\omega$ and $\Omega$. Then, each root solution with fitted values for $\Omega+\omega$ and $\omega-\Omega$ is declined as two final separate solutions, one with $(\Omega,\omega)$ and the other  with $(\Omega+180\degr,\omega+180\degr)$.

The updated orbital parameters and confident regions from the MCMC orbital fit with the  two new astrometric measurements in 2018 and 2019 
are shown in Table~\ref{tab:mcmcfit_solutions}. A sample of the orbit solutions is displayed in Fig.~\ref{fig:astrometry}, as is  the orbit given in Table~\ref{tab:mcmcfit_solutions}, which corresponds to the maximum a priori probability (MAP). By using both the prior and the reduced chi-squared ($\chi_r^2$) information, the MAP maximizes the probability given in Eq.~\eqref{eq:map} used in the MCMC:
\begin{equation}
\label{eq:map}
probability = \frac{\sin(i)} {P} \times \exp(-\chi^2_\text{r}/2)\,. 
\end{equation}

\begin{figure*}[t]
    \centering
    \includegraphics[width=\linewidth]{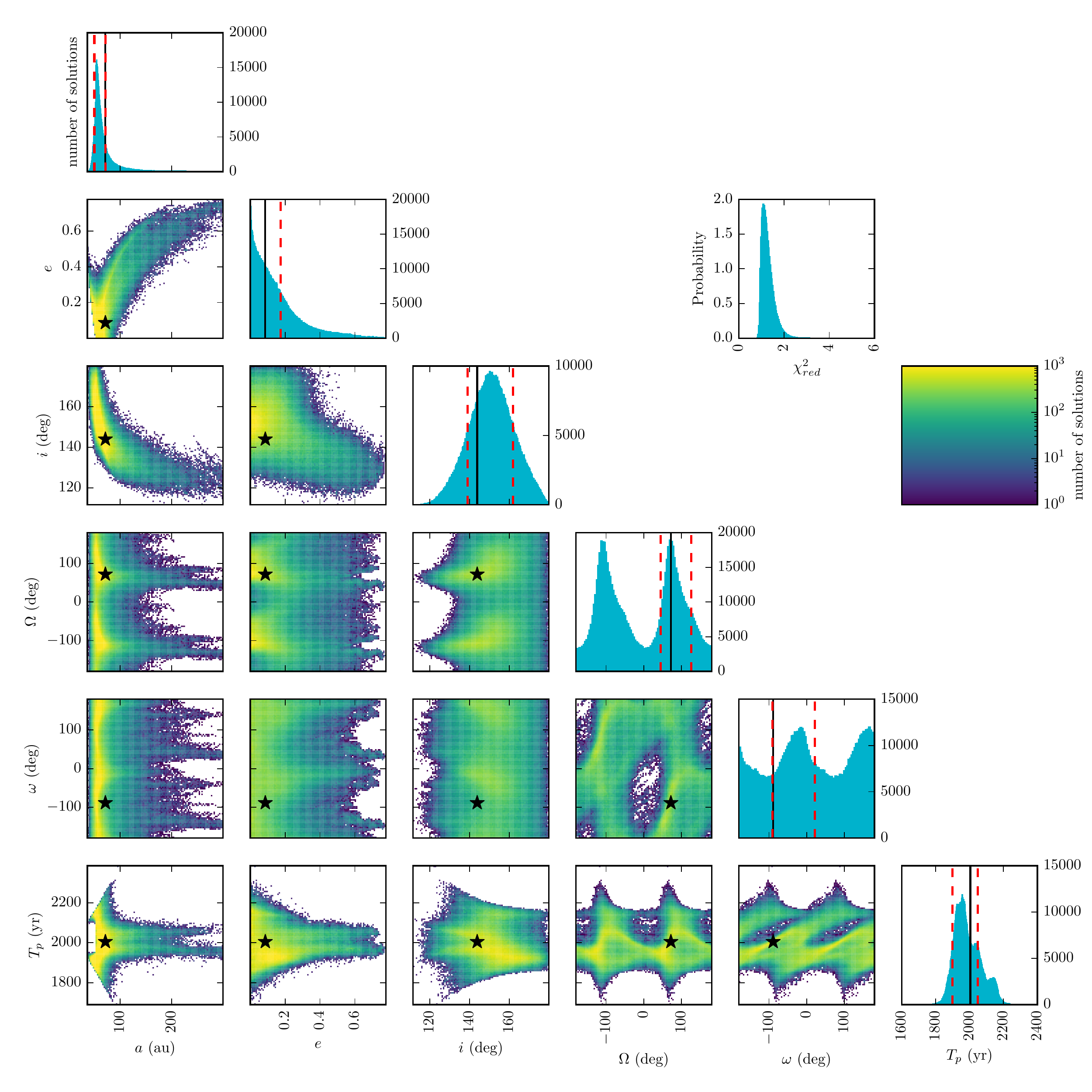}
    \caption{Results of the MCMC orbital fitting of HD 95086 based on the NaCo archive data (epochs 2012--2013) and our SPHERE-IRDIS astrometric results obtained in the K1 band with the SpeCal-TLOCI pipeline (epochs 2015--2019). 
    The orbital parameters are: the orbital period ($P$ (yr)), the eccentricity ($e$), the inclination ($i$ (\degr)), the longitude of ascending node ($\Omega$ (\degr)), the argument of periastron ($\omega$ (\degr)), the time of periastron passage ($T_p$ (yr AD)).
     The best orbital fit solution is given by the  MAP as black stars in the non-diagonal panels, and as the solid black vertical lines in the diagonal panels. The $1\,\sigma$ confidence region defined as the shortest interval comprising $68\%$ of the probability around the MAP solution is shown with the vertical dashed red lines in the diagonal panels. As $\Omega$ and $\omega$ are degenerated; confidence regions of $34$\% are given for both. If ($\Omega$, $\omega$) is a solution, then ($\Omega+180\degr$, $\omega+180\degr$) is a solution as well. 
    The color bar (on the \textit{right}) indicates the number of solutions corresponding to a given color for each subplot.} 
    \label{fig:triangle_orbital_fit}
\end{figure*}

 As the MAP gives the peak of probability in the 6D orbital parameter space, the MAP may not correspond to the peak of each 1D distribution (see black vertical lines in diagonal panels in Fig.~\ref{fig:triangle_orbital_fit}), due to the correlation between the orbital fitted elements. Nevertheless, the 2D distributions in Fig.~\ref{fig:triangle_orbital_fit} indicate 
that the MAP solution  (black star) corresponds better to the 2D peaks, which is closer to the 6D reality.

The results are consistent with the previous analysis led by \citet{Chauvin2018}. 
This is expected, first because  the fit corresponds to a linear part of the orbit, and second because it only covers  a small percent of its whole orbit. In Figure~\ref{fig:triangle_orbital_fit} we can see    that a  subsample of very eccentric solutions are found ($e\,\geq\,0.4$), but they are correlated with lower inclinations  ($i\,\leq\,140$\degr) than the inclination of the outer belt ($147$--$153$\degr). Hence, these orbital solutions could still be consistent with the double-belt architecture of the system, even though they  represent a small subsample of the MCMC orbital fit solutions.

All in all, these solutions confirm that the exoplanet HD~95086~b, located at a semimajor axis $51$--$73~\mathrm{au}$ and with a low eccentricity ($e \leq 0.18$), is likely sculpting the inner edge of the outer ring, and cannot alone sustain the large cavity observed between $10$ to $106~\mathrm{au}$ \citep{Su2015,Rameau2016}.

\begin{table}[t]
    \caption{Comparison of the MCMC and K-Stacker solutions within the 68\% confidence interval for the orbital parameters of HD~95086~b. 
    }    
    \centering \small
    \begin{tabular}{ccc} 
    \hline\hline\noalign{\smallskip}\noalign{\smallskip}
         Orbital            & MCMC solutions            & K-Stacker solutions       \\
         Parameter          & (2012-2019)               & (2016-2019)               \\ 
    \noalign{\smallskip}\noalign{\smallskip}\hline\hline\noalign{\smallskip}\noalign{\smallskip}
         $a$ (au)           &$72_{-21}^{+1}$          & $51\,\pm\,2$          \\ \noalign{\smallskip}
         $e$                & $\leq 0.18$          & $0.12\,\pm\,0.03$         \\ \noalign{\smallskip}
         $i$ (\degr)     & $144_{-4}^{+18}$             & $180\,\pm\,15$            \\ \noalign{\smallskip}
         $\Omega$ (\degr)   & $72_{-27}^{+53}\;(+\,180)$ & $-137\,\pm\,62\;(+\,180)$\\ \noalign{\smallskip}
        $\omega$ (\degr)    & $-89_{-2}^{+110}\;(\pm\,180)$  & $-43\,\pm\,62\;(\pm\,180)$ \\ \noalign{\smallskip}
        $T_p$ (yr AD)       & $2004_{-45}^{+105}$          & $2100\,\pm\,28$            \\  \noalign{\smallskip}
        star mass ($\text{M}_\odot$)&        -                & $1.56$         \\ \noalign{\smallskip}\noalign{\smallskip}\hline
    \end{tabular}
    \label{tab:mcmcfit_solutions}
    \tablefoot{  The astrometric data used regarding MCMC solutions (\textit{middle column}) are NaCo \citep{Rameau2013a} and SPHERE \citep[][and this work]{Chauvin2018} spanning from $2012$ to $2019$. As for K-Stacker solutions (\textit{right column}), only the SPHERE data acquired with satellite spots ($2016$--$2019$) were considered.
    The orbital parameters are: the semimajor axis ($a$), the eccentricity ($e$), the inclination ($i$), the longitude of ascending node ($\Omega$), the argument of periastron ($\omega$), the time of periastron passage ($T_p$). If ($\Omega$, $\omega$) is a solution, due to their degeneracy, ($\Omega\,+\,180\degr$, $\omega\,+\,180\degr$) will be a solution as well (see text for additional details). 
}
\end{table}

\footnotetext{The transmission curves are available at \url{https://www.eso.org/sci/facilities/paranal/instruments/sphere/inst/filters.html}}

By using only the four epochs imaged with SPHERE-IRDIS in the K1 band and acquired with the satellite spots enabling a better astrometric calibration (see Table \ref{tab:obslog}),
we also obtained the orbital parameters of HD~95086~b as a by-product of the  K-Stacker algorithm \citep{Lecoroller2015Kstacker}, which is used below in the search for one or two additional inner planets in the system (see Sect.~\ref{sec:spectro}). K-Stacker is an optimization algorithm that  takes advantage of the Keplerian motion of exoplanets on several epochs to then recenter the images according to their Keplerian motion. Thus, it differs from other classical orbital fitting methods, such  as MCMC, as it takes as input images from several epochs instead of derived astrometric positions, and consequently can detect objects otherwise unreachable in each observation considered individually. Hence, the coherence of the orbits found by K-Stacker (within $2\,\sigma$) illustrates its ability to constrain objects with small orbital motion between epochs ($50$ mas, i.e., $\sim1~\mathrm{FWHM}$ in the K1 band) and very faint objects as well;  about the same orbit has been found in the H band, where  HD~95086~b is hardly detectable (i.e., S/N $\leq 5$ for each epoch taken individually, see Fig. \ref{fig:mosaic_epochs_andromeda}). Within the error bars from orbital parameters similar results to the MCMC solutions are found, although the MCMC fit  uses the NaCo and SPHERE data, which  covers a longer timescale 2012--2019 (instead of 2016--2019 for K-Stacker). 
In addition, K-Stacker allows us to constrain the star mass between $1.56$ and $1.59$ \text{M}$_\odot$ (as described in \citealt{LeCoroller2020Kstacker}, Sect. 3.4).

\section{Spectral characterization \label{sec:spectro}}
\begin{figure}[t!]
    \centering
    \includegraphics[width=\columnwidth]{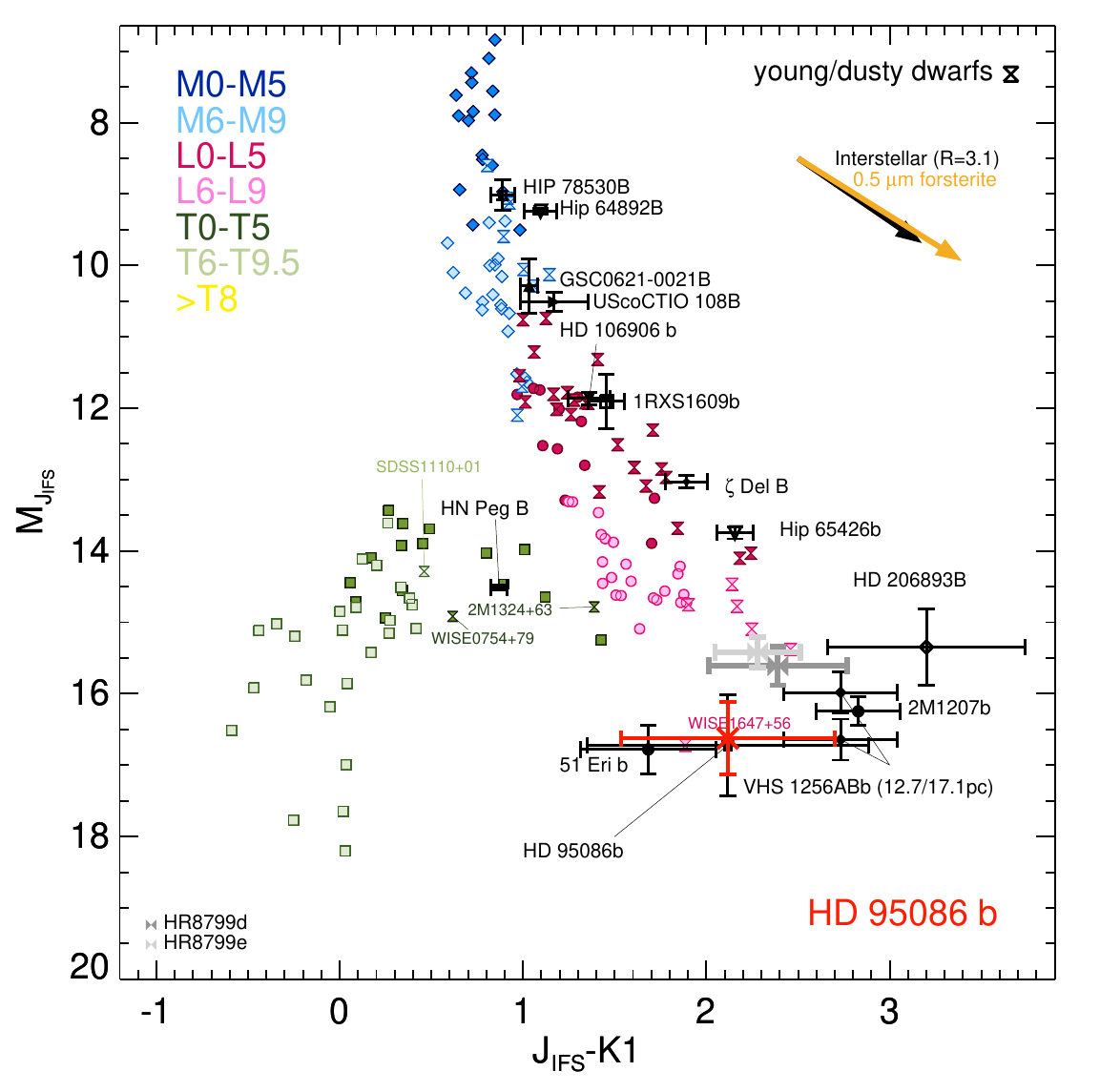}
    \caption{Color-magnitude diagram considering the SPHERE-IRDIS K1 photometry, and the J$_\text{IFS}$ photometry from 1.2 to 1.32~$\mu$m as  extracted from the SPHERE-IFS datacubes.}
    \label{fig:cmdplot}
\end{figure}

The first studies of the infrared colors of HD\,95086\,b rapidly showed that the spectral properties of the planet fall at the late L to L/T transition and that the planet is underluminous compared to the field  dwarfs of similar spectral types \citep{Galicher2014,Derosa2016,Chauvin2018}. The red colors and underluminosity, as shown in Fig.\,\ref{fig:cmdplot}, are characteristic of young L/T objects, and are often associated with the inhibited settling of dust in the upper parts of low surface gravity atmospheres. Based on photometric H ($1.5$--$1.8~\mu$m) and spectroscopic K1 ($1.9$--$2.2~\mu$m) observations of HD 95086 b with GPI, \cite{Derosa2016} confirmed the L-type dusty atmosphere, as evidenced by a featureless low-resolution spectrum and a monotonically increasing pseudo-continuum in the K1 band consistent with a cloudy atmosphere. Considering the $1.2$--$1.6~\mu$m spectrum extracted in this work, we propose below to reinvestigate the spectral properties of HD\,95086\,b. We will consider the best atmosphere models fitting current observations from $1.2$ to $3.8~\mu$m, as well as the possibility of having a circumplanetary disk around the planet b, as seen for instance in the younger Solar System analog PDS\,70, also a member of Sco-Cen for the planet c \citep{keppler2018_PDS70,Isella2019}.
   
\subsection{Comparison to models}

\begin{figure*}[p]
    \centering
    \includegraphics[width=\textwidth]{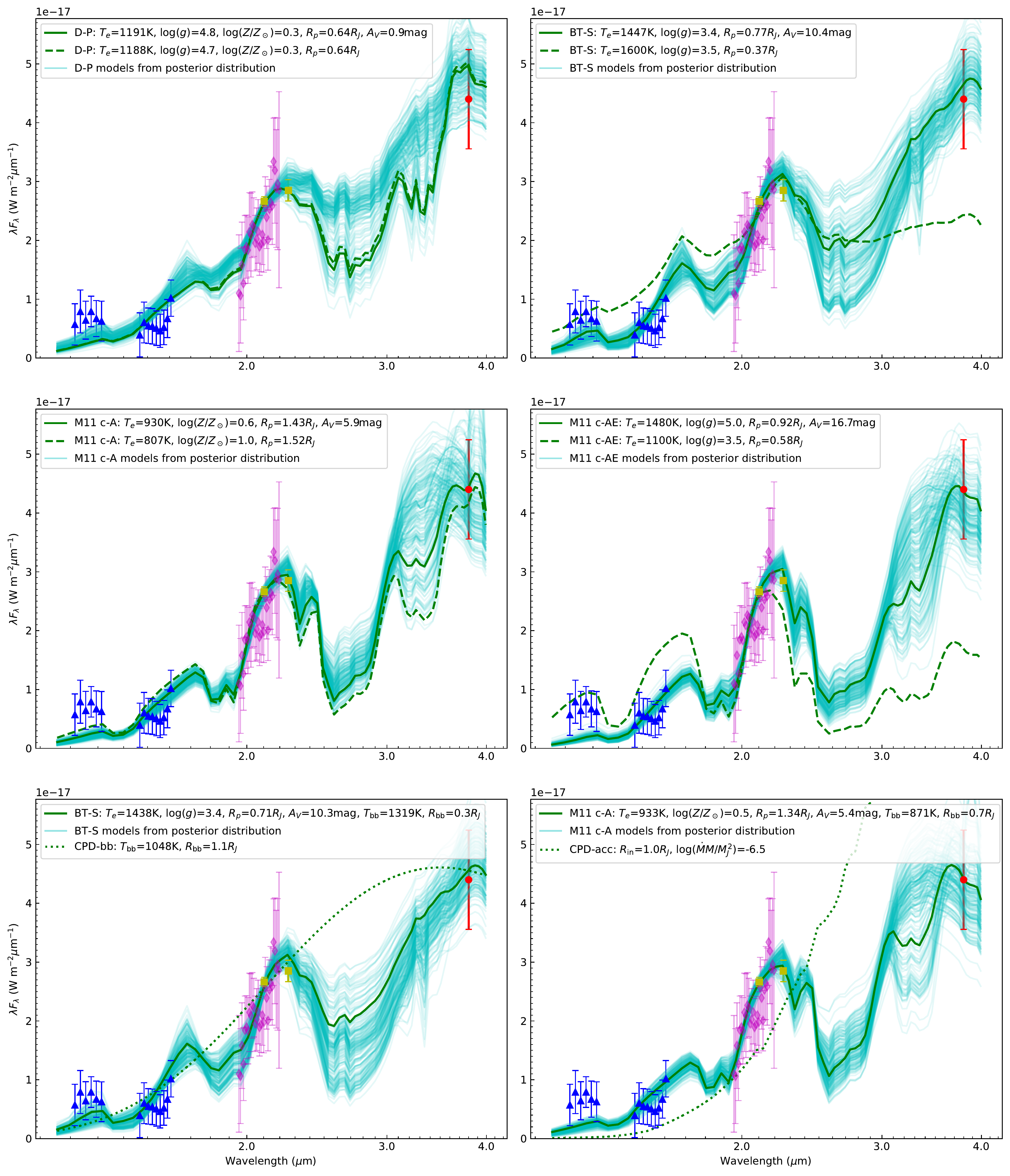}
    \caption{Measured spectrum of HD~95086~b (see Fig. \ref{fig:spectrum_complete} for  legend information) compared to the best-fit models retrieved by \texttt{special} for each atmospheric model grid considered in this work (\textit{top} four panels: DRIFT-PHOENIX, BT-SETTL, \citetalias{Madhusudhan2011} cloud-A, \citetalias{Madhusudhan2011} cloud-AE), and for different circumplanetary disk models (\textit{bottom} two panels).
    All solid lines correspond to extinction $A_V$ considered as a free parameter, while the dashed lines correspond to a fixed extinction of $A_V$ = 0 mag. 
    The circumplanetary disk models consist of either mixed atmosphere (BT-SETTL or M11 c-A) + extra blackbody models (solid lines) or CPD-only models (dotted lines). The latter correspond to either a debris CPD model (bottom left panel) or a viscous CPD model \citep[bottom right panel;][]{Zhu2015a}, where the parameter $\log(\dot{M}M/M_J^2)$ corresponds to the mass accretion rate. 
    All types of models have a similar level of support, except for the BT-SETTL and M11 c-AE models without extinction, and the viscous CPD model  
    (Table~\ref{tab:spec_results}).
    }
    \label{fig:spectral_fits}
\end{figure*}

To constrain the physical properties of the exoplanet HD~95086~b with atmospheric and circumplanetary disk models, we used the \verb+special+ package \citep{Christiaens2021}, first known as \verb+specfit+,\ a module
of the open-source python package \verb+VIP+\footnote{Available at \url{https://github.com/vortex-exoplanet/VIP}.} \citep{GomezGonzalez2017} and now available as a distinct package\footnote{Available at \url{https://github.com/VChristiaens/special}.}. The  \verb+special+ package is compatible with any atmospheric grid if a snippet function which reads the input grid files is provided. It can also fit for blackbody components (either alone or as additional component(s) to the atmosphere), the optical extinction $A_V$, the optical-to-selective extinction ratio $R_V$, or the intensity of emission lines that are provided in a dictionary.  \verb+special+ utilizes the MCMC sampler \verb+emcee+ \citep{Foreman-Mackey2013} to retrieve in a Bayesian framework the most likely physical parameters of any stellar or substellar object based on its spectral energy distribution. Models are linearly interpolated between grid points. The log-likelihood expression provided to the sampler is
\begin{equation}
\label{Eq:GoodnessOfFit}
\log \mathcal{L}(D|M) = - \frac{1}{2}\big[\mathbf{W}(\mathbf{F_{\rm obs}}-\mathbf{F_{\rm mod}})^T\big] \mathbf{C^{-1}} \big[\mathbf{W}^T(\mathbf{F_{\rm obs}}-\mathbf{F_{\rm mod}})\big]
,\end{equation}
where $F_{\rm obs}$ and $F_{\rm mod}$ are   the observed and model fluxes; $\mathbf{C}$ is the spectral covariance matrix (see Appendix~\ref{app:SpectralCovariance}); and $\mathbf{W}$ is a vector of normalized weights that are proportional to the relative width $\delta \lambda/\lambda$ of each spectral channel or photometric filter. The last prevents the fit from putting too much emphasis on the IFS points (higher density of measurements) at the expense of the photometric points, which cover a wider spectral range \citep[e.g.,][]{Ballering2013,Olofsson2016}. Nevertheless, to test their effect on the fits to the HD\,95086\,b spectrum, we show in  Appendix \ref{app:cornerplots_special} a comparison of the best-fit models obtained with and without these additional weighting coefficients.

For each fit the MCMC was run with 100 walkers until the number of steps met a criterion based on autocorrelation time   for convergence. The adopted criterion is that the number of steps is 50 times larger than the autocorrelation time for all free parameters (see, e.g., documentation of \verb+emcee+; \citealt[][]{Foreman-Mackey2013}). This resulted in $5\,000$--$20\,000$ steps for atmospheric models and $100\,000$--$250\,000$ steps for the atmospheric+circumplanetary disk models. 
We then used a ``burn-in'' factor of 0.5.

\subsubsection{Atmospheric models}

We provided the following public grids of atmospheric models as input to \verb+special+:
BT-SETTL \citep{Allard2012, Allard2014_btsettl}, DRIFT-PHOENIX \citep{Woitke2003,Woitke2004,Helling2006,Helling2008}, and the grids of A and AE forsterite cloud models with 60 $\mu$m modal grain size distribution from \citet[][hereafter \citetalias{Madhusudhan2011}]{Madhusudhan2011}.

These grids have different prescriptions for clouds, opacity, and dust. The BT-SETTL and DRIFT-PHOENIX models both rely on the PHOENIX atmosphere model \citep{Hauschildt1992}, while the \citetalias{Madhusudhan2011} models use a modified version of the TLUSTY atmosphere model \citep{Hubeny1988,Hubeny1995}.
The BT-SETTL models consider dust formation using cloud microphysics (condensation and sedimentation mixing timescales) to compute dust grain sizes in a self-consistent way. 
The DRIFT-PHOENIX models rely on the non-equilibrium cloud model DRIFT and account for the formation of dust grains through a kinetic approach (grain formation, growth, settling, advection, and evaporation). Seven different solids are considered for dust and cloud formation:
MgSiO$_3$[s], Mg$_2$SiO$_4$[s], MgO[s], SiO$_2$[s], SiO[s], Al$_2$O$_3$[s], and TiO$_2$[s].
In the \citetalias{Madhusudhan2011} grids, different distributions of the grain sizes and vertical extent are considered. The A and AE cloud models correspond to clouds extending to the top of the atmosphere and half the pressure scale height, respectively. In this work we only consider the \citetalias{Madhusudhan2011} grids calculated using forsterite (Mg$_2$SiO$_4$[s]) clouds owing to the completeness of the grids. Given the very red slope of the spectrum of HD~95086~b and the better fits obtained with the cloud-A grid (see below), we did not consider the grids with thinner and/or lower cloud distributions presented in \citetalias{Madhusudhan2011}.

All grids consider effective temperature (\teff) and surface gravity (\logg) as free parameters, except for the \citetalias{Madhusudhan2011} forsterite cloud-A models (\logg~fixed to 4.0). The DRIFT-PHOENIX and \citetalias{Madhusudhan2011} forsterite cloud-A grids also include  metallicity (\logZ) as a free parameter. The photometric radius $R_\text{phot}$ was allowed to take values between $0.1$~\rj~and $5$~\rj, the lower bound allowing for the possibility of a fraction of the emission not reaching the observer (e.g.,~in the presence of a large amount of dust around the planet). Table~\ref{tab:spec_setup} summarizes the sampling of the free parameters for each grid. For each model grid, we carried out two types of fits; the optical extinction $A_V$ was either fixed to 0 (the value estimated for the star; \citealt{Chen2012}) or left as a free parameter to account for the possible presence of dust around the planet. We considered the \citet{Cardelli1989} extinction law and allowed the value of $A_V$ to span from 0 to 20 mag. 

\begin{table}[t]
    \caption{Parameter ranges probed with the different grids of atmospheric models.}
    \label{tab:spec_setup}
    \begin{threeparttable}
        \centering \small
        \begin{tabular}{lccc}
            \hline
            \hline\noalign{\smallskip}
             Model& $T_{\rm eff}$ (K)  & $\log(g)$ & $\log(Z/Z_{\odot})$ \\
             \noalign{\smallskip}\hline\hline\noalign{\smallskip}
             D-P\tnote{a}  & 1000..2000 [100]\tnote{b} & 3.0..5.5 [0.5] & -0.3..0.3 [0.3]\\ \noalign{\smallskip}
             BT-SETTL  & 1200..4000 [100] & 2.5..5.0 [0.5] & (0)\tnote{c} \\ \noalign{\smallskip}
             \citetalias{Madhusudhan2011} c-A & 600..1700 [100] & (4.0)\tnote{c} & 0.0..1.0 [0.5]\\   \noalign{\smallskip}
             \citetalias{Madhusudhan2011} c-AE & 700..1700 [100] & 3.5..5.0 [100] & (0)\tnote{c}\\ \noalign{\smallskip}
            \hline
        \end{tabular}
        \begin{tablenotes}
        \item[a] D-P stands for DRIFT-PHOENIX.
        \item[b] Step of the grid provided in square brackets.
        \item[c] Grid only available with the value provided in parentheses.
    \end{tablenotes}
\end{threeparttable}
\end{table}

The best-fit models retrieved by \verb+special+ for each grid of atmospheric models are presented in the top four panels of Fig.~\ref{fig:spectral_fits}.
The solid and dashed green lines correspond to the maximum-likelihood samples among all posterior samples, when $A_V$ is included as a free parameter and set to 0 mag, respectively.
In order to visualize the uncertainties on the retrieved physical parameters, we also plotted 200 random samples from the posterior distribution inferred by the MCMC in the free $A_V$ case (cyan curves in Fig.~\ref{fig:spectral_fits}). Table~\ref{tab:spec_results} reports the most likely parameters for each type of model based on their marginalized distribution (some of which are shown in  Figs.~\ref{fig:corner_plots_BT-Settl_BB} to \ref{fig:corner_plot_M11cA_BB}). 

\begin{table*}[]
\caption{Physical parameters of HD ~95086~b retrieved by \texttt{special} for different atmospheric and CPD models.}
\label{tab:spec_results}
\begin{threeparttable}
\begin{tabular}{lccccccccc}
\hline
\hline\noalign{\smallskip}
Model     & $T_{\rm eff}$        & $\log(g)$                    & $R_\text{p}$                   & $\log(Z/Z_{\odot})$           & $A_V$                & $T_{\rm bb}$       & $R_{\rm bb}$        & $M_b$                & $\Delta$AIC\tnote{a} \\
                          & (K)                  &                              & (\rj)                   &                              & (mag)                & (K)                & (\rj)               & (\mj)                &              \\
\noalign{\smallskip}\hline\hline\noalign{\smallskip}
Blackbody                  & -                    &         -                     & -                       & -                            & (0)                  & $1039_{-26}^{+35}$ & $1.1 \pm 0.1$       & -                    & { 0}            \\\noalign{\smallskip}
D-P ($A_V=0$)\tnote{b}                 & $1247_{-89}^{+56}$   & $4.71_{-0.51}^{+0.46}$       & $0.61_{-0.06}^{+0.09}$  & $0.3_{-0.6}^{+0.0}$ & (0)                  & -                  & -                   & $2.9_{-2.8}^{+7.8}$  & { 0.4}          \\\noalign{\smallskip}
BT-SETTL ($A_V=0$)       & $1600_{-20}^{+3}$    & $3.50_{-0.1}^{+0.0}$         & $0.38\pm 0.01$          & (0)                          & (0)                  & -                  & -                   & $0.2_{-0.0}^{+0.0}$  & 52.5         \\\noalign{\smallskip}
\citetalias{Madhusudhan2011}-cA ($A_V=0$)\tnote{b}            & $808_{-35}^{+43}$    & (4.0)                        & $1.59_{-0.32}^{+0.07}$  & $1.0_{-0.1}^{+0.0}$ & (0)                  & -                  & -                   & $7.8_{-1.7}^{+3.0}$  & { 1.9}          \\\noalign{\smallskip}
\citetalias{Madhusudhan2011}-cAE ($A_V=0$)\tnote{b}          & $1101_{-29}^{+7}$    & $3.5 \pm 0.0$       & $0.59 \pm 0.02$         & (0)                          & (0)                  & -                  & -                   & $0.5_{-0.0}^{+0.0}$  & 98.9         \\\noalign{\smallskip}
D-P\tnote{b}              & $1549_{-353}^{+150}$ & $5.0_{-0.6}^{+0.5}$          & $0.66_{-0.08}^{+ 0.05}$ & $0.3_{-0.4}^{+0.0}$ & $1.2_{-1.2}^{+13.1}$ & -                  & -                   & $0.6_{-0.4}^{+17.7}$ & { 2.4}          \\\noalign{\smallskip}
BT-SETTL        & $1456_{-91}^{+99}$   & $3.4_{-0.9}^{+1.6}$          & $0.79_{-0.12}^{+0.11}$  & (0)                          & $11.5_{-2.1}^{+1.7}$ & -                  & -                   & $0.3_{-0.2}^{+1.7}$  & { -0.8}        \\\noalign{\smallskip}
\citetalias{Madhusudhan2011}-cA         & $977_{-113}^{+241}$  & (4.0)                        & $1.32_{-0.31}^{+0.16}$  & $0.5_{-0.3}^{+0.4}$          & $8.7_{-3.5}^{+4.5}$  & -                  & -                   & $6.9_{-3.5}^{+1.5}$  & { -0.6}         \\\noalign{\smallskip}
\citetalias{Madhusudhan2011}-cAE\tnote{b}       & $1383_{-110}^{+223}$ & $3.5_{-0.0}^{+1.4}$ & $0.91 \pm 0.12$         & (0)                          & $16.0_{-1.8}^{+1.7}$ & -                  & -                   & $1.5_{-0.9}^{+9.5}$  & { 2.2 }         \\\noalign{\smallskip}
D-P + BB       & $1600_{-271}^{+143}$ & $5.0_{-0.6}^{+0.5}$          & $0.66_{-0.07}^{+ 0.05}$ & $0.0_{-0.2}^{+0.3}$          & $1.9_{-1.9}^{+13.1}$ & $110_{-9}^{+327}$  & $0.3_{-0.3}^{+2.4}$ & $0.6_{-0.4}^{+17.5}$ & { 6.3}          \\\noalign{\smallskip}
BT-SETTL + BB  & $1451_{-86}^{+101}$  & $3.3_{-0.8}^{+1.7}$          & $0.80_{-0.14}^{+0.09}$  & (0)                          & $10.8_{-1.3}^{+2.4}$ & $108_{-8}^{+324}$  & $0.2_{-0.2}^{+2.6}$ & $0.3_{-0.2}^{+1.8}$  & { 2.9}          \\\noalign{\smallskip}
\citetalias{Madhusudhan2011}-cA + BB  & $979_{-111}^{+276}$  & (4.0)                        & $1.28_{-0.32}^{+0.19}$  & $0.5_{-0.3}^{+0.4}$          & $8.9_{-3.1}^{+5.1}$  & $187_{-87}^{+229}$ & $0.4_{-0.4}^{+2.4}$ & $6.6_{-3.5}^{+1.4}$  & { 3.2}          \\\noalign{\smallskip}
\citetalias{Madhusudhan2011}-cAE + BB  & $1421_{-128}^{+187}$ & $3.6_{-0.1}^{+1.3}$          & $0.87_{-0.10}^{+0.16}$  & (0)                          & $16.1_{-1.7}^{+1.8}$ & $156_{-56}^{+357}$ & $0.4_{-0.4}^{+2.3}$ & $1.5_{-1.0}^{+9.3}$  & { 5.5} \\\noalign{\smallskip}
\hline
\end{tabular}
\begin{tablenotes}
\item[a] $\Delta {\rm AIC} = {\rm AIC}- {\rm AIC}_{\rm bb}$. Bold font is used to highlight models with $\Delta$AIC-$\Delta$AIC$_{\rm min} < 10$ (i.e.,~models with the most support).
\item[b] The MCMC converged at the edge of the allowed range for at least one parameter; uncertainties are likely underestimated for all parameters of this model.
\end{tablenotes}
\end{threeparttable}
\end{table*}

For each type of model we computed the Akaike information criterion \citep[AIC;][]{Akaike1974} to evaluate which models reproduced  the observed spectrum better. The definition of the AIC takes into account the number of free parameters in order to avoid overfitting. Low values of AIC correspond to a good fit with a relatively small number of free parameters. For each type of model we report $\Delta$AIC = AIC$_{\rm model}$ - AIC$_{BB}$ in Table~\ref{tab:spec_results}, where AIC$_{BB}$ is the AIC obtained for a fit to a single blackbody component (no atmospheric model).
 
We show in Figs.~\ref{fig:corner_plots_BT-Settl_BB} to \ref{fig:corner_plot_M11cA_BB}.  the corner plots retrieved by \verb+special+ for some of the models with minimum AIC values, to show the degeneracy between some parameters. When $\log(g)$ is a free parameter, an additional panel is shown for the mass posterior distribution estimated from the $\log(g)$ and $R_\text{phot}$ posterior distributions (i.e.,~not a free parameter of the fit).

Inspection of Figs.~\ref{fig:spectral_fits} and \ref{fig:corner_plots_DP} and Table~\ref{tab:spec_results} reveal two types of models with high support. Depending on the assumed atmospheric grid, either the planet is found to have a relatively low effective temperature (800--1200~K), small to medium amount of extinction ($A_V \lesssim 10$ mag), and super-solar metallicity (DRIFT-PHOENIX and \citetalias{Madhusudhan2011} forsterite cloud-A models; e.g.,\, Figs.~\ref{fig:corner_plots_DP} and \ref{fig:corner_plot_M11cA}) or it has a higher effective temperature (1200--1600~K) and a high level  of extinction by surrounding dust ($A_V \gtrsim 10$ mag) for a solar metallicity (BT-SETTL and \citetalias{Madhusudhan2011} forsterite cloud-AE models; e.g.,\, Fig.~\ref{fig:corner_plots_BT-Settl_BB}). In particular, we find that only the DRIFT-PHOENIX and \citetalias{Madhusudhan2011} forsterite cloud-A models can reproduce the red slope of the spectrum when the extinction is set to $A_V = 0$ mag (dashed lines in Fig.~\ref{fig:spectral_fits}), while the BT-SETTL and \citetalias{Madhusudhan2011} cloud-AE models are unable to account for the observed spectrum without extinction ($\Delta$AIC $> 10$; \citealt{Burnham2002}). These two  solutions can also be seen in the DRIFT-PHOENIX posterior samples (top left panel of Fig.~\ref{fig:spectral_fits}) and corner plot (Fig.~\ref{fig:corner_plots_DP}) obtained when $A_V$ is set as a free parameter: two clusters of solutions can be seen corresponding to low $T_{\rm eff}$, high $\log(g)$, high $\log(Z/Z_{\odot}),$ and low $A_V$ on the one hand, and high $T_{\rm eff}$, unconstrained $\log(Z/Z_{\odot}),$ and high $A_V$ on the other hand.

Our results suggest that a large amount of dust is present, either in the upper part of the atmosphere (super-solar metallicity; see details in Sect.~\ref{discu:redorigin}) and/or around the planet (to account for the extinction). Whether circumplanetary dust could emit an additional thermal component detectable in our spectrum is further investigated in the next section.

\subsubsection{Circumplanetary disk models}
HD~95086~b is located between two debris disk belts. As it is one of the reddest substellar object known \citep{Derosa2016}, and a large amount of dust appears to be necessary to account for the observed spectrum (see previous section), we also investigated the possible presence of a circumplanetary disk (CPD) signature in our spectrum. We considered two types of CPDs:  a circumplanetary primary viscous disk, in which the exoplanet b still accretes material, and a circumplanetary debris disk consisting of heated grains and modeled by a blackbody component. 

We used the SED predictions from the grid of accreting CPD models presented in \citet{Zhu2015b}, which are characterized by two free parameters: the inner truncation radius of the CPD and the mass accretion rate, spanning $1$ to $4$~R$_{\rm Jup}$ and $10^{-4}$ to $10^{-7}$~M$_{\rm Jup}^2$~yr$^{-1}$, respectively. 
For debris CPDs we considered either  a single blackbody component (without atmospheric model), as performed recently for PDS~70~b \citep[][]{Wang2020,Stolker2020_II}, 
or an additional blackbody component besides the emission from the atmosphere.
We allowed the values of the blackbody temperature $T_{\rm bb}$ to range between $100$~K and $2000$~K and a blackbody radius between $0.1$ and $10$~\rj, with the condition that $T_{\rm bb}$ is lower than or equal to $T_{\rm eq}$, where $T_{\rm eq}$ is the equilibrium temperature corresponding to a distance of $R_{\rm bb}$ (i.e.,~the extreme case of a spherical shell of optically thick hot dust).
If the condition is not met for a particular sample, its log-likelihood is set to minus infinity.

For viscous CPD models the MCMC converged at the edge of the parameter space in terms of inner truncation radius ($1.0$~\rj) for a mass accretion rate of $\sim 10^{-6.5}$~\mj$^2$~yr$^{-1}$. However, the slope of the viscous CPD models is too red to reproduce the observed spectrum on its own (dotted line in bottom right panel of Fig.~\ref{fig:spectral_fits}), leading to a poor fit ($\Delta$AIC $\sim$ 636). This does not prevent the possibility of a combination of atmospheric and viscous CPD emission. Future measurements in the H$\alpha$ filter are required to constrain the accretion rate and definitely rule out this hypothesis.

The single-blackbody model leads to a satisfactory fit for a temperature of $1039_{-26}^{+35}$~K and a photometric radius of $1.1 \pm 0.1$~\rj~(bottom left panel of Fig.~\ref{fig:spectral_fits} and Fig.~\ref{fig:corner_plots_BT-Settl_BB}). It is one of the models that minimizes the $\Delta$AIC.
For atmospheric+debris CPD models, we find that the addition of a blackbody component can also reproduce the observed spectrum without significantly increasing the $\Delta$AIC value. The minor improvement for the maximum-likelihood atmospheric+debris CPD models is such that $\Delta$AIC is still lower than 10, hence implying a similar level of support as models with fewer free parameters (see bold values of $\Delta$AIC in Table~\ref{tab:spec_results}), including super-solar metallicity atmosphere models. The corner plots associated with atmosphere+debris CPD models (e.g.,~Fig.~\ref{fig:corner_plot_M11cA_BB} for \citetalias{Madhusudhan2011} c-A+debris CPD) show that a number of solutions correspond to negligible contribution from the additional blackbody. However, some high-likelihood solutions involve values of $T_{\rm bb} \sim$ 800--1300 K and $R_{\rm bb} \sim$ 0.3--0.7~\rj~ that are comparable to $T_{\rm eff}$ and $R_\text{phot}$, respectively. This is the case in particular for the maximum-likelihood BT-SETTL+CPD and \citetalias{Madhusudhan2011}c-A+CPD models, shown as a green curve in the bottom two panels of Fig.~\ref{fig:spectral_fits}. Considering the blackbody temperature to be the equilibrium temperature of the dust, $T_{\rm bb} \approx 1319$~K (resp. $871$~K) for $T_{\rm eff} \approx 1438$~K (resp. $933$~K) would imply that the heated dust is located near the top of the atmosphere in either case.

\subsubsection{Evolutionary models}

\begin{figure*}[t!]
    \centering
    \includegraphics[width=\textwidth]{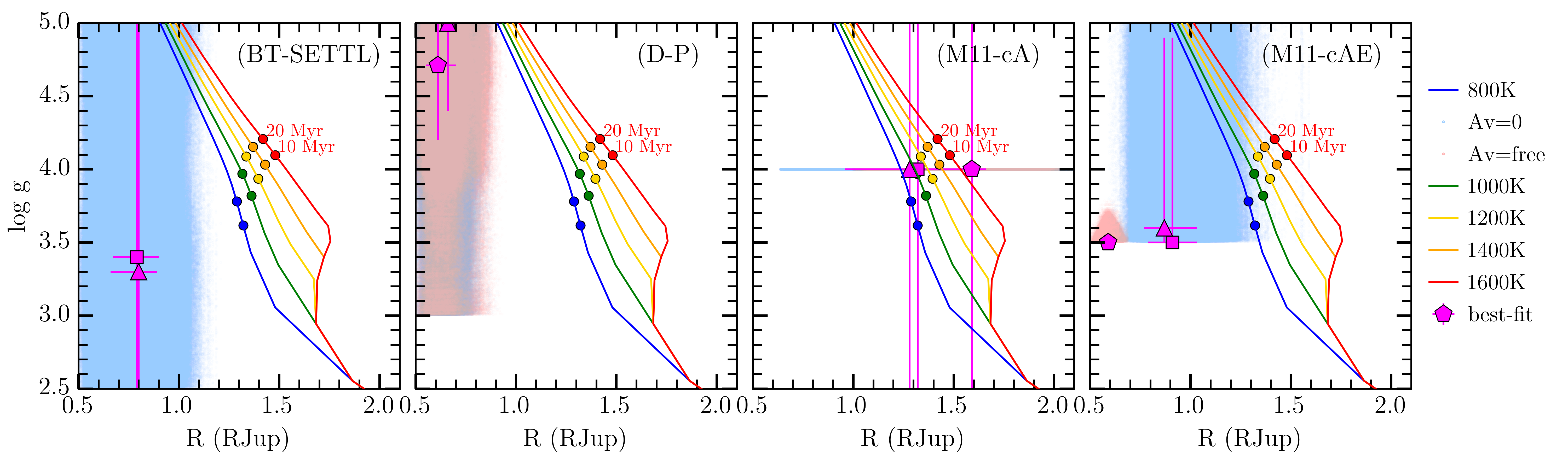}
    \includegraphics[width=\textwidth]{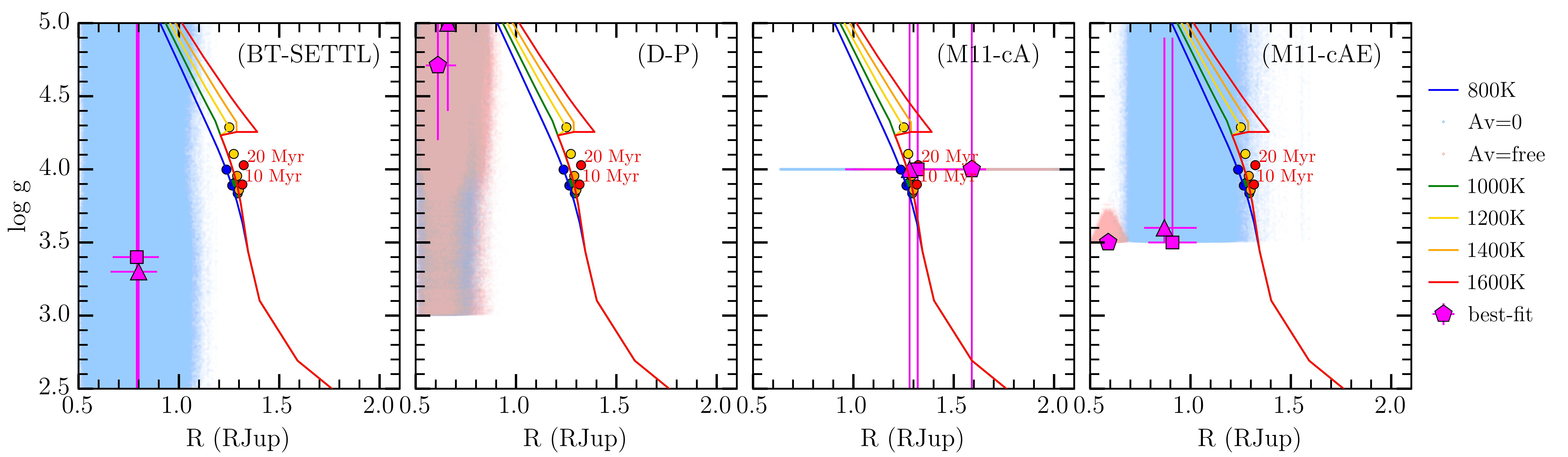}
    \caption{Comparison of the solutions for the different models of atmosphere (\textit{left:} BT-SETTL, \textit{middle left:}  DRIFT-PHOENIX, \textit{middle right:}   \citetalias{Madhusudhan2011} cloud-A, and \textit{right:}    \citetalias{Madhusudhan2011} cloud-AE) and models of evolution (\textit{top panel}: BEX-Hot and \textit{bottom panel:} BEX-Warm). As for the  atmospheric models, the best-fit solutions are shown in pink, with a pentagon if the optical extinction $A_V = 0$, a square  if the extinction is a free parameter, and a triangle   if there is an additional blackbody fitted; 
    the samples of solutions are in red if the extinction $A_V = 0$ and in blue if the extinction is a free parameter. For the  evolutionary models, the solid curves represent the surface gravity as a function of the planetary radius given different effective temperature (blue for the coolest to red for the hottest). The colored  circles represent the expected log g and R at a given age (10 or 20 Myr): hence lower and upper values expected for HD 95086 aged  13.3 Myr.}
    \label{fig:consistence_atm_models_vs_evolutive_models}
\end{figure*}

Atmosphere modeling of exoplanets and brown dwarfs consists in describing the physical and chemical processes at play in substellar atmospheres using radiative-convective equilibrium models, which can include non-equilibrium chemistry processes, the effect of stellar irradiation, cloud formation, dust settling, and/or mixing  to simulate spectra. Even so, their results must be compared to predictions of evolutionary models, which give the evolution of the internal structure of exoplanets and brown dwarfs in time, to exclude non-physical solutions. To do so, based on the apparent photometry, age, and distance of HD~95086~b, we used the Bern EXoplanet cooling tracks \citep[BEX, ][]{Marleau2019a} with the AMES-COND atmospheres \citep{Baraffe2003}, corresponding to hot or warm start initial conditions, to derive the predicted bulk properties  of
the planet (luminosity, mass, effective temperature, surface gravity, and radius). Following a similar approach to \cite{Delorme2017_HD206893}, we compared the regime of solutions between the two atmosphere and evolutionary models for the predicted surface gravity and radius, as shown in Fig.\,\ref{fig:consistence_atm_models_vs_evolutive_models}. 
At the age of HD\,95086, BEX models predict for effective temperatures between $800$ and $1600$\,K, typical values of $1.2$--$1.5$~\rj~and $\log(g)$ between $3.6$ and $4.2$, with a predicted mass of $3$--$12$~\mj~for HD\,95086\,b.

From the best-fit solutions of the atmosphere models shown in Fig. \ref{fig:consistence_atm_models_vs_evolutive_models} and reported in Table\,\ref{tab:spec_results}, the only model consistent with the BEX predictions is \citetalias{Madhusudhan2011} with forsterite cloud-A models. The spectrum of HD~95086~b is well reproduced by a super-solar metallicity 800--1200~K atmosphere, with only a small to medium amount of additional extinction required. The atmospheric fits lead to values of radii ($1.32_{-0.31}^{+0.16}$~\rj) comparable to the physical radius predicted by the BEX models. For all other grids of models, the favored values of photometric radius (0.6--1.0\,\rj) appear smaller than expected, which may either suggest that the \citetalias{Madhusudhan2011} forsterite cloud-A models are the most appropriate for the case of HD~95086~b or that a fraction of the atmospheric flux is obscured by circumplanetary dust, as we  discuss below. 

\subsection{Origin of the red spectral slope of b} \label{discu:redorigin}

\subsubsection{Super-solar metallicity atmosphere}
The good fit to high-metallicity atmospheric models suggests that HD~95086~b could be somewhat similar to the  unusually red L dwarfs \citep[e.g.,][and references therein]{Gizis2012,Marocco2014}. \citet{Looper2008} and \citet{Stephens2009}   suggested that a high metallicity was indeed responsible for the unusually red slope of the field L dwarfs in their samples (for which a low gravity appeared unlikely). A high metallicity facilitates the production of dust grains in the atmosphere, hence clouds.
To reproduce the spectrum of unusually red L dwarfs, \citet{Marocco2014} tested different extinction laws corresponding to different dust compositions, namely corundum (Al$_2$O$_3$), enstatite (MgSiO$_3$), iron, and ISM-like  ($R_V=3.1$). They found that dereddening with any of these extinction laws (including ISM) makes the spectra consistent with that of standard field L dwarfs. Their findings also corroborate our two categories of best-fit models. It appears observationally difficult from the near-infrared spectrum alone to constrain where the dust is located, either in the upper atmosphere (super-solar metallicity enhancing cloud formation) or around the planet (high circumplanetary extinction).

\subsubsection{Circumplanetary disk}

The first evidence for the presence of a viscous circumplanetary disk was presented 
in \citet{Christiaens2019_PDS70} and \citet{Isella2019} for the case of protoplanet PDS~70~b. This is consistent with the estimated young age of the system ($\sim$ 5 Myr) and the presence of a large amount of gas in the protoplanetary disk. Since the HD~95086 system is older ($\sim13.3~\mathrm{Myr}$), and a low amount of CO has been observed \citep{2019Booth}, the viscous disk is not favored. Our best-fit viscous CPD model also appears too red on its own to account for the observed near-infrared spectrum.  
Nonetheless, considering that \citet{Chinchilla2021_Halpha_accretion} and \citet{Eriksson2020} find evidence for ongoing accretion onto a planetary-mass object with a main estimated age of $25$\,Myr and $30$--$40$\,Myr, respectively, future measurements in the H$\alpha$ filter are required to constrain the accretion rate of HD~95086~b, and definitively rule out the possibility of a combined  atmosphere+viscous CPD model. We note that
the comparison with the studied system by \citet{Eriksson2020} could be nuanced as they studied a binary system of two M stars that could keep the gas-rich disk longer  \citep[e.g., known as the  Peter Pan disks, ][]{Silverberg2020_Peterpan}, and whose  age is poorly constrained and debated. It could be younger, with a recent averaged age estimated at about $20$\,Myr \citep[and an age range of $3$--$65$\,Myr, ][]{Ujjwal2020}. Finally, if H-alpha measurements are necessary to measure the accretion rate, they could also be a sign of chromospheric activity as it is a common feature in  late M and early L objects \citep{Chinchilla2021_Halpha_accretion}.

An alternative explanation for the red slope of HD~95086~b is the presence of circumplanetary dust causing  high extinction. In this scenario we can  expect the dust located the closest to the planet to be heated to high enough temperatures to show a signature corresponding to an IR excess comparable to the atmospheric emission alone. 
We find in Sect.~\ref{sec:spectro} that the addition of a second blackbody component to model a circumplanetary debris disk around HD~95086~b can also reproduce the observed spectrum with a similar level of support to models with fewer free parameters (i.e.,~without an extra blackbody component).
In particular, the solution from the model of atmosphere \citetalias{Madhusudhan2011} with cloud-A and an extra blackbody component and optical extinction $A_V$ is consistent with predictions from evolutionary models. This model  favors a super-solar metallicity ($0.5^{+0.4}_{-0.3}$) with a medium level of extinction ($A_V = 8.9^{+5.1}_{-3.1}\rm\,mag$). Hence, both a circumplanetary disk and a super-solar metallicity could account for the red spectral slope of b, and both could be present together;  if there is a debris CPD, some of this debris might have been accreted onto the planet's atmosphere, which naturally increases the metallicity. 

A small subset of the debris disk CPD solutions from models of atmosphere is expected to have a non-negligible signature at near-IR wavelengths (solutions with large $T_{\rm bb}$ and $R_{\rm bb}$ similar to $R_\text{phot}$, as shown in the bottom panels of Fig.~\ref{fig:spectral_fits}). In these cases, considering the blackbody temperature to correspond to the equilibrium temperature at the separation of the dust implies that the heated dust would be located near the top of the atmosphere (within $\sim$ 1 \rj~distance above the atmosphere). However, since a significant fraction of posterior samples corresponds to no significant excess at near-IR wavelengths, it is also possible that only cold circumplanetary dust is present. 
We note that \citet{Perez2019a} used ALMA 1.3\,mm observations from \cite{Su2017} to search for the presence of a circumplanetary disk around HD\,95086\,b, and derived an upper limit of 30\,$\mu$Jy at the planet location.
All the models retrieved by \texttt{special} are consistent with this non-detection, being over three orders of magnitude fainter than their upper limit.

\begin{figure*}[t!]
    \centering
    \includegraphics[width=16cm]{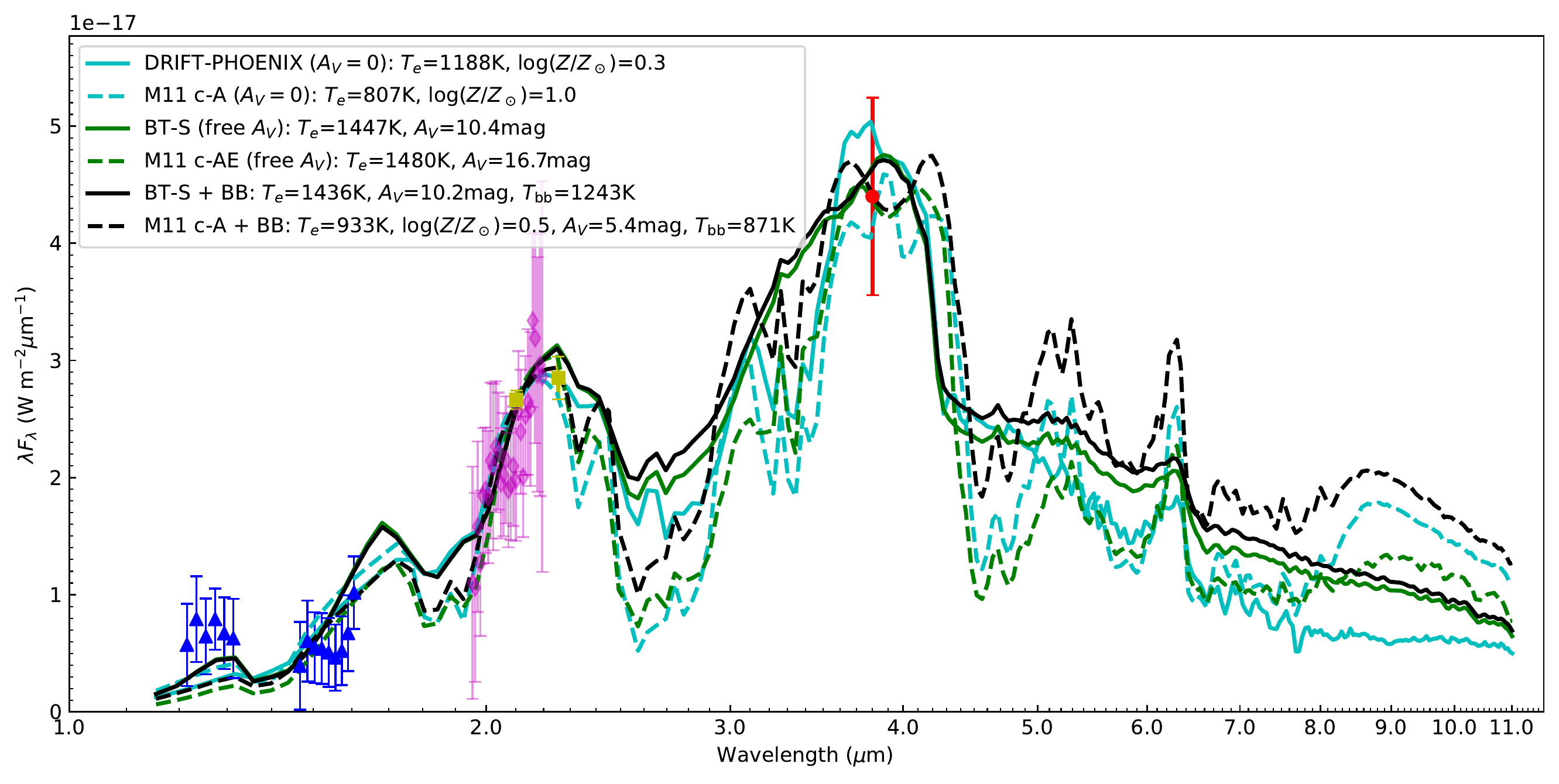}
    \caption{Spectrum of HD~95086~b compared to the most likely models retrieved by \texttt{special} with the highest-likelihood atmospheric and CPD models, extended up to 11 $\mu$m. Cyan curves show models with high metallicity but no extinction: DRIFT-PHOENIX (solid line) and \citetalias{Madhusudhan2011} cloud-A (dashed line).
    Green curves show models with mid to high extinction: BT-SETTL (solid line) and \citetalias{Madhusudhan2011} cloud-AE (dashed line).
    Black curves show models with both extinction and an extra blackbody component: BT-SETTL (solid line) and \citetalias{Madhusudhan2011} cloud-A (dashed line). There is no atmospheric model+viscous CPD reported here as their best model does not represent a good enough fit  to the data with respect to the other models.
    \label{fig:LongWLpredictions}}
    
\end{figure*}

To test whether longer infrared wavelength observations may allow us to distinguish between high-metallicity cold circumplanetary dust and hot+cold circumplanetary dust, we show in Fig.~\ref{fig:LongWLpredictions} the predictions at longer wavelengths for the highest-likelihood models of each type. The predicted spectra for six of the most likely models are reported up to 11 $\mu$m: two models with high metallicity only and no extinction (in cyan), two models with mid to high extinction (i.e., cold circumplanetary dust), and two models with both extinction and an extra blackbody component (i.e., both hot and cold circumplanetary dust). We see that multiple accurate measurements at longer wavelengths, possibly including  a spectrum, would be required to distinguish between the different scenarios since the predicted fluxes are still relatively similar for the different scenarios. 

On the other hand, observations at short wavelengths in the visible and the near-infrared may also reveal the presence of a CPD. If there is a similar disk around HD\,95086\,b, it should be polarized, and also detectable at short wavelengths based on Mie's theory, for example by ZIMPOL and SPHERE-IRDIS in dual-polarization imaging mode \citep{vanHolstein2021_polarization_survey}. In particular, we  note the excess of flux at the shorter wavelengths (below $1.4~\mu$m), with respect to the best-fit models (within $1$--$2~\sigma$) in Fig. \ref{fig:LongWLpredictions}, which would be consistent with a debris disk. However, this CPD should be modeled with a more tuned model than the additional blackbody component used in this work to account for the observed spectrum.

Future ground-based observations with ERIS at the VLT, METIS at the ELT, and JWST in space, should soon enable us to unambiguously confirm the origin of the very red spectrum of HD\,95086\,b. This is indeed highlighted by synthetic observations of CPD from \citet[][for the ERIS instrument]{Szulagyi2019_CPD_ERIS_NACO} and \citet[][for the JWST and ELT telescopes]{ChenSzulagyi2021_CPD_JWST_ELT}.

\section{Searching for planet c \label{sec:detection_limit}}

The architecture of the young planetary system HD\,95086 offers an interesting comparison case with two emblematic systems HR\,8799 \citep{2016A&A...592A.147G,Su2015} and PDS\,70 \citep{keppler2018_PDS70}, and more generally with the interpretation that these multiple-belt debris disks are young analogs to our Solar System. The imaged giant planets would be responsible for the dynamical clearing of the debris disks and the formation of observed multiple-belt architecture, as suggested by \cite{2014MNRAS.444.3164K} and \cite{2016MNRAS.462L.116S}. For HD\,95086 the observed planet--belt architecture composed of a warm and relatively narrow inner belt at $\sim8$~au, a broad cavity from typically $10$ to $100$~au inside which  the massive ($4$--$5$~\mj, $a\sim53\pm16$\,au and $e\leq0.17$, see Table\,\ref{tab:mcmcfit_solutions}) planet HD\,95086\,b orbits,  and finally, a cold outer belt lying between $106$ to $320$~au, suggests that probably more than one giant planet is orbiting in this system \citep{Su2017,Rameau2016,Chauvin2018}. 

Applying Eqs. 4 and 5 of \cite{2016MNRAS.462L.116S} to the case of HD\,95086 ($1.6$~M$_{\star}$, $14.3\pm2$~Myr, cavity from $10$ to $100$~au), \cite{Chauvin2018} derived a minimum mass of the planets in the cavity of $0.35$~\mj~and a typical number of required planets of $2.4$ (i.e., $2$ to $3$ giant planets depending on their respective separation). Comparing these results to the outcome of HARPS and SPHERE combined detection limits (up to 2017) in the context of a planet--disk coplanar configuration, they found that there might still be room for two additional stable planets c and d in the cavity in addition to b with typical masses between $0.35$~\mj~(dynamical clearing constraint) and $6$~\mj~for a semimajor axis between $10$ and $30$~au or $0.35$~\mj~and $5$~\mj~beyond $30$~au.

Considering the most recent SPHERE multi-epoch observations (up to May 2019) and including the deepest high-contrast images obtained so far with SPHERE (in early 2018), 
we revisit here the search for planet c (and d) beyond the inner warm narrow belt at $\sim8$~au, that is beyond $\sim100$~mas for the HD\,95086 system.
We first push the current exploration of the close environment to search for c using the K-Stacker  algorithm. With no clear detection, we then set new upper limits on the potential masses of inner giant planets in the system.

\subsection{K-Stacker exploration}


The existence of HD\,95086\,c has been investigated by \cite{LeCoroller2020Kstacker} using the K-Stacker algorithm \citep{2018A&A...615A.144N}. The code was applied to a combination of six IFS observations (median over Y-H wavelengths) obtained at different epochs (from February 2015 to May 2017). Following a similar strategy, the complete set of observations reported in Table\,\ref{tab:obslog} is now considered for both IFS and IRDIS. They represent a total of ten epochs between February 2015 and May 2019. Based on the detection limits achieved with SpeCal-PCAPad and SpeCal-TLOCI in the H and K1 bands with IFS and IRDIS, respectively (see Fig. \ref{fig:contrast_ifs_irdis}), we selected the eight best ones, rejecting February 2015 and January 2016. For IFS, the choice of H band (median over the channels between $1.47$ and $1.59~\mu$m) over Y and J bands was motivated by deeper sensitivities down to masses of $2$~\mj~given the very red colors (mid-L to L/T types) expected for the detectable planets according to the SPHERE detection limits. As fixed parameters for K-Stacker, the recent GAIA-DR2 distance and the primary stellar mass derived in this work (see Sect. \ref{sec:orbit}) were used, together with an exploration range of orbital parameters for HD\,95086\,c compatible with stable dynamical orbits considering the system architecture and the presence of HD\,95086\,b (see Table \ref{parameters}). 

For both K-Stacker runs, IRDIS at the K1 band and IFS at the H band, we do not detect any clear point-source signal with a signal-to-noise ratio higher  than 5, which could indicate the probable presence of a closer-in planet. Further characterization to improve the detection limits in the $100$--$300$~mas regime using either reference differential imaging with star-hopping \citep{2021A&A...648A..26W} or molecular mapping techniques \citep{2021A&A...648A..59P} will be needed.

\begin{table}[t]
    \centering
    \caption{Orbital parameter ranges for the research of  exoplanet c. The origin of times is set at May 5, 2015.}
    \begin{small}{}
    \begin{tabular}{cccc}
    \noalign{\smallskip}\hline\hline\noalign{\smallskip}
        Orbital & Interval in K & Interval in H & Distribution \\
        Parameter & Range & Range & Type \\
    \noalign{\smallskip}\hline\hline\noalign{\smallskip}
        $a$ (au) & [10.5, 14] & [9.5, 14] & uniform\\
        $e$ & [0, 0.6] & [0, 0.6] & uniform \\
        $inc.$ (rad) & $[0, \pi]$ & $[0, \pi]$ & uniform\\
        $\Omega$ (rad) & $[-\pi, \pi]$ & $[-\pi, \pi]$ & uniform \\
        $\omega$ (rad) & $[-\pi, \pi]$ & $[-\pi, \pi]$ & uniform\\
        $t_0$ (years) & [0, 42] & [0, 42] & uniform \\
        Stellar mass (M$_\odot$) & 1.54 & 1.54 & fixed value  \\
        Star distance (pc) & 86.2 & 86.2 & fixed value \\
        \noalign{\smallskip}\hline\noalign{\smallskip}
    \end{tabular}
    \end{small}
    \label{parameters}
\end{table}

\subsection{Detection probabilities}
Without any clear detection of additional planets orbiting HD\,95086 on individual epochs or using K-Stacker and all the best epochs available, we now explore the completeness of previous HARPS and new SPHERE observations
 using the pyMESS2 code (A.-M. Lagrange, priv. comm.), a Pythonized version of the MESS2 code \citep{2017A&A...603A..54L}, together with the Exoplanet Detection Map Calculator \citep[Exo-DMC][]{Bonavita2020}\footnote{\url{https://github.com/mbonav/Exo_DMC}}. Both codes are the latest 
versions of the Multi-purpose Exoplanet Simulation System  \citep[MESS;][]{Bonavita2012}, a Monte Carlo tool for the statistical analysis of direct imaging survey results. 
In a similar fashion to its predecessors, both codes combine the information on the target stars with the instrument detection limits to estimate the probability of detection of a given synthetic planet population, ultimately generating detection probability maps.

\begin{figure*}[t]
    \centering
   \includegraphics[width=0.33\linewidth]{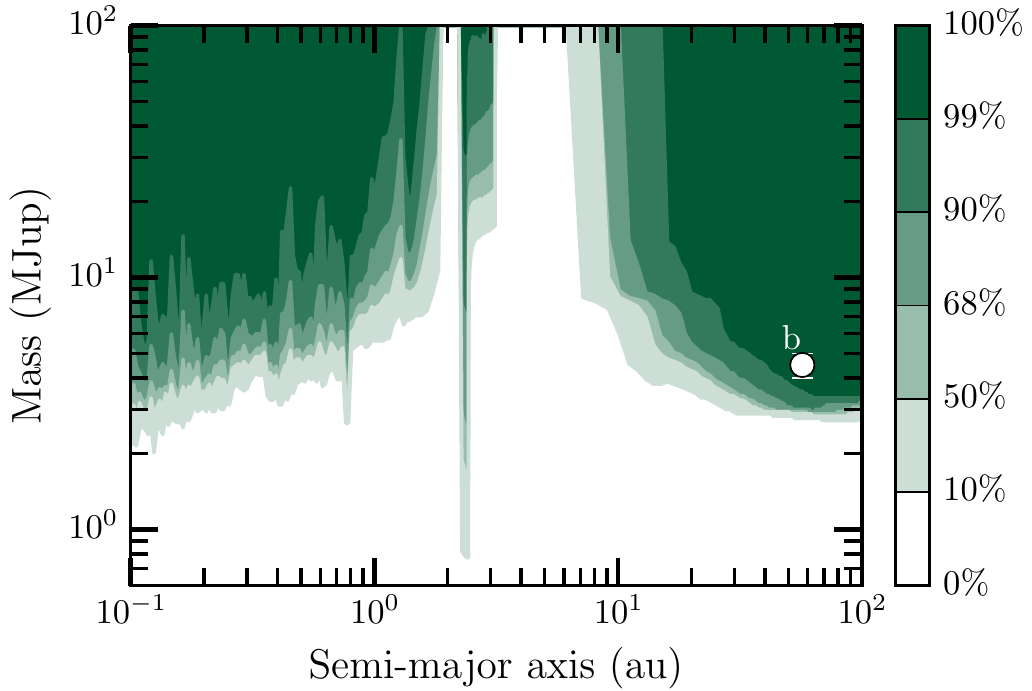}
   \includegraphics[width=0.33\linewidth]{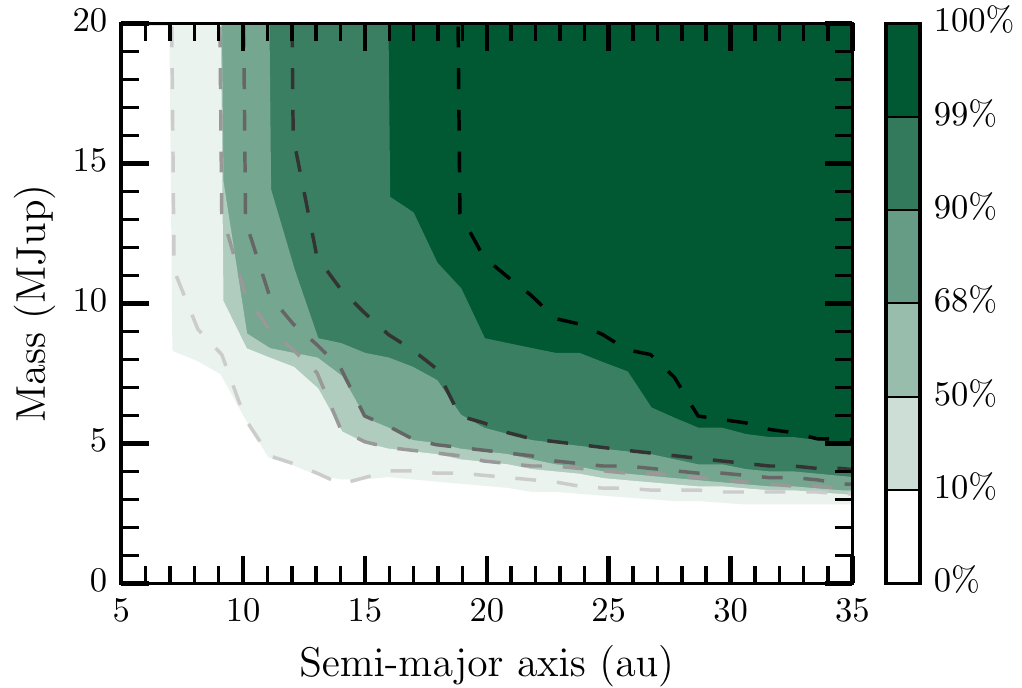} 
   \includegraphics[width=0.33\linewidth]{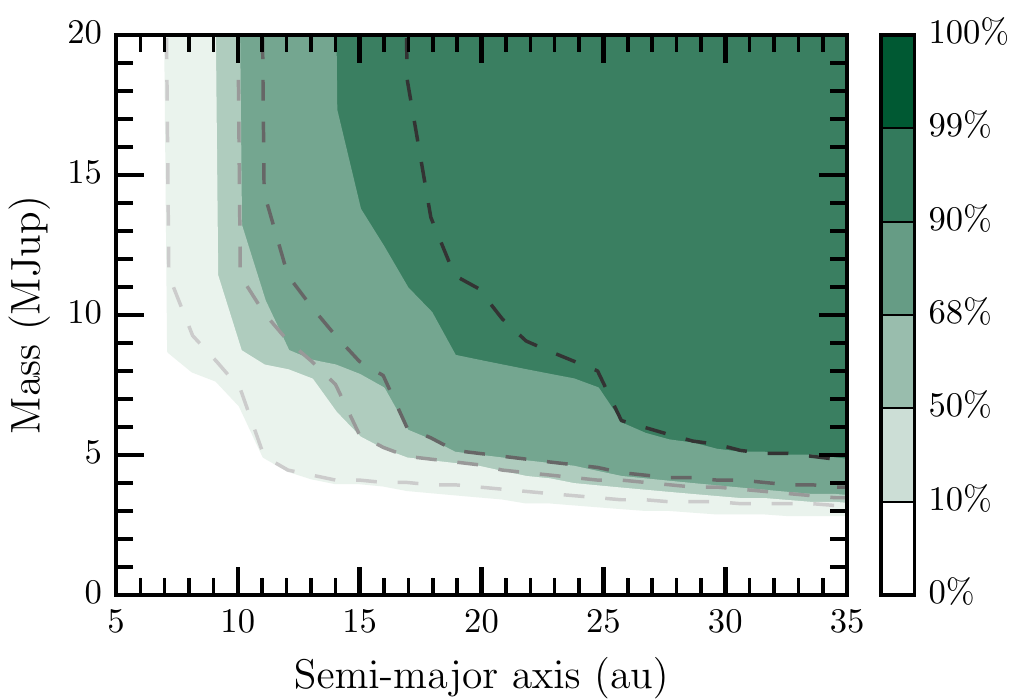}
    \caption{Probability of companion detection as a function of the distance to the star from pyMESS2. 
    On the \textit{left}, overview of the detection limit results based on radial velocity ($\leq 5$ au) and direct imaging (all SPHERE+NaCo observations, $\geq 5$ au) data for the whole HD 95086 system. In the \textit{middle} and \textit{right}, direct imaging results are zoomed between $5$ and $35$ au. In the \textit{left} and \textit{middle}, coplanarity is assumed between the researched exoplanets and the outer belt (i.e., with $i=-30\pm3\degr$ and $\Omega=97\pm3\degr$), but  not on the \textit{right}. The results without the last five epochs imaged with SPHERE (in 2018 and 2019) are shown in gray to black contours for a given probability of $10$\%, $50$\%, $68$\%, $90$\%, and if coplanarity is  assumed $99$\%. The color bars show the levels of probability detection.
    }
    \label{fig:detection_limits_mass}
\end{figure*}

For each star in the sample, they generate a grid of masses and physical separations of synthetic companions, then estimate the probability of detection given the provided detection limits at each epoch. The default setup uses a flat distribution in log space for both the mass and semimajor axis, but in a similar fashion to their predecessors, they allow for a high level of flexibility in terms of possible assumptions on the synthetic planet population to be used for the determination of the detection probability.  
For each point in the mass--semimajor axis grid, pyMESS2 and DMC generate a fixed number of sets of orbital parameters. By default all the orbital parameters are uniformly distributed except for the eccentricity, which is generated using a Gaussian eccentricity distribution with $\mu =0$ and $\sigma = 0.3$ (constraint: $e \geq  0$), following the approach by \cite{Hogg2010} \citep[see][for details]{Bonavita2013}.
This allows us to  properly take into account  the effects of projection when estimating the detection probability using the contrast limits in Fig.~\ref{fig:contrast_ifs_irdis}. They calculate the projected separations corresponding to each orbital set for all the values of the semimajor axis in the grid \citep[see][for a detailed description of the method used for the projection]{Bonavita2012}. This allows us to estimate the probability that each synthetic companion is truly   in the instrument field of view and therefore that it will be detected if the value of the mass is higher than the limit.
In this specific case, we chose to restrict the inclination and the longitude of the node of each orbital set to make sure that all companions in the population would lie in the same orbital plane as the disk (i.e., with $i=-30\pm3\degr$ and $\Omega=97\pm3\degr$) for one run, and consider no a priori information on the system's orientation for a second run. 
We also assumed a uniform distribution for the eccentricity, with a maximum value of 0.6. 

While MESS was limited in its use to direct imaging, both DMC and pyMESS2 can also be used to draw similar constraints using other kinds of datasets, including radial velocity (RV) data. Given the provided RV time series, both codes use the local power analysis (LPA) approach described by \cite{meunier2012} to estimate, for each mass and separation in the grid, for what fraction of the generated orbital sets the signal generated by the companion would be compatible with the data. DMC at this stage is limited to an independent determination of the detection probability for each technique that is  later combined. The two sets of maps are merged by considering, for each point in the grid, the best value of the probability. With pyMESS2 (and MESS2), the detection probability directly combines the detectability of each generated planet by checking if each planet can be detectable at least in one DI epoch or in the RV epochs. The probability is then derived by counting, for each [mass,a], how many of the generated planets are detected with either technique. In  the end, combining the two methods  allows a more accurate determination of the detection probability.

Figure\,\ref{fig:detection_limits_mass} (\textit{top left}) illustrates the advantages of combining HARPS radial velocity observations with direct imaging, as shown by \cite{Chauvin2018}, but here updating their results with the latest SPHERE measurements. The detection probability gain with the new SPHERE epochs is shown in the \textit{bottom left} (coplanar case) and \textit{bottom  right} (no orbital constraints) panels for the specific region of interest between 5 and 35\,au, where the presence of inner giant planets is suspected. The contrast gain of about 1.0\,mag at typically 100--200\,mas with both IFS and IRDIS presented in Fig.\,\ref{fig:contrast_ifs_irdis}, with the multi-epoch combination, enables us to nail down the detection probability map by a fraction of Jupiter mass and au. A further relevant gain would likely imply   using alternative observing strategies, such as  star-hopping \citep{2021A&A...648A..26W} or molecular mapping \citep{2018A&A...617A.144H,2021A&A...648A..59P}, before the arrival of the extremely large telescopes \citep{2018arXiv181002031C}. 

\section{Conclusion}

In this work we presented and analyzed five new observations from SPHERE ($2018$--$2019$) on the young Solar System analog HD~95086, as well as a re-analysis of the five previous SPHERE observations ($2015$--$2017$), and the  results from archival data from the  GPI and NaCo instruments. We reported an in-depth characterization of the system HD~95086:

   \begin{enumerate}
      \item Regarding the exoplanet HD~95086~b, we extracted for the first time its spectrum in the J and H bands by combining six epochs imaged with SPHERE-IFS to maximize the signal-to-noise ratio of the planet as it is hardly detectable in these bands.
      \item We constrained the physical properties of HD~95086~b  providing spectroscopic and photometric measurements in J, H, K1, K2, and L' bands from the  SPHERE-IFS, GPI, SPHERE-IRDIS, and NaCo instruments. We obtained two types of solutions: for a surface gravity $\log$($g$) between $3.3$--$5.0$, either the exoplanet seems to have a high effective temperature between $1400$--$1600$ K and a significant extinction ($\gtrsim 10$), or lower temperatures between $800$--$1200$ K with a small to medium level of extinction ($\lesssim 10$) and a super-solar metallicity. Both of these solutions reveal the presence of significant dust, which can  explain the redness of the exoplanet b. The dust could be present either in the upper layers of the atmosphere, explaining the super-solar metallicity atmosphere found for some atmospheric model best solutions, or around the exoplanet HD~95086~b with the presence of a circumplanetary disk, explaining  the high extinction found for the other atmospheric models. 
      \item Additional modeling combining at the same time models of atmospheres and circumplanetary disks confirm the possibility of the  presence of a debris circumplanetary disk around the exoplanet HD~95086~b since the solution is as likely as the other solutions found without it based on the Akaike information criterion. Considering the age of the system, the nature of this possible circumplanetary disk suggested by our modeling is more likely to  be a debris disk  than a viscous disk.  Nevertheless, future measurements sufficiently precise in the H$\alpha$ filter are required to constrain the accretion rate of HD~95086~b and rule out definitively the possibility of a viscous circumplanetary disk combined with specific atmospheric models. Other future measurements, particularly in M and N bands,  are necessary to discriminate between our remaining best-fit atmospheric models.
      \item We updated the orbital parameters for the exoplanet b, adding two additional monitoring years from SPHERE. Our best orbital solutions are consistent with previous published orbital parameters.
      \item As for additional exoplanets in the system, we pushed the detection performance by combining the best epochs using the  K-Stacker algorithm, which combines the best observations through different Keplerian motions to correct for the orbit of any additional exoplanet. We also  applied several post-processing algorithms, but  we did not find any robust candidates. 
      Nonetheless, we put new constraints on the masses and locations of putative additional exoplanets in the system, and we ruled out any other $5$\,$\mathrm{M}_\text{Jup}$ inner planet in the system located at a  distance greater than $17$\,au at a $50$\% confidence level (or $9$\,$\mathrm{M}_\text{Jup}$ inner planet at a distance greater than $10$\,au at a $50$\% confidence level).
   \end{enumerate}

Future observations with the JWST (GTO target), at the VLT/I with GRAVITY, ERIS, SPHERE and its potential upgrade, and with the first light instruments of the ELT, should enable us to understand the global architecture and origin of HD\,95086, and its commonality with our own Solar System.

\begin{acknowledgements}

We would like to thank Kate Su for sharing the ALMA continuum image used in Fig.\,1 together with the SPHERE-IRDIS high-contrast image at K1-band.
We acknowledge financial support from the Programme National de
Planétologie (PNP) and the Programme National de Physique Stellaire
(PNPS) of CNRS-INSU. 
The project is supported by CNRS, by the Agence Nationale de la
Recherche (ANR-14-CE33-0018). This work is partly based on data
products produced at the SPHERE Data Center hosted at OSUG/IPAG,
Grenoble. 
SPHERE is an instrument designed and built by a consortium consisting of IPAG
(Grenoble, France), MPIA (Heidelberg, Germany), LAM (Marseille,
France), LESIA (Paris, France), Laboratoire Lagrange (Nice, France),
INAF–Osservatorio di Padova (Italy), Observatoire de Genève
(Switzerland), ETH Zurich (Switzerland), NOVA (Netherlands), ONERA
(France) and ASTRON (Netherlands) in collaboration with ESO. SPHERE
was funded by ESO, with additional contributions from CNRS (France),
MPIA (Germany), INAF (Italy), FINES (Switzerland) and NOVA
(Netherlands). SPHERE also received funding from the European
Commission Sixth and Seventh Framework Programmes as part of the
Optical Infrared Coordination Network for Astronomy (OPTICON) under
grant number RII3-Ct-2004-001566 for FP6 (2004–2008), grant number
226604 for FP7 (2009–2012) and grant number 312430 for FP7
(2013–2016). This project has received funding from the European Research Council (ERC) under the European Union's Horizon 2020 research and innovation programme (COBREX; grant agreement n$^\circ$ 885593). V.C. acknowledges funding from the Australian Research Council via DP180104235. A.V. and G.P.P.L.O. acknowledge funding from the European Research Council (ERC) under the European Union's Horizon 2020 research and innovation programme (grant agreement No.~757561). Finally, we would like
to thank the anonymous referee and the editor for their helpful comments.
\end{acknowledgements}

\bibliographystyle{aa}



\newpage

\begin{appendix}
\onecolumn
\section{Observational conditions}

\begin{figure*}[h!]
    \centering
    \includegraphics[height=18cm]{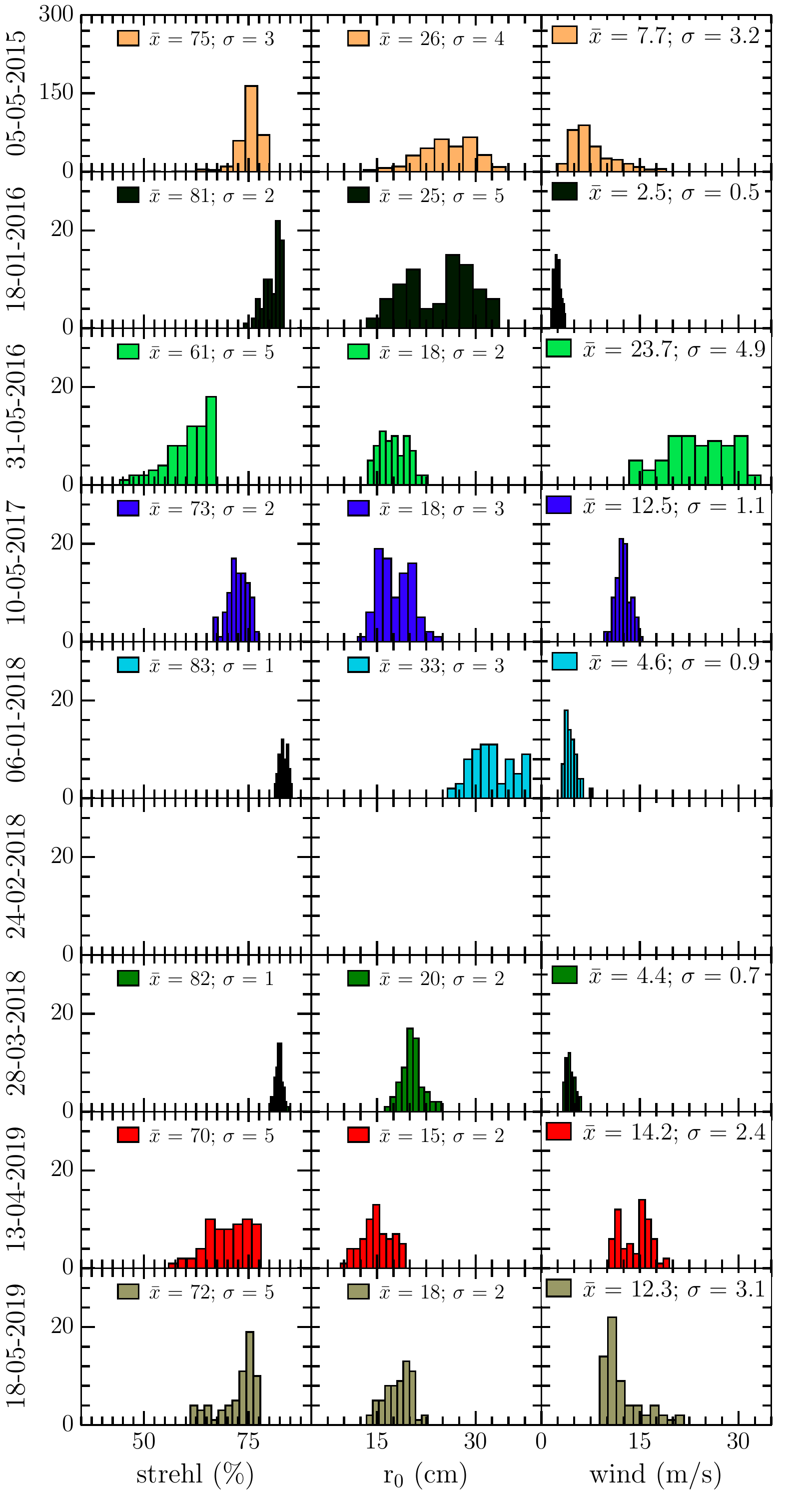} 
    \includegraphics[height=18cm]{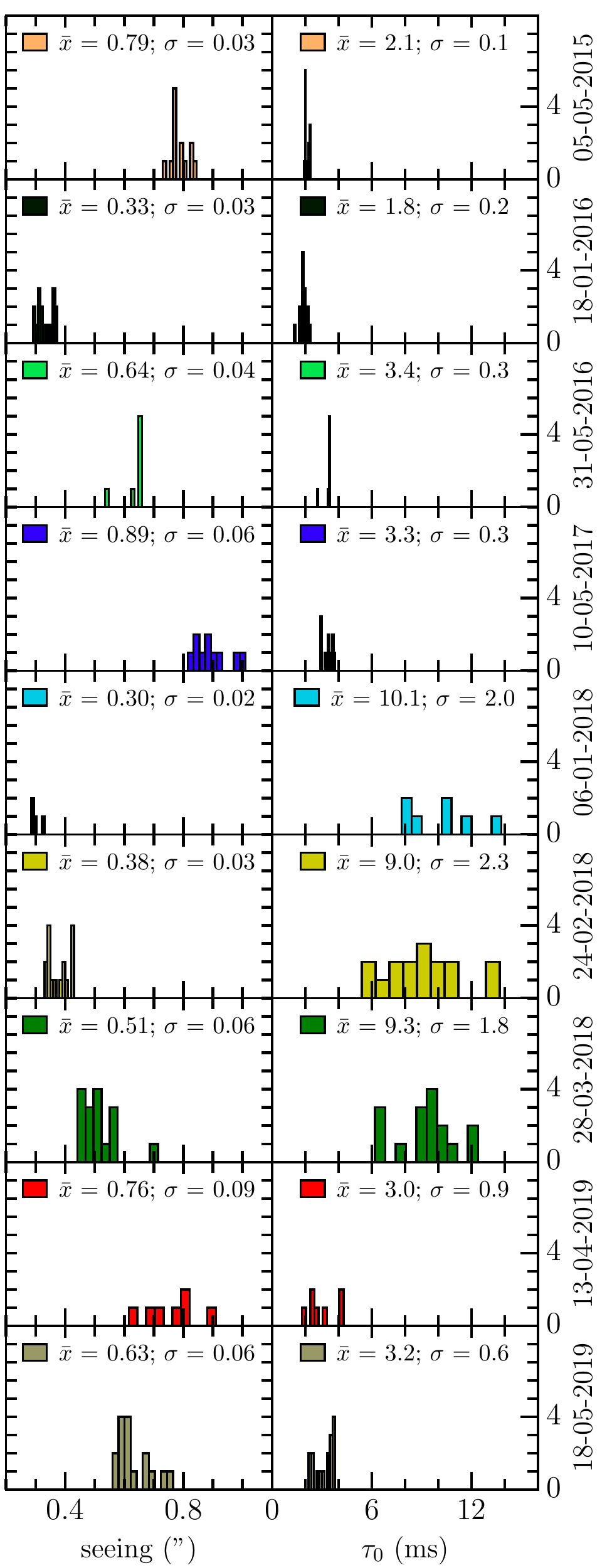}
   \caption{Distribution of the observational conditions (when available) for all the epochs: Strehl ratio, atmospheric coherence length $r_0$, wind, seeing, and atmospheric coherence time $\tau_0$. The Strehl ratio, atmospheric coherence length, and wind were all measured by SPARTA, the computer of the adaptive system of SPHERE, while both the seeing and atmospheric coherence time were measured by the DIMM telescope at Paranal. A wind~$\leq 3~\mathrm{m\cdot s^{-1}}$ indicates the presence of low wind effect \citep{Milli2017}, while a coherence time~$\leq 3~\mathrm{ms}$ indicates the presence of wind-driven halo \citep{Cantalloube2020_WDH}. 
   }
    \label{fig:histo_sparta_dimm}
\end{figure*}

Figure \ref{fig:histo_sparta_dimm} shows the observation conditions for all the available SPHERE epochs used.

\clearpage

\section{Reduced images with SpeCal}

\begin{figure*}[h]
    \centering
    \includegraphics[width=1\linewidth]{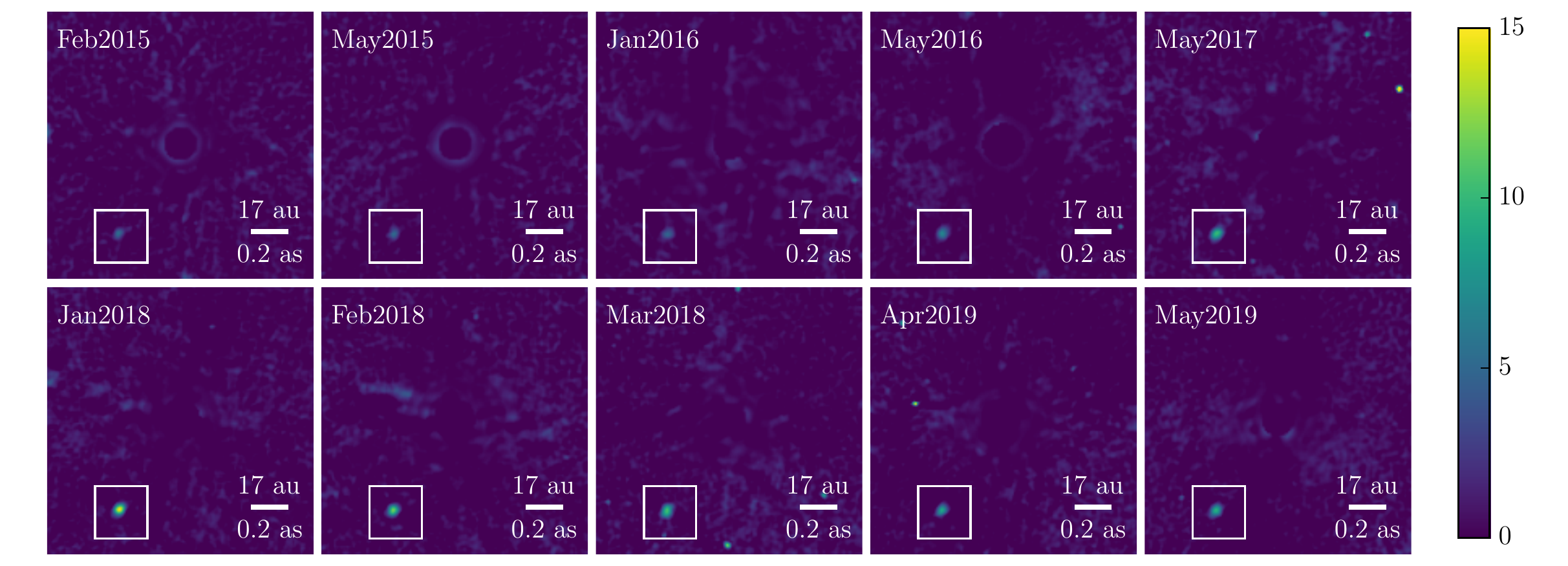} 
    \includegraphics[width=1\linewidth]{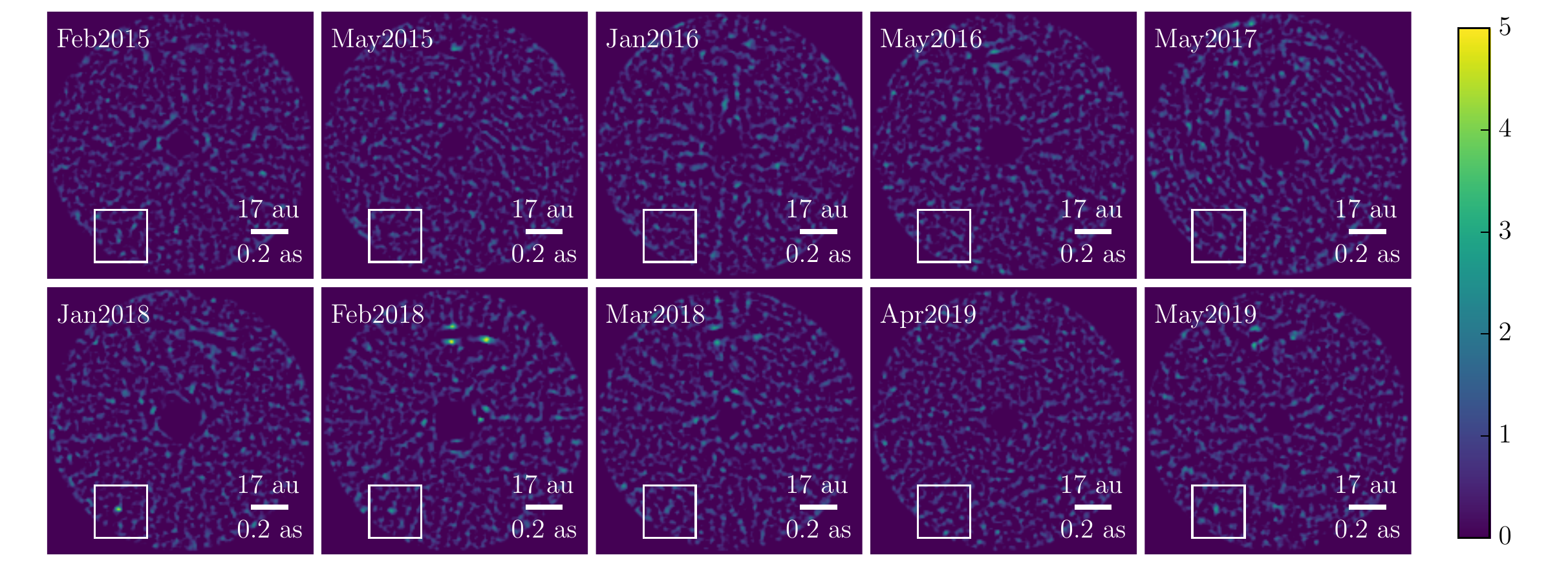}
    \caption{Signal-to-noise ratio  maps for all the SPHERE epochs (\textit{top}: IRDIS in the K1 band reduced by the pipeline SpeCal-TLOCI, and  \textit{bottom}: IFS in the YJH bands) reduced by the pipeline SpeCal-PCAPad (PCA ADI+SDI). The color bar corresponds to the signal-to-noise ratio. The region below the inner working angle of the coronagraph is masked.  \label{fig:mosaic_epochs_specal}}
\end{figure*}

Figure \ref{fig:mosaic_epochs_specal} shows the images reduced with the pipelines SpeCal-TLOCI and SpeCal-PCAPad for the IRDIS and IFS data, respectively.

\clearpage

\twocolumn

\section{Spectral correlation matrix}\label{app:SpectralCovariance}

\begin{figure}[h!]
    \centering
    \includegraphics[width=0.8\columnwidth]{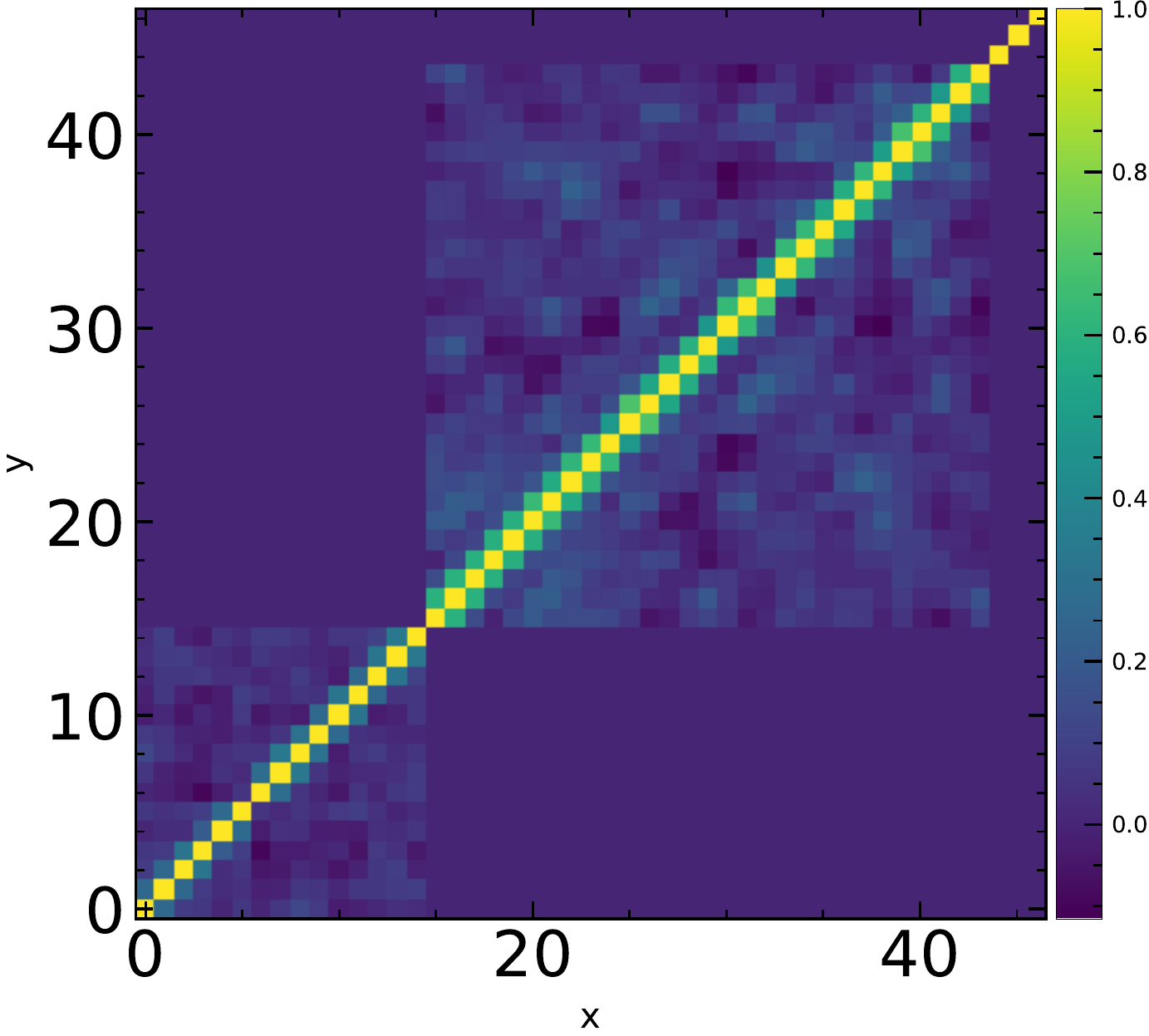}
    \caption{Spectral correlation matrix computed from our SPHERE/IFS (first 15 points) and GPI/IFS (points  15--43) data. For IRDIS and NACO, $C_{ij} = \delta_{ij}$, where $\delta_{ij}$ is the Kroenecker symbol.}
    \label{fig:covar_matrix}
\end{figure}

Figure~\ref{fig:covar_matrix} shows the spectral correlation matrix ($\phi$) empirically estimated from our IFS datasets, as in \citet{Greco2016} and \citet{Derosa2016}. The spectral covariance matrix $C$  used for spectral fits is obtained as $C_{ij} = \phi_{ij} \sigma_i \sigma_j$, where $\sigma_i$ and $\sigma_j$ correspond to the reported uncertainties in flux measurements for points $i$ and $j$.

\clearpage

\section{Corner plots of the posteriors of the atmospheric models\label{app:cornerplots_special}}

\begin{figure*}[p]
    \centering
 
     \includegraphics[width=\textwidth]{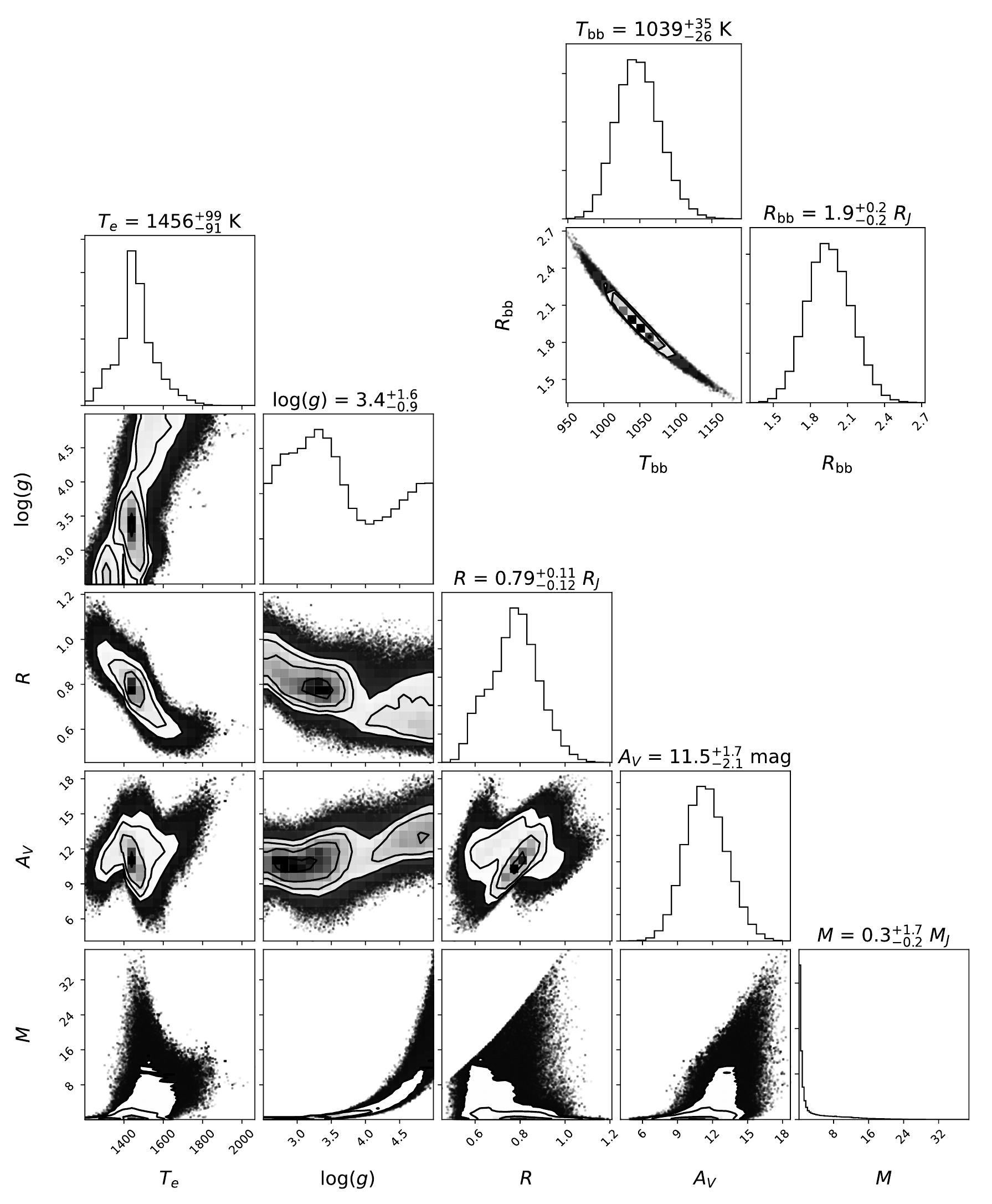}
    \caption{ Corner plots retrieved by \texttt{special} for the single-blackbody model at the \textit{top right}, BT-SETTL  with free extinction $A_V$ model at the \textit{bottom left}.  }
    \label{fig:corner_plots_BT-Settl_BB}
\end{figure*}

\begin{figure*}[p]
    \centering
     \includegraphics[width=\textwidth]{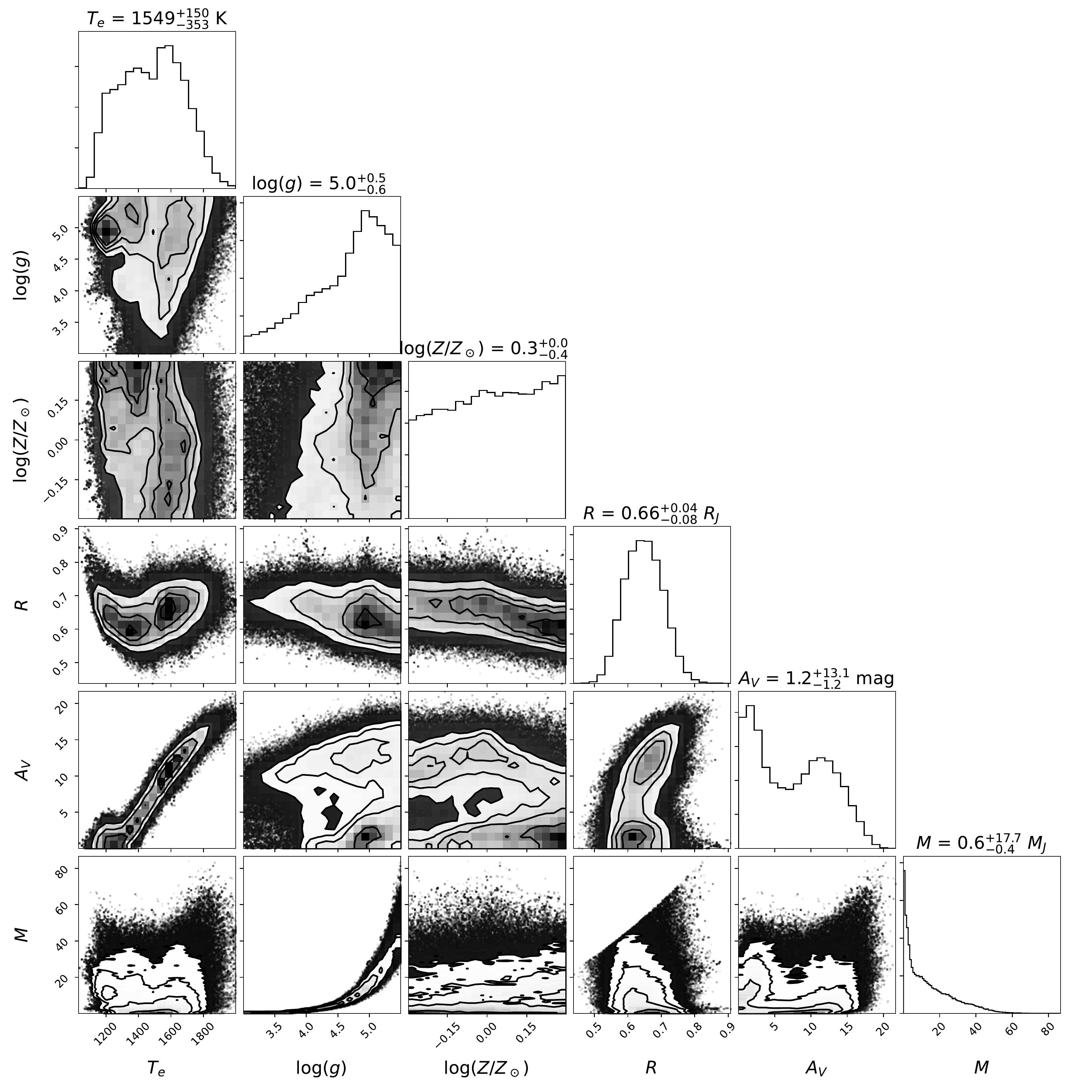}
    \caption{ Corner plot retrieved by \texttt{special} for the DRIFT-PHOENIX model with free extinction $A_V$. }
    \label{fig:corner_plots_DP}
\end{figure*}

\begin{figure*}[p]
    \centering
     \includegraphics[width=\textwidth]{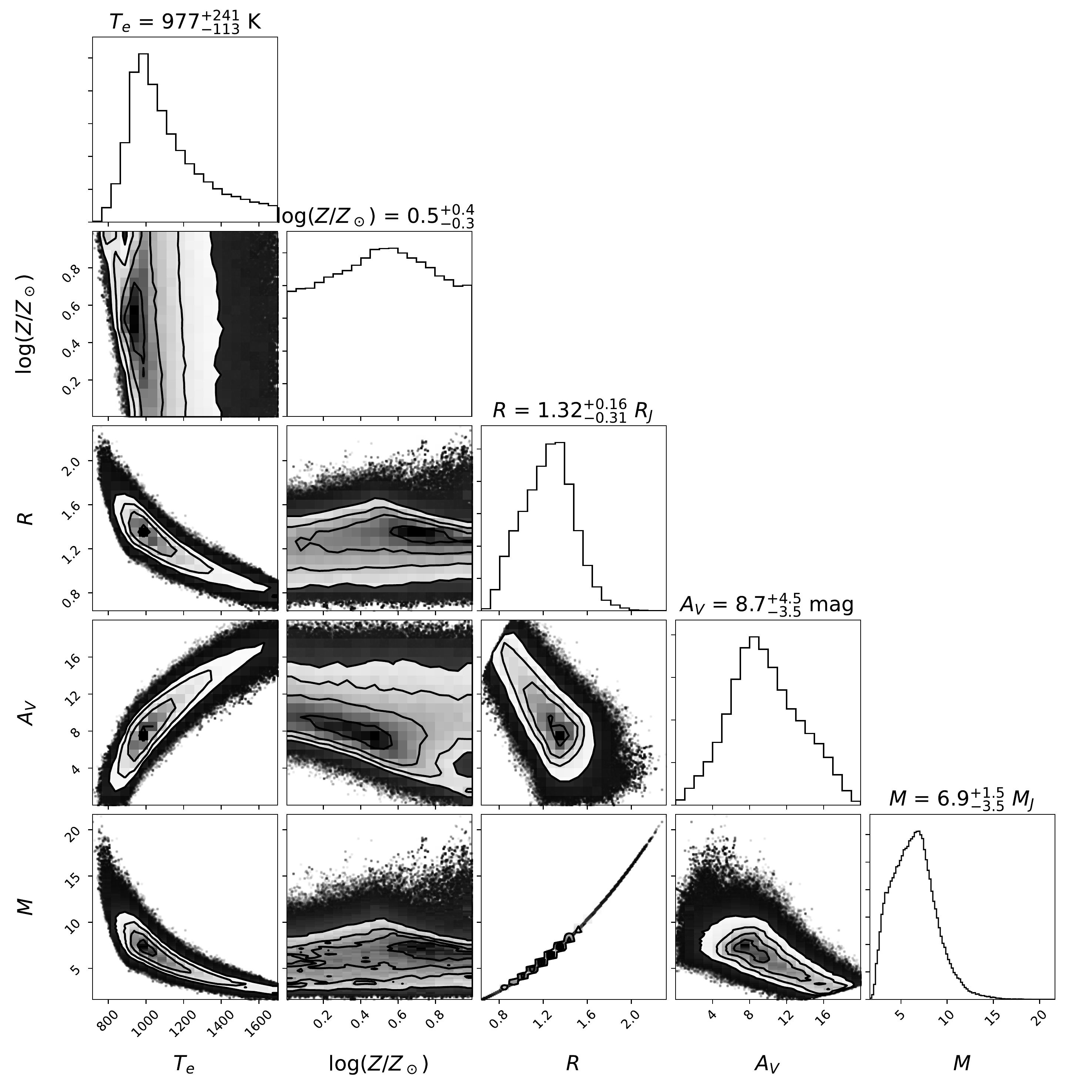}
    \caption{ Corner plot retrieved by \texttt{special} for the \citetalias{Madhusudhan2011} cloud-A model with free extinction $A_V$. }
    \label{fig:corner_plot_M11cA}
\end{figure*}

\begin{figure*}[p]
    \centering
     \includegraphics[width=\textwidth]{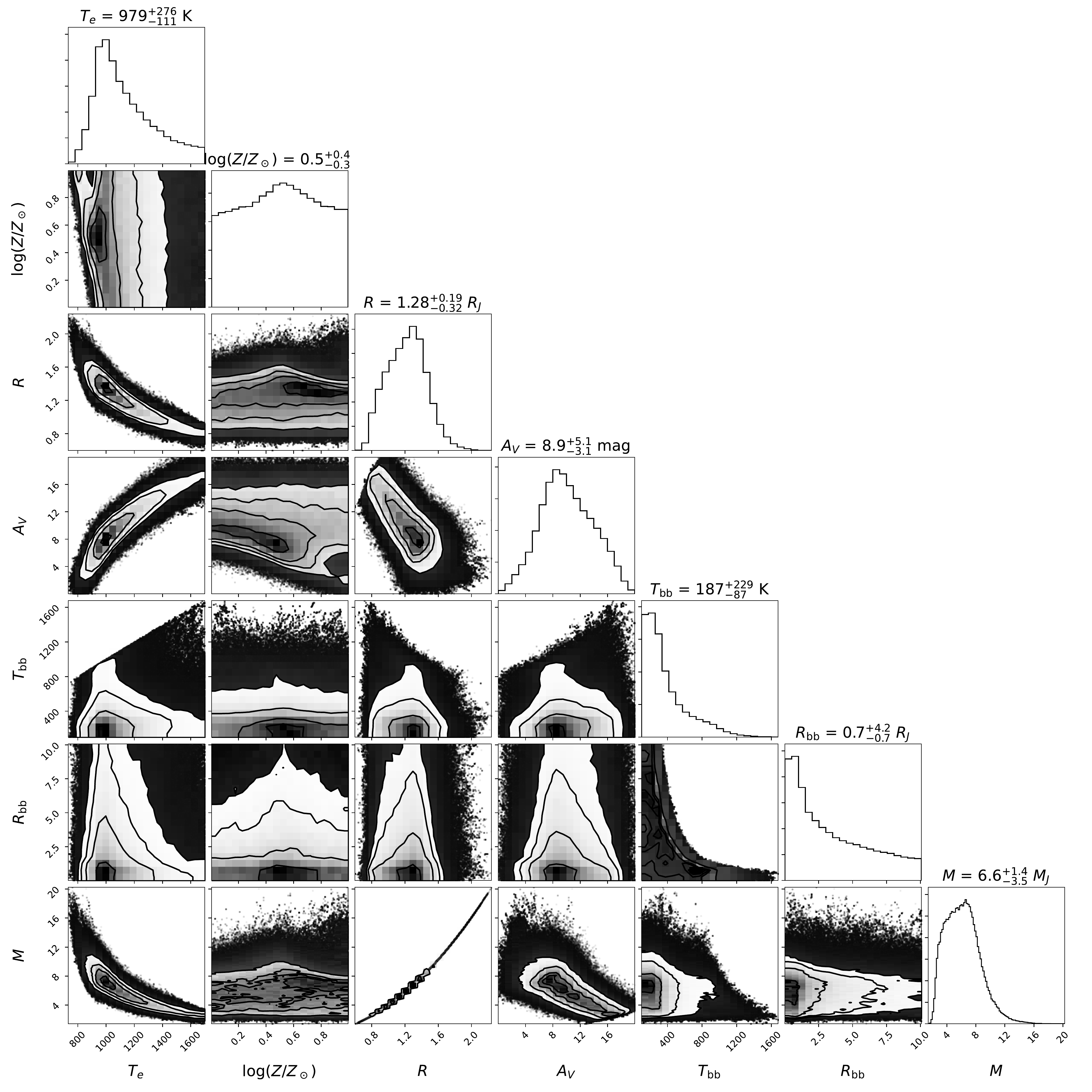}
    \caption{ Corner plot retrieved by \texttt{special} for the \citetalias{Madhusudhan2011} cloud-A model with free extinction $A_V$ and with an extra blackbody component. }
    \label{fig:corner_plot_M11cA_BB}
\end{figure*}

{ In this appendix we show the posteriors of the atmospheric models retrieved by \texttt{special} for some of the best models at reproducing the observed spectrum: the single-blackbody model and the BT-SETTL model with free extinction (both in Fig.~\ref{fig:corner_plots_BT-Settl_BB}); the DRIFT-PHOENIX model with free extinction (Fig.~\ref{fig:corner_plots_DP}); and  the \citetalias{Madhusudhan2011} cloud-A model with free extinction, without and with an extra blackbody component (Figs.~\ref{fig:corner_plot_M11cA} and \ref{fig:corner_plot_M11cA_BB}, respectively). 
In each case the mass is not a free parameter of the fit, but is calculated from the $\log(g)$ and $R_\text{p}$ posterior distributions.}

We note that in Eq. \eqref{Eq:GoodnessOfFit}, which represents the log-likelihood expression provided to the sampler, in addition to  the spectral covariance matrix, additional weighting coefficients are used. These coefficients can account for the fact that absolute flux calibration is performed instrument per instrument (i.e.,~it is not specific to each spectroscopic point), whose individual uncertainties and covariances are not properly accounted for in the fit. In the hypothetical case where the uncertainty on absolute flux calibration is the same for the spectrograph and the photometers involved in acquiring the combined spectrum, but dominates over individual statistical uncertainties, the absence of additional weights would typically pin the absolute flux calibration to that of the spectrograph,  given the significantly larger number of data points.

This absolute flux calibration uncertainty can be difficult to assess. It was noted when comparing Spitzer/IRS versus Spitzer/MIPS measurements, a  case for which the use of additional weights was suggested \citep[e.g.,][]{Ballering2013,Chen2014}. However, it is known to still be an issue when comparing measurements obtained by the latest generation of high-contrast imagers. The amplitude of the systematic bias between the overlapping SPHERE and GPI spectra of HD\,206893\,B is on the same order as the total estimated (statistical+systematic) uncertainties \citep[][]{Kammerer2021}. This kind of bias has also been noted when comparing the spectrum of PDS 70 b obtained by SINFONI and SPHERE \citep{Christiaens2019_PDS70}.

Assigning weights such that their sum is 1 for all measurements obtained by a given instrument (i.e., 1 per photometric measurement and 1/N\_spec for each spectrograph measurement, where N\_spec is the number of channels) would correspond to assigning the same level of confidence for the absolute flux calibration of each instrument. However, this would too severely undermine the significance of the spectroscopic data in the fit. 
It would indeed make all measurements of a given spectrograph as significant as a single photometric measurement, while spectrographs provide more than just an absolute flux measurement;  they also provide information about the shape of the spectrum.

Therefore, as a compromise we opted for weights that are proportional to the spectral bandwidth, as in \citet{Ballering2013} and \citet{Olofsson2016}.
Some estimated model parameters are dependent on the absolute flux calibration (e.g.,~photometric radius), on the features, and/or shape of the spectrum (e.g.,~surface gravity), or on both together (effective temperature and extinction). Therefore, a certain choice of weights may in principle change the estimated best-fit parameters. 

\begin{figure*}[h!]
    \centering
     \includegraphics[width=\textwidth]{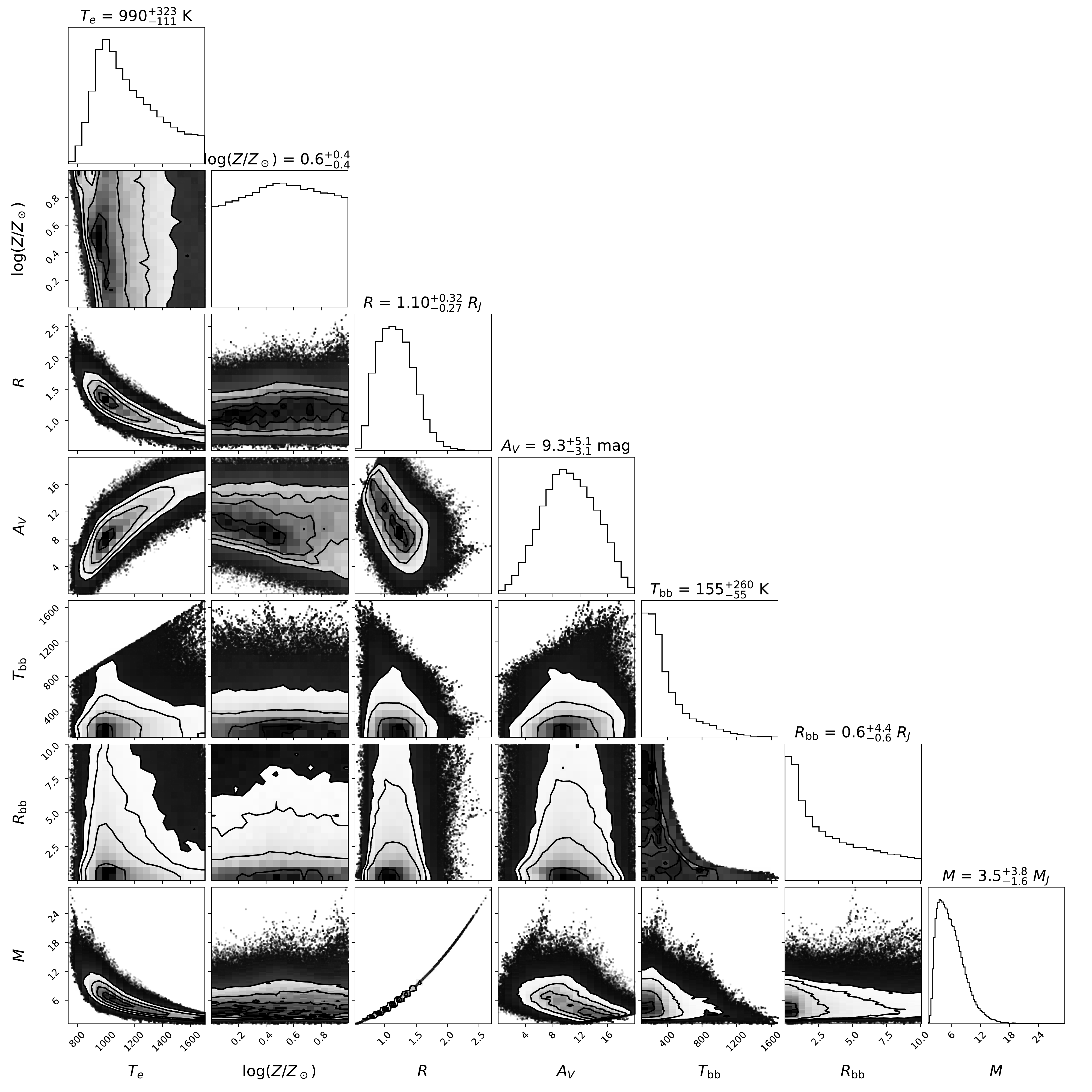}
    \caption{ Corner plot retrieved by \texttt{special} for the \citetalias{Madhusudhan2011} cloud-A model with free extinction $A_V$ and with an extra blackbody component without additional weighting coefficients. }
    \label{fig:corner_plot_M11cA_BB_no_weight}
\end{figure*}

To test in practice the effect of the additional weighting coefficients on the fits to the HD 95086 b spectrum, we show in Figs.~\ref{fig:corner_plot_M11cA_BB} and \ref{fig:corner_plot_M11cA_BB_no_weight} the best-fit solutions obtained with and without the additional weighting coefficients, respectively. The best-fit parameters show only minor differences, which may have been expected given the good match between most near-IR spectroscopic and photometric points of HD 95086 b (Fig. \ref{fig:spectrum_complete}). 
The largest relative differences correspond to the estimated radius and mass of the planet, although all parameters still remain within the  1$\sigma$ error bars of the estimates obtained with and without weights.

\end{appendix}

\end{document}